\documentclass[a4paper,reqno]{amsart}

\usepackage[T1]{fontenc}

\usepackage[
  colorlinks,
  linkcolor = blue,
  citecolor = blue,
  urlcolor = blue]{hyperref}

\usepackage{amsmath,amsthm,amssymb,amsfonts}

\usepackage{thmtools}

\usepackage{cleveref}

\newtheorem*{theorem*}{Theorem}
\newtheorem{theorem}{Theorem}[section] 
\newtheorem{proposition}[theorem]{Proposition} 
\newtheorem{lemma}[theorem]{Lemma}
\newtheorem{definition}[theorem]{Definition}
\newtheorem{corollary}[theorem]{Corollary}
\newtheorem{remark}[theorem]{Remark}
\newtheorem{example}[theorem]{Example}

\crefname{lemma}{Lemma}{Lemmas}
\crefname{definition}{Definition}{Definitions}
\crefname{theorem}{Theorem}{Theorems}
\crefname{conjecture}{Conjecture}{Conjectures}
\crefname{section}{Section}{Sections}
\crefname{claim}{Claim}{Claims}
\crefname{appendix}{Appendix}{Appendices}
\crefname{figure}{Fig.}{Figs.}
\crefname{table}{Table}{Tables}
\crefname{proposition}{Proposition}{Propositions}
\crefname{corollary}{Corollary}{Corollaries}
\crefname{example}{Example}{Examples}
\crefname{remark}{Remark}{Remarks}

\usepackage[shortlabels]{enumitem}

\usepackage{makecell}
\usepackage{multirow}
\usepackage[table,xcdraw]{xcolor}
\usepackage{hhline}
\usepackage{diagbox}

\usepackage{caption}

\usepackage[group-separator={,},group-minimum-digits=4]{siunitx}

\usepackage{mathtools}

\DeclarePairedDelimiter{\set}{\lbrace}{\rbrace}
\DeclarePairedDelimiter{\abs}{\lvert}{\rvert}

\DeclarePairedDelimiter{\floor}{\lfloor}{\rfloor}

\DeclarePairedDelimiter{\of}{\lparen}{\rparen}
\DeclarePairedDelimiter{\sof}{\lbrack}{\rbrack}
\DeclarePairedDelimiter{\ip}{\langle}{\rangle}

\newcommand{\defeq}{\vcentcolon=}
\newcommand{\eqdef}{=\vcentcolon}
\renewcommand{\leq}{\leqslant}
\renewcommand{\geq}{\geqslant}

\newcommand{\bra}[1]{\langle{#1}\rvert}
\newcommand{\ket}[1]{\lvert{#1}\rangle}

\newcommand{\ketbra}[2]{\ket{#1}\bra{#2}}
\newcommand{\proj}[1]{\ketbra{#1}{#1}}

\newcommand{\mx}[1]{\begin{pmatrix}#1\end{pmatrix}}

\newcommand{\ct}{^{\dagger}}
\newcommand{\tp}{^{\mathsf{T}}}
\newcommand{\ptp}{^{\mathsf{\Gamma}}}

\newcommand{\+}{\oplus}
\newcommand{\x}{\otimes}
\newcommand{\xp}[1]{^{\otimes #1}}
\newcommand{\ctxp}[1]{^{\dagger\otimes #1}}

\newcommand{\Vin}{V_\text{in}}
\newcommand{\Vout}{V_\text{out}}

\newcommand{\C}{\mathbb{C}} 
\newcommand{\R}{\mathbb{R}} 
\newcommand{\T}{\mathrm{T}} 
\newcommand{\size}{\mathrm{size}} 
\newcommand{\cont}{\mathrm{cont}} 
\newcommand{\len}{\mathrm{len}} 
\newcommand{\B}{\mathcal{B}} 
\newcommand{\A}{\mathcal{A}} 
\newcommand{\CA}{\mathcal{U}} 
\newcommand{\Z}{\mathcal{Z}} 
\newcommand{\X}{\mathcal{X}} 

\newcommand{\I}{\mathcal{I}} 
\renewcommand{\O}{\mathcal{O}} 
\renewcommand{\P}{\mathcal{P}} 
\newcommand{\F}{\mathcal{F}} 

\newcommand{\Irr}{\operatorname{Irr}} 
\newcommand{\Paths}{\operatorname{Paths}} 
\newcommand{\Res}{\operatorname{Res}} 
\newcommand{\loops}{\operatorname{loops}} 
\newcommand{\0}{\varnothing}

\let\S\relax
\DeclareMathOperator{\S}{S} 
\DeclareMathOperator{\CS}{\mathbb{C}S} 
\DeclareMathOperator{\U}{U} 
\DeclareMathOperator{\D}{D} 
\DeclareMathOperator{\GL}{GL} 
\DeclareMathOperator{\End}{End} 
\DeclareMathOperator{\Tr}{Tr} 
\DeclareMathOperator{\spn}{span} 
\DeclareMathOperator{\diag}{diag} 

\newcommand{\SWAP}{\mathrm{SWAP}}

\newcommand{\pt}{\mathbin{\vdash}} 

\newcommand{\w}{0.5cm}

\newcommand{\bx}[3]{
  \draw[fill = white] #3 (#1*\w-\w/2,-#2*\w-\w/2) rectangle (#1*\w+\w/2,-#2*\w+\w/2);
}

\newcommand{\yd}[2][0.4]{%
  \begin{tikzpicture}[scale = #1, baseline={([yshift=-0.6ex]current bounding box.center)}]
    \foreach \li [count = \y] in {#2} {
      \foreach \x in {1,...,\li} {
        \bx{\x}{\y}{}
      }
    }
  \end{tikzpicture}%
}

\usepackage{epigraph} 

\usepackage[backend=biber,style=alphabetic,maxnames=9,maxalphanames=3,isbn=false,backref,backrefstyle=two]{biblatex}
\usepackage{tikz}
\usetikzlibrary{backgrounds}

\usepackage[foot]{amsaddr}

\title{Linear programming with unitary-equivariant constraints}

\author{Dmitry Grinko\textsuperscript{1}\hspace{-.25em}}
\address{\textsuperscript{1}Institute for Logic, Language, and Computation, University of Amsterdam and QuSoft, Amsterdam, The Netherlands}
\email{dmitry.grinko@cwi.nl}

\author{Maris Ozols\textsuperscript{1,2}\hspace{-.25em}}
\address{\textsuperscript{2}Korteweg-de Vries Institute for Mathematics and Institute for Theoretical Physics, University of Amsterdam, The Netherlands}
\email{marozols@gmail.com}

\thanks{This research was supported by an NWO Vidi grant (Project No.~VI.Vidi.192.109).}

\begin{document}

\begin{abstract}
Unitary equivariance is a natural symmetry that occurs in many contexts in physics and mathematics. Optimization problems with such symmetry can often be formulated as semidefinite programs for a $d^{p+q}$-dimensional matrix variable that commutes with $U\xp{p} \x \bar{U}\xp{q}$, for all $U \in \U(d)$.
Solving such problems naively can be prohibitively expensive even if $p+q$ is small but the local dimension $d$ is large.
We show that, under additional symmetry assumptions, this problem reduces to a linear program that can be solved in time that does not scale in $d$, and we provide a general framework to execute this reduction under different types of symmetries.
The key ingredient of our method is a compact parametrization of the solution space by linear combinations of walled Brauer algebra diagrams.
This parametrization requires the idempotents of a Gelfand--Tsetlin basis, which we obtain by adapting a general method \cite{doty2019canonical} inspired by the Okounkov--Vershik approach.
To illustrate potential applications of our framework, we use several examples from quantum information:
deciding the principal eigenvalue of a quantum state,
quantum majority vote, asymmetric cloning and transformation of a black-box unitary.
We also outline a possible route for extending our method to general unitary-equivariant semidefinite programs.
\end{abstract}

\maketitle

\tableofcontents

\setlength{\epigraphwidth}{6cm}
\epigraph{It is only slightly overstating the case to say that physics is the study of symmetry.}{\textit{Philip W.~Anderson} \cite{Anderson}}

\section{Introduction}\label{sec:Introduction}

Symmetry plays a fundamental role in physics, particularly in quantum mechanics and particle physics,
where thanks to Noether's theorem symmetry is responsible for conservation laws.
At the same time, symmetry is also the focus of a major area of mathematics---group theory,
whose applications range from physics to biology, chemistry and the arts.
``Symmetry argument'' is a common problem solving technique both in physics and mathematics:
observing the symmetries inherent to a problem often allows simplifying the problem.
In the absence of symmetry, difficult problems can often be made more tractable by assuming some form of symmetry.
Generally speaking, the more symmetry a problem has the more tractable it is.

\subsection{Schur--Weyl duality}\label{sec:SW duality intro}

The unitary group of symmetries plays a special role
in the context of quantum mechanics and quantum information,
and, for systems consisting of many identical parts,
the unitary symmetry becomes intertwined with permutational symmetry.
This intimate connection between unitary and permutation groups
acting on several identical quantum systems
is expressed via the so-called \emph{Schur--Weyl duality}.
In its simplest form, it states that the two-qubit \emph{singlet state}
$\ket{\psi^-} \defeq (\ket{01} - \ket{10}) / \sqrt{2}$
is the unique (up to a global phase) state
that is invariant under identical local unitary rotations:
$(U \x U) \ket{\psi^-} \sim \ket{\psi^-}$
for any $U \in \U(2)$,
as well as the unique anti-symmetric state:
$\SWAP \ket{\psi^-} = -\ket{\psi^-}$.
A similar dual characterization in terms of unitary and permutational symmetries
applies not only to $\ket{\psi^-}$ but also its orthogonal complement,
allowing to decompose the whole two-qubit space $\C^2 \x \C^2$ into invariant subspaces.
This duality extends also to $(\C^d)\xp{p}$ where each of the $p$ systems has dimension $d$.

\subsection{Applications of Schur--Weyl duality in quantum computing}

A natural situation where Schur--Weyl duality arises in quantum computing is when dealing with many copies of some quantum state $\rho$.
The total state $\rho\xp{p}$ is then invariant under permutations and
transforms in a straightforward way under simultaneous unitary basis change on each of the $p$ systems.
This scenario is very common in quantum information theory
where Schur--Weyl duality has become an important tool \cite{HarrowThesis,Botero}.
Its algorithmic manifestations,
weak Schur sampling \cite{SchurSampling}
and quantum Schur transform
\cite{HarrowThesis,bch2006quantumschur,kirby2018practical,krovi2019quantumschur},
play a similarly important role in quantum algorithm design \cite{WrightThesis}.
Specific quantum algorithmic tasks where Schur--Weyl duality is used include
quantum spectrum \cite{SpectrumEstimation}
and entropy \cite{EntropyEstimation} estimation,
quantum spectrum testing \cite{o2015quantum},
state tomography
\cite{keyl2006quantumstateestimation,
haah2017optimaltomography,
o2016efficient,o2016efficient2},
and quantum majority vote \cite{Majority}.

\subsection{Group-covariant quantum channels}

\newcommand{\rep}[1]{\phi_\text{#1}} 

Transformations that preserve symmetry are common in quantum computing.
A quantum channel $\Phi$ is called \emph{$G$-covariant}, for some group $G$,
if there exist two unitary representations of $G$,
$\rep{in}$ and $\rep{out}$, such that
$\Phi\of[\big]{\rep{in}(g) \, \rho \, \rep{in}(g)\ct}
= \rep{out}(g) \, \Phi(\rho) \, \rep{out}(g)\ct$,
for any group element $g \in G$ and state $\rho$.
The structure of group-covariant quantum channels can be much simpler
than the structure of general channels
\cite{mozrzymas2017structure}.
For example, while perfect universal programming of general quantum channels is impossible
\cite{NoProgramming},
covariant channels can be programmed
\cite{Programmability},
even in infinite dimensions
\cite{gschwendtner2021infiniteprogrammability}.

\subsection{Group covariance in different contexts}

Group covariance is important in many contexts.
Let us briefly illustrate two that are less obvious: quantum error correction and machine learning.

\subsubsection{Group covariance in quantum error correction}

Group covariance is particularly important in the context of quantum error correction and fault tolerance.
Many quantum error correcting codes are covariant with respect to the Clifford group and thus allow for simple or so-called ``transversal'' implementation of Clifford gates.
An even higher degree of symmetry, namely possessing a universal set of transversal gates, would be ideal for devising schemes that can manipulate encoded quantum data.
However, codes with such continuous symmetries are ruled out by the well-known Eastin--Knill theorem \cite{EastinKnill}.
The interplay between continuous symmetries and quantum error correction has received revived attention in the context of holography and quantum gravity where an approximate version of the Eastin--Knill theorem was recently established \cite{SymmetriesInQEC}.
Group-covariant quantum codes with continuous symmetries are also closely related to the notion of a quantum reference frame \cite{CovariantQEC,UniversalQEC}.

\subsubsection{Equivariance in machine learning}

A special case of group covariance is \emph{equivariance}, which means that the representations $\rep{in}$ and $\rep{out}$ are either identical or related to each other in some simple way.
Intuitively, equivariance says that applying some transformation on the input is equivalent to applying the same transformation on the output.
This is a natural condition that occurs in many contexts.
For example, in machine learning, the structure of neural networks should respect the symmetries of the problem at hand, such as translations and rotations when dealing with images.
Such group-equivariant neural networks can have substantially increased expressive capacity without the need to increase their number of parameters \cite{CohenThesis,CohenWelling,bronstein2021geometric}.
In particular, unitary-equivariant neural networks that capture symmetries of many-body quantum systems have recently found applications to quantum chemistry \cite{qiao2021unite}.
More generally, coordinate-independent convolutional networks on Riemannian manifolds require equivariance under local gauge transformations \cite{weiler2021coordinate}.

In quantum machine learning, group-equivariant convolutional quantum circuits have been proposed for speeding up learning of quantum states \cite{zheng2021speeding}.
In general, equivariant gatesets can be used to exploit symmetry in variational quantum algorithms \cite{SymmetryQML}.
A general framework for group-invariant and equivariant quantum machine learning was recently outlined in \cite{larocca2022group}.

\subsection{Local unitary equivariance}\label{sec:equivariance intro}

A particularly natural special case of group covariance is \emph{local unitary equivariance}
which corresponds to the case when $\Phi$ is a quantum channel from $p$ to $q$ systems,
each of dimension $d$,
the symmetry group $G$ is the full unitary group $\U(d)$,
and the two unitary representations are given by
$\phi_p(U) \defeq U\xp{p}$ and
$\phi_q(U) \defeq U\xp{q}$.
In other words, applying the same unitary $U \in \U(d)$
on each of the $p$ input systems of $\Phi$
is equivalent to applying $U$ on each of the $q$ output systems.
This is concisely captured by the condition that the Choi matrix $X^\Phi$ of $\Phi$ obeys the symmetry
$\sof[\big]{
  X^\Phi,
  U\xp{p} \x \bar{U}\xp{q}
} = 0$,
for all $U \in \U(d)$.

The classic Schur--Weyl duality mentioned in \cref{sec:SW duality intro} admits various generalizations \cite{Berg,MarvianSpekkens,Benkart,bchlls,doty2007new}.
In particular, the above setting is sometimes referred to as \emph{mixed Schur--Weyl duality} since $X^\Phi$ can be thought of as a ``mixed'' $(p+q)$-tensor with two types of indices \cite{halverson1996characters,nikitin2007centralizer}.
The regular Schur--Weyl duality is then the special case when either $p = 0$ or $q = 0$, namely when there are either no input systems (such as in state preparation) or no output systems (such as in quantum measurement).
We are interested in the general case of arbitrary $p$, $q$ and local dimension $d$.

The most common scenario is when either $p = 1$ or $q = 1$.
Such symmetry naturally occurs in many quantum tasks
with a single input or a single output system, such as
asymmetric quantum cloning
\cite{AsymmetricCloningCerf,AsymmetricCloning},
port-based teleportation
\cite{IshizakaHiroshima,mozrzymas2018optimal,studzinski2017port,leditzky2020optimality,christandl2021asymptotic,studzinski2021degradation,MinimalPortBased},
quantum majority vote \cite{Majority},
or $\U(d)$-covariant quantum error correction
\cite{kong2021chargeconserving,kong2021nearoptimal}.
It also occurs in situations that involve a partial transpose on a single system,
such as in entanglement detection
\cite{EggelingWerner,collins2018family,huber2021positive}.

Symmetries with general values of $p > 1$ and $q > 1$
correspond to scenarios with multiple input and output systems \cite[Section~7]{Keyl}, such as
quantum state purification \cite{Purification}
and cloning \cite{Cloning,Cloning2},
multi-port-based teleportation
\cite{kopszak2020multiport,studzinski2020efficient,mozrzymas2021optimal}.
Such symmetries also occur in situations that involve the partial transpose on several systems, such as in
extendability problem
\cite{JohnsonViola,jakab2022extendibility},
entanglement detection
\cite{EntanglementDetection,balanzójuandó2021positive},
or universality of qudit gate sets
\cite{sawicki2021check,dulian2022matrix,słowik2022calculable}.
Universality of quantum circuits with two-local $\U(d)$-equivariant gates has recently been considered in \cite{Marivan,Circuits,Circuits2} from the perspective of conservation laws.
Finally, this class of symmetries is of independent interest also in
high-energy physics \cite{Branes,Candu}
and the study of quantum spin systems \cite{RyanPhD,HeisenbergModels}.

\subsection{Optimization subject to local unitary equivariance}\label{sec:intro optimization}

Taking problem's symmetry into account is a good idea for almost any problem,
including optimization problems, since this can significantly reduce the number of parameters.
In particular, this is  the case in semidefinite optimization \cite{invariantSDPs,RepLAB}.
Given the wide range of problems in quantum information with local unitary equivariance symmetry,
the main focus of our work is on linear and semidefinite optimization
under a local $(p,q)$-unitary equivariance constraint.
Typically this means optimizing over unitary-equivariant quantum channels or other tensors with this symmetry.

An early example of using $U \x \bar{U}$ symmetry to reduce a semidefinite optimization problem to a linear one is the work of Rains \cite{Rains} on entanglement distillation under completely positive-partial-transpose preserving operations.
He characterizes the optimal distillation fidelity by a semidefinite program and then exploits the symmetry
\begin{equation}
    (U \x \bar{U}) \sum_i \ket{i} \x \ket{i} = \sum_i \ket{i} \x \ket{i}
    \label{eq:UbarU}
\end{equation}
of the canonical maximally entangled state to reduce this to a linear program
(see \cref{ex:pq1d2} for more details).
This work has inspired a long sequence of results \cite{AudenaertPlenioEisert,LeungMatthews,XinPhD,XinWilde}.
However, generalizing them to PPT-extendible channels \cite{HoldsworthVishalWilde} requires taking advantage of more complicated symmetries.

The closely related $U \x U$ symmetry appears in the so-called \emph{quantum max cut} problem that has recently received significant attention
\cite{QMC-Anshu,QMC-UniqueGames,QMC-Lasserre,QMC-Streaming,QMC-ParekhThompson,QMC-Lee,QMC-King}.
This problem is concerned with approximating the ground state and ground energy of a local Hamiltonian on a graph whose vertices are qubits and edges are assigned the projector $\proj{\psi^-}$ onto the singlet state
$\ket{\psi^-} \defeq (\ket{01} - \ket{10}) / \sqrt{2}$
(see \cref{sec:SW duality intro}).
The $U \x U$ symmetry of $\ket{\psi^-}$ is one of the basic cases captured by our framework (see \cref{ex:p2d2}), while the general case of $U\xp{p}$ is captured by Schur--Weyl duality (see \cref{sec:Schur-Weyl}).

Other instances of semidefinite optimization problems with $U\xp{p} \x \bar{U}\xp{q}$ symmetry appearing in quantum computing are
quantum majority vote and basis-independent evaluation of Boolean functions
\cite{Majority},
black-box transformations of quantum gates
\cite{quintino2019reversing,quintino2019probabilistic,quintino2021deterministic,yoshida2021universal,yoshida2022reversing,ComplexConjugation},
asymmetric cloning
\cite{AsymmetricCloning},
and entanglement witnesses
\cite{huber2021dimensionfree}.

Previously each of these problems has been approached individually and by ad hoc methods that work only for restricted choices of $p,q,d$, such as $p = 1$, $q = 1$, or $d = 2$.
Our goal is to provide a general framework for solving unitary-equivariant semidefinite optimization problems for a wide range of values of $p,q,d$.
More specifically, our aim is to answer the following two questions:
\begin{enumerate}
    \item \textit{How to efficiently eliminate the irrelevant degrees of freedom from a $\U(d)$-equivariant optimization problem?}
    \item \textit{Can a $\U(d)$-equivariant optimization problem be solved in time that does not scale in $d$?}
\end{enumerate}
We provide answers to these questions using representation-theoretic and diagrammatic methods.

\subsection{Summary of our main result}

\newcommand{\ftr}{m_1}
\newcommand{\ptr}{m_2}

Consider the following general semidefinite program (SDP) for a Hermitian matrix variable $X$:
\begin{equation}
    \begin{aligned}
        \max_X \quad & \Tr(CX) \\
        \textrm{s.t.} \quad
        & \Tr(A_k X) \leq b_k, & \forall k &\in [\ftr], \\
        & \Tr_{S_k} (X) = D_k, & \forall k &\in [\ptr], \\
        &   \sof[\big]{X, U\xp{p} \x \bar{U}\xp{q}} = 0, & \forall U &\in \U(\C^d), \\
        & X \succeq 0,
    \end{aligned}
    \label{eq:input SDP intro}
\end{equation}
where $C, A_k, D_k$ are fixed Hermitian matrices,
$b_k \in \R$ are fixed scalars,
$\ftr$ and $\ptr$ denote the number of constraints that involve full trace and partial trace,
and $S_k \subseteq [p+q]$ denote subsets of systems that are traced out.

Note that all matrices in \cref{eq:input SDP intro} are of size $d^{p+q} \times d^{p+q}$. For this problem to have an efficient description,
we assume that $C, A_k, D_k$ are $s$-sparse,
i.e., can be specified as a linear combination of at most $s$ walled Brauer algebra diagrams (see \cref{sec:wba}) and elementary rank-$1$ matrices $\ketbra{i}{j}$ where $i,j \in [d]^{p+q}$.
Our main result, \cref{thm:main}, provides an efficient way of converting the above semidefinite program to an equivalent linear program (LP) when the matrix variable $X$ is subject to one of the following additional symmetries:
$\S_p \times \S_q$ symmetry,
walled Brauer algebra symmetry,
or Gelfand--Tsetlin symmetry
(see \cref{sec:symmetries} for more details).

\begin{theorem*}[Informal]
Assuming one of the above additional symmetries on $X$,
the SDP~\eqref{eq:input SDP intro} can be converted to an equivalent LP with $n \leq N$ variables and $\ftr + \ptr N + n$ constraints where $N \defeq (p+q)!$.
\end{theorem*}

Our approach has the advantage that $d$ can be arbitrary\footnote{In fact, $d$ can even be symbolic!} and the computational resources are tied only to the value of $p+q$.
For example, let $(p,q) = (2,3)$ be small constants and $d = 1000$ be very large.
In this regime, naively solving the above SDP is impossible since it has $d^{2(p+q)} = 1000^{10} = 10^{30}$ scalar variables.
However, the complexity of our method scales only in the parameter $N = (p+q)!$, which in this case is $(2+3)! = 5! = 120$.

Our method can provide an advantage also when $d$ is small.
For example, if $d = 2$ then $d^{2(p+q)} = 2^{2(2+3)} = 2^{10} = 1024$ variables are needed naively while only $42$ suffice with our method (see \cref{table:SDP} in \cref{apx:numeric tables}).

Even though our method provides a significant improvement in terms of $d$, its complexity still scales as $(p+q)!$, so in practice we can only deal with relatively small values of $p$ and $q$.
However, numerical solutions to small problem instances are still valuable since they may reveal structures that can help to tackle larger instances.
For example, numerical insights can lead to a refined ansatz with fewer parameters that scales up more easily.
Good ansatz, even if suboptimal, can still be used to obtain numerical bounds.
If the ansatz is sufficiently simple, it might even be amenable to analytic methods.
In this way, our method can potentially be used to bootstrap from small problem instances to much larger ones.

We have implemented our method in \emph{SageMath} and made our code available \cite{github}.
To illustrate potential applications of our method, in \cref{sec:applications} we provide several examples from quantum information:
deciding the principal eigenvalue of a quantum state,
quantum majority vote, asymmetric cloning and transformation of a black-box unitary operation.
We have implemented the calculations for these examples as \emph{Wolfram Mathematica} notebooks that are also available \cite{github}.

\subsection{Intuition}

The main idea behind our result is as follows.
While the naive semidefinite program~\eqref{eq:input SDP intro} has $d^{2(p+q)}$ scalar variables, the unitary equivariance condition alone reduces this down to $\dim(\A_{p,q}^d)$, where $\A_{p,q}^d$ is the matrix algebra of partially transposed permutations. This observation already gives us a $d$-independent upper bound on the number of parameters since $\dim(\A_{p,q}^d) \leq (p+q)!$.
While generally this bound is loose for small $d$, it saturates when $d \geq p + q$.

This reduction in the number of parameters happens thanks to a generalized or ``mixed'' Schur--Weyl duality (see \cref{sec:mixed SW duality}).
Together with Schur's lemma, this duality implies that any unitary-equivariant matrix variable $X$ can be written as
\begin{align}
    X \cong
      \bigoplus_{\lambda \in \Irr(\A^d_{p,q})}
      \sof*{
        X_\lambda \x I_{m_\lambda}
      }
    \label{eq:X blocks}
\end{align}
in some basis, where the size of each block $X_\lambda$ is independent of $d$.
The main difficulty then lies in obtaining an ansatz that captures all relevant degrees of freedom for such $X$, in a way that does not scale in the local dimension $d$.
In particular, we cannot afford to apply the ``mixed'' Schur transform that implements the basis change in \cref{eq:X blocks} since the underlying space has dimension $d^{p+q}$.

For simplicity, in this work, we only consider the special case when each $X_\lambda$ is diagonal.
This assumption is justified when $X$ is subject to some additional symmetry (see \cref{sec:symmetries} for more details).
For example, under certain symmetry $X$ is diagonal in the so-called Gelfand--Tsetlin basis and can thus be written as a linear combination of primitive idempotents of the partially transposed permutation algebra $\A^d_{p,q}$.
Since this algebra is closely related to the walled Brauer algebra $\smash{\B^d_{p,q}}$, which is multiplicity-free, we can adapt a general framework of \cite{doty2019canonical} to compute its idempotents.
A crucial ingredient in this computation are so-called Jucys--Murphy elements.
In \cref{thm:main technical}, which is our main technical contribution, we show that these elements can be lifted from $\smash{\B^d_{p,q}}$ to $\smash{\A^d_{p,q}}$.
This allows us to perform the entire computation within the walled Brauer algebra $\smash{\B^d_{p,q}}$.
In contrast to $\smash{\A^d_{p,q}}$, $\smash{\B^d_{p,q}}$ is diagrammatic and hence we do not need to manipulate any matrices of size $d^{p+q}$. This is precisely why the complexity of our approach does not scale in $d$.
In particular, we do not require explicit knowledge of the mixed Schur transform.

\subsection{Open problems}

Our work raises several interesting open questions.
\begin{enumerate}
    \item Derive an explicit basis in which our ansatz for $X$ is diagonal under the $\S_p \times \S_q$ symmetry when $d = 2$ or $\min(p,q) = 2$.
    \item We believe that our method can be extended to solve general unitary-equivariant SDPs.
    This requires removing the additional symmetry assumption that reduces such SDP to an LP.
    This could be done along the lines outlined in \cref{apx:blocks of Apq}, but it remains to formally verify this.
    \item What is the role of representation theory in our approach?
    Could there be a more direct way of doing this without the use of representation theory?
    \item Characterize the kernel of the map $\psi^d_{p,q}$ defined in \cref{eq:psi} that embeds the walled Brauer algebra $\smash{\B^d_{p,q}}$ into the partially-transposed permutation matrix algebra $\smash{\A^d_{p,q}}$.
    This would allow removing the technical restriction~\eqref{eq:d big} in \cref{thm:main} that $d$ must be sufficiently large.
    Moreover, this could also provide additional speedups for small $d$ since all calculations could be performed using a linearly independent diagram basis of size much smaller than $(p+q)!$.
    One possible method for computing $\ker \psi^d_{p,q}$ is sketched in \cref{rem:kernel}.
    \item Given a solution $X$ of a unitary-equivariant SDP that describes a Choi matrix, we would like to find an efficient quantum circuit that implements the corresponding quantum channel.
    Note that \cite{Majority} achieves this when $d = 2$, $p = 2k+1$ and $q = 1$.
    An efficient implementation of the mixed Schur transform $U_{\textnormal{Sch}(p,q)}$ (see \cref{rm:Schur}) for general values of $p,q,d$ would be a useful subroutine for achieving the general case.
    \item It should be possible to treat the local dimension $d$ symbolically and deduce the asymptotic scaling of the solution as $d \to \infty$.
    \item The applications we provide in \cref{sec:applications} are only for illustrative purposes. We expect that one should be able to go much further by bearing the full weight of our method. For example, concerning the application in \cref{sec:quantum_comb}, we would like to find, for any given $d$, how many copies of a black-box unitary $U$ are needed to implement $f(U)$ exactly and deterministically via a sequential superchannel.
    \item Our approach is an instance of a general philosophy outlined in \cite{invariantSDPs} for solving SDPs with $*$-matrix algebra symmetries. Other instances of this setting are also useful in quantum information~\cite{CliffordSW} and hence worthwhile investigating.
\end{enumerate}

\subsection{Outline}

Our paper is structured as follows.
In \cref{sec:preliminaries}, we introduce the main preliminary concepts, such as quantum channels and their symmetries, and semidefinite programming.
In \cref{sec:mixed SW duality}, we discuss Schur--Weyl duality and its generalization to mixed tensors.
Here we also introduce walled Brauer algebras and their matrix analogs ---partially transposed permutation algebras---that play a central role in mixed Schur--Weyl duality.
In \cref{sec:DLS}, we review a general construction of primitive central idempotents for any multiplicity-free family of algebras due to \cite{doty2019canonical}.
In \cref{sec:A idempotents}, we specialize this construction to walled Brauer algebras and adapt it to their matrix analogues (which can be of independent interest).
Finally, in \cref{sec:SDP} we derive our main result---a general meta-algorithm for reducing unitary-equivariant semidefinite programs to significantly smaller linear programs.
This reduction is possible under certain natural permutation symmetry assumptions.
We conclude in \cref{sec:applications} by providing several applications of our framework to quantum computing.

\section{Preliminaries}\label{sec:preliminaries}

Throughout the paper, we fix a local dimension $d \geq 2$ and let $V \defeq \C^d$.
For any integers $p,q \geq 0$, let
$V^p \defeq V^{\otimes p}$ denote the $p$-fold tensor product of $V$, and let
$V^{p,q} \defeq V^{\otimes p} \otimes V^{* \otimes q}$
denote the \emph{mixed} tensor product where $V^*$ is the dual space of $V$.
While
$V^* \cong V = \spn_\C \set{\ket{1}, \dotsc, \ket{d}}$,
we make the distinction to emphasize that $V^*$ is subject to the \emph{dual} action (see below).

Let $\End(V)$ denote the set of all complex $d \times d$ matrices.
We denote by $\U(V)$ all \emph{unitary} matrices on $V$, i.e., all $U \in \End(V)$ such that $U\ct U = I_V$,
where $U\ct \defeq \bar{U}\tp$ denotes the conjugate transpose of $U$ and $I_V$ is the identity matrix on $V$.
The \emph{dual action} of $U \in \U(d)$ on $V^*$ is given by the entry-wise complex conjugate $\bar{U}$.

We denote by $\D(V)$ the set of all \emph{density matrices} on $V$, i.e., all matrices $\rho \in \End(V)$ such that $\rho \succeq 0$ ($\rho$ is \emph{positive semidefinite}) and $\Tr(\rho) \defeq \sum_{i=1}^d \bra{i} \rho \ket{i} = 1$.

A \emph{representation} of a group $G$ on a complex vector space $V$ is a homomorphism $R\colon G \to \U(V)$, i.e., $R(gh) = R(g) R(h)$ for all $g,h \in G$.
The representation is \emph{faithful} if $R$ is an injection, i.e., different group elements are represented by different matrices.
A representation $R$ is \emph{reducible}\footnote{Technically such representation is called \emph{decomposable} and the notion of irreducibility is weaker (this distinction will become important when we consider representations of algebras instead of groups, see \cref{sec:algebras}). However, in the context of group representations the two notions are equivalent since all representations we consider are unitary.}
if there exists a basis change $U \in \U(V)$ and two representations $R_1$ and $R_2$ of dimension at least one such that
$U R(g) U\ct = R_1(g) \+ R_2(g)$ for all $g \in G$.
Otherwise $R$ is called \emph{irreducible}.

An \emph{algebra} is a vector space equipped with a bilinear product.
If $G$ is a group, its corresponding \emph{group algebra}
$\C G \defeq \spn_\C \set{\ket{g} : g \in G}$
is an algebra that extends the group operation of $G$ by linearity:
$\ket{g} \cdot \ket{h} = \ket{gh}$, for all $g,h \in G$.
If $V$ is a vector space, a \emph{matrix algebra} on $V$ is a linear subspace of $\End(V)$ that is closed under matrix multiplication.
The \emph{centralizer} of a matrix algebra $\A$ in $\End(V)$ is the set of all matrices acting on $V$ that commute with $\A$:
\begin{equation}
    \End_\A(V) \defeq \set{B \in \End(V) : [A,B] = 0 \text{ for every } A \in \A},
    \label{eq:centralizer}
\end{equation}
where $[A,B] \defeq AB - BA$ denotes the \emph{commutator} of $A$ and $B$.
For more background on algebras and their representations see \cref{sec:algebras}.

For any integer $n \geq 1$, we let $[n] \defeq \set{1,\dotsc,n}$.
We write $\lambda \pt p$ to mean that $\lambda$ is a \emph{partition} of an integer $p \geq 0$, i.e., $\lambda = (\lambda_1, \dotsc, \lambda_k)$ is a tuple of $k$ integers, for some $k \geq 1$, such that $\lambda_1 \geq \dotsb \geq \lambda_k \geq 0$ and $\lambda_1 + \dotsb + \lambda_k = p$. We also think of $\lambda$ as a \emph{Young diagram}, i.e., a collection of $p$ square cells arranged in $k$ rows with $\lambda_i$ of them in the $i$-th row. For example,
\begin{equation}
  \yd[1]{5,3,1}
\end{equation}
depicts the partition $(5,3,1)$.
We denote by $\len(\lambda) \defeq k$ the \emph{length} of $\lambda$, i.e., the length of the first column (or the number of rows).
We denote by $\size(\lambda) \defeq p$ the \emph{size} of $\lambda \pt p$, i.e., the number of cells in the Young diagram.
The \emph{content} of cell $(i,j) \in \lambda$ is $j-i$,
where $i$ denotes the row and $j$ denotes the column of the corresponding cell.
The \emph{content} of partition $\lambda$ is the total content of all its cells:
$\cont(\lambda) \defeq \sum_{(i,j) \in \lambda} (j-i)$,
where the sum runs over all cells in the diagram.

\subsection{Quantum channels}

Let $\Vin \defeq \C^{d_\text{in}}$ and $\Vout \defeq \C^{d_\text{out}}$ be finite-dimensional complex Euclidean spaces.
A \emph{quantum channel} $\Phi\colon \End(\Vin) \to \End(\Vout)$ is a completely positive and trace-preserving linear map.
\emph{Complete positivity} means that, for any reference space $V_\text{ref}$ and state $\rho \in \D(\Vin \x V_\text{ref})$, we have $(\Phi \x \mathrm{I}_\text{ref})(\rho) \succeq 0$ where $\mathrm{I}_\text{ref}$ denotes the identity channel on $V_\text{ref}$.
\emph{Trace preservation} means that
$\Tr\of[\big]{\Phi(\rho)} = \Tr(\rho)$,
for all $\rho \in \D(\Vin)$.

A convenient way of representing a quantum channel is by its \emph{Choi matrix} $X^\Phi \in \End(\Vin \x \Vout)$ defined as
\begin{equation}
    X^\Phi \defeq \sum_{i,j=1}^{d_\text{in}} \ketbra{i}{j} \x \Phi\of[\big]{\ketbra{i}{j}}
\end{equation}
where $\set{\ket{1}, \dotsc, \ket{d_\text{in}}}$ is an orthonormal basis for $\Vin$.
The action of $\Phi$ on $\rho \in \D(\Vin)$ can be recovered from its Choi matrix $X^\Phi$ as follows:
\begin{equation}
    \Phi(\rho) = \Tr_{\Vin} \sof[\big]{X^\Phi (\rho\tp \x I_\text{out})}.
    \label{eq:J output}
\end{equation}
A given matrix $X \in \End(\Vin \x \Vout)$ describes a quantum channel if and only if
\begin{align}
    X &\succeq 0, &
    \Tr_{\Vout} (X) &= I_\text{in}.
    \label{eq:Choi constraints}
\end{align}
For more background on quantum information theory see~\cite{watrous}.

\subsection{Unitary equivariance}

Our main motivating problem is the optimization of a linear function over unitary-equivariant quantum channels.
Recall that $V \defeq \C^d$ for some $d \geq 2$ and $V^p \defeq V\xp{p}$.

\begin{definition}\label{def:equivariance}
We say that $\Phi\colon \End(V^p) \to \End(V^q)$ is a \emph{$p \to q$ channel}. Such channel is \emph{locally $\U(V)$-equivariant} or simply \emph{unitary-equivariant} if
\begin{equation}
    \Phi\of[\big]{U\xp{p} \, \rho \, U\ctxp{p}}
    = U\xp{q} \, \Phi(\rho) \, U\ctxp{q}
    \label{eq:Phi equivariance}
\end{equation}
for every $U \in \U(V)$ and $\rho \in \D(V^p)$.
\end{definition}

To optimize over such channels, we need to understand their structure or, equivalently, the structure of the associated Choi matrices.
The following is a well-known characterization of the unitary equivariance of $\Phi$ in terms of its Choi matrix.

\begin{proposition}\label{propl:Choi equivariance}
Let $X^\Phi \in \End(V^{p,q})$ be the Choi matrix of a $p \to q$ channel $\Phi$. Then $\Phi$ is unitary-equivariant if and only if
\begin{equation}
    \sof[\big]{
        X^\Phi,
        U\xp{p} \x \bar{U}\xp{q}
    } = 0,
    \qquad \forall U \in \U(V).
    \label{eq:Choi equivariance}
\end{equation}
\end{proposition}

\noindent
Note that the last $q$ registers of $X^\Phi$ are subject to the dual action of $\U(V)$.

\subsection{Semidefinite and linear programming}

\emph{Semidefinite programming} is an important subfield of optimization \cite{handbookSDP} that has numerous applications in quantum information theory \cite{SDPs,watrous}. A typical formulation of a \emph{semidefinite program} (SDP) has the form \cite[Section~1.1]{handbookSDP}:
\begin{equation}
  \begin{aligned}
    \max_X \quad & \Tr \of{C\tp X} \\
    \textrm{s.t.} \quad & \Tr \of{A_i\tp X} = b_i, & \forall i \in [m], \\
    & X \succeq 0,
  \end{aligned}
\end{equation}
where $X$ is a Hermitian matrix variable,
$C$ and $A_i$ are constant Hermitian matrices, and
$b_i$ are real constants.
A special case of SDPs are \emph{linear programs} (LPs)
which correspond to the case when all matrices involved are diagonal.
Any LP can be formulated in the standard form
\begin{equation}
  \begin{aligned}
    \max_x \quad & c\tp x \\
    \textrm{s.t.} \quad & a_i\tp x = b_i, & \forall i \in [m], \\
    & x \geq 0,
  \end{aligned}
\end{equation}
where $x$ is a real vector variable,
$c$ and $a_i$ are constant real vectors, and
$b_i$ are real constants.
In practice LPs are much faster to solve than SDPs.
Therefore being able to reduce a given SDP to an LP under additional symmetry assumptions is often desirable.

\subsection{Motivating problem}

Consider the following motivating semidefinite optimization problem.
Fix a quantum state $\rho \in \D(V^p)$ and an arbitrary Hermitian matrix $H \in \End(V^q)$,
and assume we want to find a unitary-equivariant $p \to q$ channel $\Phi$ that maximizes the linear function $\Tr\sof{\Phi(\rho) H}$.
For example, if $H = \proj{\psi}$ for some pure state $\ket{\psi} \in V^q$ then $\Tr\sof{\Phi(\rho) H} = \bra{\psi} \Phi(\rho) \ket{\psi}$ is the fidelity between the output state $\Phi(\rho)$ and the desired target state $\ket{\psi}$.

According to \cref{eq:J output},
$\Tr\sof{\Phi(\rho) H} = \Tr \sof[\big]{X^\Phi (\rho\tp \x H)}$,
so the constraints on $X^\Phi$ from \cref{eq:Choi constraints} give us the following SDP:
\begin{equation}
    \max_{X^\Phi \in \End(V^{p,q})} \Tr \sof[\big]{X^\Phi (\rho\tp \x H)}, \quad
    \Tr_{V^q} (X^\Phi) = I_{V^p}, \quad
    X^\Phi \succeq 0.
    \label{eq:basic SDP}
\end{equation}
This problem has a trivial solution---a channel that ignores its input $\rho$ and prepares the principal eigenvector of $H$ as output.
To make the problem non-trivial, consider instead a collection of $n$ pairs $(\rho_i, H_i)$,
with the goal of maximizing the smallest value of $\Tr\sof{\Phi(\rho_i) H_i}$, $i = 1,\dotsc,n$.
This is captured by the following SDP:
\begin{equation}
    \max_{\substack{X^\Phi \in \End(V^{p,q}) \\ c \in \R}} c, \quad
    \forall i\colon
    \Tr \sof[\big]{X^\Phi (\rho_i\tp \x H_i)} \geq c, \quad
    \Tr_{V^q} (X^\Phi) = I_{V^p}, \quad
    X^\Phi \succeq 0.
    \label{eq:modified SDP}
\end{equation}
This problem no longer admits a trivial solution where $\Phi$ ignores the input state.

Motivated by applications to problems mentioned in \cref{sec:intro optimization}, we would like to incorporate the unitary-equivariance constraint \eqref{eq:Phi equivariance} on the channel $\Phi$ in the SDPs~\eqref{eq:basic SDP} and~\eqref{eq:modified SDP}.
Note that using \cref{propl:Choi equivariance} in a naive way would result in an optimization problem with an uncountable number of linear constraints and a matrix variable of dimension $d^{p+q}$ that scales badly in $d$ even for constant $p$ and~$q$.
Our main contribution is an efficient method that can deal with both of these issues simultaneously.
Under additional symmetry assumptions, it reduces the above SDPs to finite linear programs whose size does not scale in $d$ (we sketch in \cref{apx:blocks of Apq} a possible way to remove the symmetry assumption).

\subsection{Optimization under unitary equivariance}

Symmetry is a powerful tool for simplifying almost any type of problem, including problems in semidefinite optimization \cite{invariantSDPs,RepLAB}.
Our goal is to investigate semidefinite optimization for a matrix variable subject to a unitary equivariance constraint.
To simplify the problem even further, we assume one of several additional symmetries that reduce the semidefinite program to a linear program.

A naive way of imposing unitary equivariance on $\Phi$ in our motivating problem in \cref{eq:modified SDP} is by including \cref{eq:Choi equivariance} as an extra linear constraint.
However, this technically constitutes an uncountably infinite set of constraints.
To get around this issue, we could instead demand that
\begin{equation}
  \int_{U \in \U(d)}
  \of*{U\xp{p} \x \bar{U}\xp{q}}
  X^\Phi
  \of*{U\xp{p} \x \bar{U}\xp{q}}\ct
  dU
  = X^\Phi
  \label{eq:Haar}
\end{equation}
where $dU$ denotes the Haar measure on $\U(d)$.
This integral can in principle be evaluated using Weingarten calculus \cite{Weingarten,CollinsSniady}, producing a single linear constraint on $X^\Phi$.
The resulting SDP has finite size and can be supplied to a standard solver.

However, there is another serious issue that can prevent the SDP from being solvable in practice.
Namely, $X^\Phi$ is a matrix of size $d^{p+q} \times d^{p+q}$.
Since each matrix entry of $X^\Phi$ is represented by a separate scalar variable in the SDP, the total number of variables is $d^{2(p+q)}$, which is prohibitive even for moderate values of $d$.

Motivated by this issue, our main goal is to understand whether optimization problems with a $\U(d)$-equivariance constraint can be solved in time that does not scale in $d$.
In this work, we focus on linear programming as a special case of semidefinite programming and answer the above question in the affirmative (see \cref{thm:main} for our main result).

In the context of unitary-equivariant channels, linear programs occur naturally when additional symmetries are imposed on~$\Phi$.
Indeed, appropriately chosen symmetries guarantee that the Choi matrix $X^\Phi$ is diagonal in a certain basis, allowing the semidefinite constraint $X^\Phi \succeq 0$ to be replaced by scalar inequalities.

\subsection{Additional symmetries}\label{sec:symmetries_intro}

One natural example of additional $p \to q$ channel symmetries is invariance under permutations of the $p$ input and $q$ output systems,
where each type of system is permuted separately.

The \emph{symmetric group} $\S_p$ on $p$ elements acts naturally on $p$ qudits by permuting them.
This can be captured by a representation $\psi^d_p\colon \S_p \to \End(V^p)$ defined on simple tensors $\ket{i_1} \x \dotsb \x \ket{i_p}$ with $i_1, \dotsc, i_p \in \set{1,\dotsc,d}$ as
\begin{equation}
    \psi^d_p(\pi) \of[\big]{\ket{i_1} \x \dotsb \x \ket{i_p}} \defeq
    \ket{i_{\pi^{-1}(1)}} \x \dotsb \x \ket{i_{\pi^{-1}(p)}},
    \label{eq:psi basic}
\end{equation}
for all $\pi \in \S_p$, and extended linearly to all vectors in $V^p$.

\begin{definition}\label{def:sym channel}
A $p \to q$ channel $\Phi$ is \emph{input-symmetric} if
$\Phi\of[\big]{\psi^d_p(\pi) \, \rho \, \psi^d_p(\pi)\ct} = \Phi(\rho)$,
for every $\rho \in \D(V^p)$ and $\pi \in \S_p$.
Similarly, $\Phi$ is \emph{output-symmetric} if
$\Phi(\rho) = \psi^d_q(\sigma) \, \Phi(\rho) \, \psi^d_q(\sigma)\ct$,
for every $\rho \in \D(V^p)$ and $\sigma \in \S_q$.
We call $\Phi$ \emph{symmetric} if it is both input- and output-symmetric:
\begin{equation}
    \Phi\of[\big]{\psi^d_p(\pi) \, \rho \, \psi^d_p(\pi)\ct}
    = \psi^d_q(\sigma) \, \Phi(\rho) \, \psi^d_q(\sigma)\ct.
\end{equation}
if for every $\rho \in \D(V^p)$ and every pair of permutations $(\pi,\sigma) \in \S_p \times \S_q$.
\end{definition}

Similar to \cref{propl:Choi equivariance}, this symmetry of $\Phi$ can also be expressed in terms of its Choi matrix $X^\Phi$.

\begin{proposition}\label{lem:choi_perm}
A $p \to q$ channel $\Phi$ is symmetric if and only if
its Choi matrix $X^\Phi$ satisfies
\begin{align}
    \sof[\big]{X^\Phi, \psi^d_p(\pi) \x \psi^d_q(\sigma)} = 0,
    \qquad
    \forall  (\pi,\sigma) \in \S_p \times \S_q.
\end{align}
\end{proposition}

In \cref{sec:symmetries} we consider two additional types of symmetries and discuss when a unitary-equivariant SDP reduces to an LP under such symmetries.

The structure of unitary-equivariant quantum channels can be described using a generalization of Schur--Weyl duality to mixed tensor products.

\section{Mixed Schur--Weyl duality}\label{sec:mixed SW duality}

To understand the interplay between unitary equiariance and permutational or other symmetries, it is very useful to know about the relationship between their representations.
Indeed, the irreducible representations of $\U(d)$ and $\S_p$ on the $p$-fold tensor product space $V^p = (\C^d)\xp{p}$ are intimately related because the actions of $U\xp{p}$, $U \in \U(d)$ and $\psi^d_p(\pi)$, $\pi \in \S_p$ on this space mutually commute.
This relationship, known as \emph{Schur--Weyl duality}, is an important tool in quantum information theory, see \cite[Chapter~6]{HarrowThesis} and \cite{Botero}.
We will need a generalization of this duality to the \emph{mixed} tensor product space $V^{p,q} = V^p \x V^{*q}$.

\subsection{Schur--Weyl duality}\label{sec:Schur-Weyl}

Let us first review the classical Schur--Weyl duality on $V^p$ where $V = \C^d$.
For more background, see \cite[Section~5.3]{HarrowThesis}.

A natural way for $\U(d)$ to act on $p$ qudits is by applying an identical unitary transformation on each of them.
This is captured by a representation
$\phi^d_p \colon \U(d) \to \End(V^p)$ defined as
\begin{equation}
  \phi^d_p (U) \defeq U\xp{p},
  \label{eq:phi}
\end{equation}
for all $U \in \U(d)$.
Let us denote the algebra generated by the image of $\U(d)$ under $\phi^d_p$ by
\begin{equation}
  \CA^d_p \defeq \spn_\C \set{\phi^d_p(U) : U \in \U(d)}.
  \label{eq:Up}
\end{equation}

Let $\CS_p \defeq \spn_\C \S_p$ denote the group algebra of the symmetric group.
We extend the representation $\psi^d_p \colon \S_p \to \End(V^p)$ in \cref{eq:psi basic} from $\S_p$ to $\CS_p$ by linearity
and denote the resulting algebra of $p$-qudit permutations by
\begin{equation}
  \A^d_p \defeq \psi^d_p(\CS_p).
  \label{eq:Ap}
\end{equation}

Recall from \cref{eq:centralizer} that $\End_{\A^d_p}(V^p)$ denotes the centralizer of $\A^d_p$ in $\End(V^p)$, i.e., all matrices in $\End(V^p)$ that commute with $\A^d_p$.
One way to state the Schur--Weyl duality is that $\CA^d_p$ is the centralizer of $\A^p_d$, and vice versa.

\begin{theorem}[Schur--Weyl duality]\label{thm:schur-weyl}
The algebra $\CA^d_p$ is the centraliser algebra of $\A^d_p$ in $\End(V^p)$ and vice versa, i.e.,
\begin{align}
    \CA^d_p &= \End_{\A^d_p}(V^p), &
    \A^d_p &= \End_{\CA^d_p}(V^p).
\end{align}
Moreover, when $d \geq p$ the representation $\psi^d_p$ is faithful, i.e., $\A^d_p \cong \CS_p$.
\end{theorem}

One of the simplest instances of this duality is in the case of two qubits.

\begin{example}[$p = 2$ and $d = 2$]\label{ex:p2d2}
Since $\S_2 = \set{(), (12)}$, the algebra $\A^2_2$ is generated by the identity matrix and $\SWAP$:
\begin{equation}
  \A^2_2 \defeq \set*{
    x \mx{
        1 & 0 & 0 & 0 \\
        0 & 1 & 0 & 0 \\
        0 & 0 & 1 & 0 \\
        0 & 0 & 0 & 1
    } +
    y \mx{
        1 & 0 & 0 & 0 \\
        0 & 0 & 1 & 0 \\
        0 & 1 & 0 & 0 \\
        0 & 0 & 0 & 1
    } :
    x,y \in \C
  },
  \label{eq:A22}
\end{equation}
whereas $\CA^2_2$ is generated by tensor squares of unitary matrices:
\begin{equation}
  \CA^2_2 \defeq \spn_\C \set*{
    \mx{a & b \\ c & d} \x
    \mx{a & b \\ c & d} :
    \mx{a & b \\ c & d} \in \U(2)
  }.
\end{equation}
Using the two-qubit Schur transform \cite[p.~121]{HarrowThesis}
\begin{equation}
  U \defeq \frac{1}{\sqrt{2}}
  \mx{
    0 & 1 & -1 & 0 \\
    \sqrt{2} & 0 & 0 & 0 \\
    0 & 1 & 1 & 0 \\
    0 & 0 & 0 & \sqrt{2}
  },
\end{equation}
we can simultaneously block-diagonalize both algebras:
\begin{align}
  U \A^2_2 U\tp
 ={}& \set*{
      \left(
      \begin{array}{@{}c|ccc@{}}
        x-y & 0 & 0 & 0 \\ \hline
        0 & x+y & 0 & 0 \\
        0 & 0 & x+y & 0 \\
        0 & 0 & 0 & x+y
      \end{array}
      \right) : x,y \in \C
    }, \\
  U \CA^2_2 U\tp
  = \spn_\C
  & \set*{
      \left(
      \begin{array}{@{}c|ccc@{}}
        ad-bc & 0 & 0 & 0 \\ \hline
        0 & a^2 & \sqrt{2} a b & b^2 \\
        0 & \sqrt{2} a c & a d+b c & \sqrt{2} b d \\
        0 & c^2 & \sqrt{2} c d & d^2
      \end{array}
      \right) : \mx{a & b \\ c & d} \in \U(2)
    }.
  \label{eq:U22}
\end{align}
These algebras centralize each other since
\begin{align}
  U \A^2_2 U\tp &= \C \+ \C I_3, &
  U \CA^2_2 U\tp &= \C \+ \End(\C^3),
  \label{eq:A22 and U22}
\end{align}
where the second equality follows from Burnside's theorem \cite{LR04}
since the $3 \times 3$ block in \cref{eq:U22} is irreducible.
\end{example}

\begin{example}[Unfaithfulness of $\psi^d_3$]\label{ex:faithfulness}
It is important to observe that $\psi^d_p$ has a non-trivial kernel when the local dimension $d$ is smaller than $p$. For example,
$\psi^d_3 \of*{\sum_{\sigma \in \S_3} \operatorname{sign}(\sigma) \sigma}$
vanishes when $d = 2$ and does not vanish when $d \geq 3$.
For this reason we will often have to make a distinction between small and large local dimensions $d$.
\end{example}

\subsection{Walled Brauer algebra}\label{sec:wba}

\newcommand{\BrauerTikZStyle}{\tikzset{
  every path/.style = {semithick}, 
  looseness = 0.7, 
  dot/.style = {shift only, radius = 1.3pt, fill = black}, 
  l/.style = {out = -90, in =  90}, 
  u/.style = {out = -90, in = -90}, 
  n/.style = {out =  90, in =  90}  
}}

\newcommand{\curlybrace}[3]{
  \def\r{0.2} 
  \def\e{0.1} 
  \draw[radius = \r] (#1-#3-\e,#2-\r)
    arc[start angle = 180, end angle =  90] -- (#1-\r,#2) arc[start angle = -90, end angle = 0]
    arc[start angle = 180, end angle = 270] -- (#1+#3-\r+\e,#2)
    arc[start angle =  90, end angle =   0];
}

To analyze the Choi matrix of unitary-equivariant quantum channel,
we need a generalization of Schur--Weyl duality for the space $V^{p,q}$ of mixed tensor products.
This generalization requires the notion of walled Brauer algebras
\cite{turaev1989operator,koike1989decomposition,bchlls,Benkart,nikitin2007centralizer,Bulgakova},
a restricted version of Brauer algebras
(see \cite{Panorama} for a survey on Brauer and other diagram algebras).

Let $p,q \geq 0$ and $\delta \in \C$.
The \emph{walled Brauer algebra} $\B^\delta_{p,q}$ consists of formal complex linear combinations of diagrams,
where each diagram has two rows of $p+q$ nodes each, with a vertical ``wall'' between the first $p$ and the last $q$ nodes.
These nodes are connected up in pairs, with each pair subject to the following restriction:
if both nodes are in the same row, they must be on different sides of the wall,
while if they are in different rows, they must be on the same side of the wall.
For example, a diagram in $\B^\delta_{3,2}$ may look like this:
\begin{equation}
  \begin{tikzpicture}[baseline = 1cm]
    \BrauerTikZStyle
    \curlybrace{2.0}{1.4}{1.0} \node at (2.0,1.9) {$p=3$};
    \curlybrace{4.5}{1.4}{0.5} \node at (4.5,1.9) {$q=2$};
    \foreach \i in {1,...,5} {
      \fill (\i,1) circle [dot] coordinate (A\i);
      \fill (\i,0) circle [dot] coordinate (B\i);
    }
    \draw (A1) to [l] (B1);
    \draw (A2) to [l] (B3);
    \draw (A4) to [l] (B5);
    \draw (A3) to [u] (A5);
    \draw (B2) to [n] (B4);
    \draw [dashed] (3.5,1.2) -- (3.5,-0.2);
  \end{tikzpicture}
\end{equation}

Multiplication in $\B^\delta_{p,q}$ corresponds to concatenation of diagrams, with the bottom row of the first diagram identified with the top row of the second diagram. If some loops appear in this process, one should erase them and multiply the resulting diagram with the scalar $\delta^{\#\loops}$:
\begin{equation}
  \begin{tikzpicture}[baseline = 1cm]
    \BrauerTikZStyle

    \begin{scope}
      \node[anchor = east] at (0.9,1.5) {$\rho =$};
      \node[anchor = east] at (0.9,0.5) {$\sigma =$};
      \foreach \i in {1,...,5} {
        \fill (\i,2) circle [dot] coordinate (A\i);
        \fill (\i,1) circle [dot] coordinate (B\i);
        \fill (\i,0) circle [dot] coordinate (C\i);
      }
      \draw (A1) to [l] (B1);
      \draw (A2) to [l] (B3);
      \draw (A4) to [l] (B5);
      \draw (A3) to [u] (A5);
      \draw (B2) to [n] (B4);
      \draw (B1) to [l] (C3);
      \draw (B3) to [l] (C1);
      \draw (B5) to [l] (C5);
      \draw (B2) to [u] (B4);
      \draw (C2) to [n] (C4);
      \draw [dashed] (3.5,2.2) -- (3.5,-0.2);
    \end{scope}

    \begin{scope}[xshift = 6.6cm, yshift = 0.5cm]
      \node[anchor = east] at (0.9,0.5) {$\rho \, \sigma = \delta \;\cdot$};
      \foreach \i in {1,...,5} {
        \fill (\i,1) circle [dot] coordinate (D\i);
        \fill (\i,0) circle [dot] coordinate (E\i);
      }
      \draw (D1) to [l] (E3);
      \draw (D2) to [l] (E1);
      \draw (D4) to [l] (E5);
      \draw (D3) to [u] (D5);
      \draw (E2) to [n] (E4);
      \draw [dashed] (3.5,1.2) -- (3.5,-0.2);
    \end{scope}

  \end{tikzpicture}
\end{equation}

The group algebra of the permutation group $\S_p \times \S_q$
forms a subalgebra of the walled Brauer algebra $\B^\delta_{p,q}$
consisting only of those diagrams where no edge goes across the wall.
In fact, the two algebras are isomorphic when $p = 0$ or $q = 0$, i.e.,
$\CS_p \cong \B^\delta_{p,0} \cong \B^\delta_{0,p}$ for any value of $\delta$.
The walled Brauer algebra $\B^\delta_{p,q}$ itself is a subalgebra of the full \emph{Brauer algebra} $\B^\delta_{p+q}$ that is defined similarly but without the wall and with no restrictions on which pairs of nodes can be connected.
This algebra was originally introduced by Brauer \cite{Brauer} for studying Schur--Weyl-like dualities of orthogonal and symplectic groups.

For any diagram $\sigma \in \B^\delta_{p,q}$, the \emph{partial transpose} $\sigma\ptp$ is obtained by exchanging the last $q$ nodes of both rows
(i.e., the nodes on the right-hand side of the wall):
\begin{equation}
  \Bigg(\;
  \begin{tikzpicture}[baseline = 0.4cm]
    \BrauerTikZStyle

    \begin{scope}
      \foreach \i in {1,...,5} {
        \fill (\i,1) circle [dot] coordinate (A\i);
        \fill (\i,0) circle [dot] coordinate (B\i);
      }
      \draw (A1) to [l] (B1);
      \draw (A2) to [l] (B3);
      \draw (A4) to [l] (B5);
      \draw (A3) to [u] (A5);
      \draw (B2) to [n] (B4);
      \draw [dashed] (3.5,1.2) -- (3.5,-0.2);
    \end{scope}
  \end{tikzpicture}
  \;\Bigg)\ptp \quad = \quad
  \begin{tikzpicture}[baseline = 0.5cm]
    \BrauerTikZStyle
    \begin{scope}[xshift = 6.6cm]
      \foreach \i in {1,...,5} {
        \fill (\i,1) circle [dot] coordinate (A\i);
        \fill (\i,0) circle [dot] coordinate (B\i);
      }
      \draw (A1) to [l] (B1);
      \draw (A2) to [l] (B3);
      \draw (A5) to [l] (B4);
      \draw (A3) to [l] (B5);
      \draw (A4) to [l] (B2);
    \end{scope}
  \end{tikzpicture}
  \label{eq:diagram transpose}
\end{equation}
Note that $\sigma \in \B^\delta_{p,q}$ is a walled Brauer algebra diagram if and only if
$\sigma\ptp \in \S_{p+q}$.
In particular, $\dim(\B^\delta_{p,q}) = (p+q)!$ since $(\B^\delta_{p,q})\ptp = \CS_{p+q}$ as vector spaces.

The walled Brauer algebra $\B^\delta_{p,q}$ is generated by \emph{transpositions} $\sigma_i$ that swap the $i$-th and $(i+1)$-th node of the two rows,
where $i \in \set{1, \dotsc, p+q-1} \setminus \set{p}$,
and a \emph{contraction} $\bar{\sigma}_p$ between the $p$-th and $(p+1)$-th node within each row.
For example, $\B^\delta_{2,2}$ is generated by
\begin{equation}
  \begin{tikzpicture}[baseline = 0.5cm, scale = 0.8]
    \BrauerTikZStyle
    \tikzset{looseness = 1}

    \begin{scope}[yshift = 1.7cm]
      \node[anchor = east] at (0.9,0.5) {$\sigma_1 =$};
      \foreach \i in {1,...,4} {
        \fill (\i,1) circle [dot] coordinate (A\i);
        \fill (\i,0) circle [dot] coordinate (B\i);
      }
      \draw (A1) to [l] (B2);
      \draw (A2) to [l] (B1);
      \draw (A3) to [l] (B3);
      \draw (A4) to [l] (B4);
      \draw [dashed] (2.5,1.2) -- (2.5,-0.2);
    \end{scope}

    \begin{scope}
      \node[anchor = east] at (0.9,0.5) {$\bar{\sigma}_2 =$};
      \foreach \i in {1,...,4} {
        \fill (\i,1) circle [dot] coordinate (C\i);
        \fill (\i,0) circle [dot] coordinate (D\i);
      }
      \draw (C1) to [l] (D1);
      \draw (C2) to [u] (C3);
      \draw (D2) to [n] (D3);
      \draw (C4) to [l] (D4);
      \draw [dashed] (2.5,1.2) -- (2.5,-0.2);
    \end{scope}

    \begin{scope}[yshift = -1.7cm]
      \node[anchor = east] at (0.9,0.5) {$\sigma_3 =$};
      \foreach \i in {1,...,4} {
        \fill (\i,1) circle [dot] coordinate (E\i);
        \fill (\i,0) circle [dot] coordinate (F\i);
      }
      \draw (E1) to [l] (F1);
      \draw (E2) to [l] (F2);
      \draw (E3) to [l] (F4);
      \draw (E4) to [l] (F3);
      \draw [dashed] (2.5,1.2) -- (2.5,-0.2);
    \end{scope}

  \end{tikzpicture}
\end{equation}

For completeness, below is an algebraic definition of the walled Brauer algebra $\B^\delta_{p,q}$
in terms of relations obeyed by the transpositions $\sigma_i$ and contraction $\bar{\sigma}_p$.
One can easily verify that the corresponding diagrams obey these relations, while the completeness of these relations is less obvious \cite{nikitin2007centralizer}.

\begin{definition}[\cite{brundan2012gradings}]
Let $p,q \geq 0$ and $\delta \in \C$.
The \emph{walled Brauer algebra} $\B^\delta_{p,q}$ is a finite associative algebra over $\C$ with generators $\bar{\sigma}_p$ and $\sigma_i$, with $i \in \set{1,\dotsc,p+q-1} \setminus \set{p}$, and the following relations between them:
\begin{align}
  \sigma_i^2 &= 1, &
  \sigma_i \sigma_{i+1} \sigma_i &= \sigma_{i+1} \sigma_i \sigma_{i+1}, &
  \sigma_i \sigma_j &= \sigma_j \sigma_i \quad (|i-j| > 1), \\
  \bar{\sigma}_p^2 &= \delta \bar{\sigma}_p, &
  \bar{\sigma}_p \sigma_{p \pm 1} \bar{\sigma}_p &= \bar{\sigma}_p, &
  \bar{\sigma}_p \sigma_i &= \sigma_i \bar{\sigma}_p \quad (i \neq p \pm 1),
\end{align}
\vspace{-16pt}
\begin{align}
  \bar{\sigma}_p \sigma_{p+1} \sigma_{p-1} \bar{\sigma}_p \sigma_{p-1} &= \bar{\sigma}_p \sigma_{p+1} \sigma_{p-1} \bar{\sigma}_p \sigma_{p+1}, \\
  \sigma_{p-1} \bar{\sigma}_p \sigma_{p+1} \sigma_{p-1} \bar{\sigma}_p &= \sigma_{p+1} \bar{\sigma}_p \sigma_{p+1} \sigma_{p-1} \bar{\sigma}_p.
\end{align}
\end{definition}

Walled Brauer algebras have a natural notion of trace and partial trace.
The \emph{trace}
$\Tr \colon \B^\delta_{p,q} \to \C$
of a walled Brauer algebra diagram $\sigma$ is defined as
\begin{equation}
  \Tr(\sigma) \defeq \delta^{\loops(\sigma)},
  \label{eq:full trace}
\end{equation}
where $\loops(\sigma)$ denotes the number of loops formed by connecting all nodes in the top row of $\sigma$ to the corresponding nodes in the bottom row.
This definition is extended to the whole of $\B^\delta_{p,q}$ by linearity.
For any subset $S \subseteq [p+q]$, the corresponding \emph{partial trace}
$\Tr_S \colon \B^\delta_{p,q} \to \B^\delta_{p',q'}$
is defined similarly, except we connect only those pairs of nodes in $\sigma$ that are indicated by $S$:
\begin{equation}
  \Tr_S(\sigma) \defeq \delta^{\loops_S(\sigma)} \sigma',
  \label{eq:partial trace}
\end{equation}
where $\loops_S(\sigma)$ denotes the number of loops formed in this way and $\sigma' \in \B^\delta_{p',q'}$ denotes the smaller diagram left after erasing the loops.
Note that $p + q = p' + q' + |S|$ where
$p' \defeq p - |S \cap [p]|$ and
$q' \defeq q - |(S-p) \cap [q]|$.

\newcommand{\diagCycles}[2][0.5]{\,
  \tikz[scale = #1, baseline = (O)]{
    \BrauerTikZStyle
    \tikzset{looseness = 1}
    \path (1.5,0.35) coordinate (O);
    \foreach \i in {1,...,5} {
      \fill (\i,1) circle [dot] coordinate (A\i);
      \fill (\i,0) circle [dot] coordinate (B\i);
    }
    #2
    \draw (A1) to [l] (B1);
    \draw (A2) to [l] (B3);
    \draw (A3) to [l] (B2);
    \draw (A4) to [u] (A5);
    \draw (B4) to [n] (B5);
    \draw [dashed] (4.5,1.2) -- (4.5,-0.2);
  }\,
}

\newcommand{\diagCyclesSmall}[2][0.5]{\,
  \tikz[scale = #1, baseline = (O)]{
    \BrauerTikZStyle
    \tikzset{looseness = 1}
    \path (1.5,0.35) coordinate (O);
    \foreach \i in {1,...,2} {
      \fill (\i,1) circle [dot] coordinate (A\i);
      \fill (\i,0) circle [dot] coordinate (B\i);
    }
    #2
    \draw (A1) to [l] (B1);
    \draw (A2) to [l] (B2);
    \draw [dashed] (1.5,1.2) -- (1.5,-0.2);
  }\,
}

\newcommand{\close}[3]{
  \def\r{0.3}
  \draw (#2) to [controls = +(90:\r) and +(90:\r)] ++(#1\r,0)
    to [out = -90, in = 90] ++(0,-1)
    to [controls = +(-90:\r) and +(-90:\r)] (#3);
}

\begin{example}
The trace of a diagram in $\B^\delta_{4,1}$:
\begin{equation}
  \Tr \of[\Big]{\diagCycles{}}
  = \diagCycles{
      \close{-}{A1}{B1}
      \close{-}{A2}{B2}
      \close{+}{A3}{B3}
      \close{-}{A4}{B4}
      \close{+}{A5}{B5}
    }
  = \delta^3.
\end{equation}
The partial trace of the same diagram over $S = \set{2,3,4}$:
\begin{equation}
  \Tr_S \of[\Big]{
    \diagCycles{
      \foreach \i in {1,...,5} {
        \path (A\i)+(0,0.5) node {\scriptsize\textnormal{\i}};
      }
    }
  }
  = \diagCycles{
      \close{-}{A2}{B2}
      \close{+}{A3}{B3}
      \close{-}{A4}{B4}
    }
  = \delta \cdot \diagCyclesSmall{}.
\end{equation}
\end{example}

\subsection{Matrix algebra of partially transposed permutations}\label{sec:Apq}

\newcommand{\p}[1]{\underline{#1}} 

One of the central ideas of our approach is to represent matrices by diagrams and to use graphical methods to perform linear algebra operations such as matrix multiplication and partial traces.
For this purpose we consider a matrix representation of the walled Brauer algebra $\B^d_{p,q}$ by extending \cref{eq:psi basic} from a representation of $\S_p$ to a representation $\psi^d_{p,q}\colon \B^d_{p,q} \to \End(V^{p,q})$ of $\B^d_{p,q}$.
Denoting the nodes in the first row of a diagram $\sigma \in \B^d_{p,q}$ by $1,\dotsc,p+q$ and in the second row by $\p{1},\dotsc,\p{p+q}$,
the action of $\psi^d_{p,q}(\sigma)$ on the standard basis tensors of $V^{p,q}$ is given by
\begin{equation}
  \psi^d_{p,q}(\sigma) \of[\big]{ \ket{i_1} \x \dotsb \x \ket{i_{p+q}} }
  = \sum_{1 \leq i_{\p{1}}, \dotsc, i_{\p{p+q}} \leq d}
    \sigma^{i_1, \dotsc, i_{p+q}}_{i_{\p{1}}, \dotsc, i_{\p{p+q}}} \,
    \ket{i_{\p{1}}} \x \dotsb \x \ket{i_{\p{p+q}}},
  \label{eq:psi}
\end{equation}
for all $i_1, \dotsc, i_{p+q} \in \set{1,\dotsc,d}$,
where the coefficients are given by
\begin{equation}
  \sigma^{i_1, \dotsc, i_{p+q}}_{i_{\p{1}}, \dotsc, i_{\p{p+q}}} \defeq
  \begin{cases}
    1 & \text{if $i_r = i_s$ for all connected pairs of vertices} \\
      & \text{$r,s \in \set{1,\dotsc,p+q,\p{1},\dotsc,\p{p+q}}$ of $\sigma$}, \\
    0 & \text{otherwise}.
  \end{cases}
\end{equation}
Equivalently,
\begin{equation}
  \sigma^{i_1, \dotsc, i_{p+q}}_{i_{\p{1}}, \dotsc, i_{\p{p+q}}} =
  \prod_{(r,s) \in \sigma} \delta_{i_r,i_s},
  \label{eq:sigma}
\end{equation}
where the product is over all pairs $(r,s)$ of nodes that are connected in $\sigma$.

\newcommand{\diagRand}[2][0.5]{\,
  \tikz[scale = #1, baseline = (O)]{
    \BrauerTikZStyle
    \path (1.5,0.35) coordinate (O);
    \foreach \i in {1,...,5} {
      \fill (\i,1) circle [dot] coordinate (A\i);
      \fill (\i,0) circle [dot] coordinate (B\i);
    }
    #2
    \draw (A1) to [l] (B1);
    \draw (A2) to [l] (B3);
    \draw (A4) to [l] (B5);
    \draw (A3) to [u] (A5);
    \draw (B2) to [n] (B4);
    \draw [dashed] (3.5,1.2) -- (3.5,-0.2);
  }\,
}

\newcommand{\Labels}{
    \foreach \i in {1,...,5} {
        \path (A\i)+(0, 0.5) node {$x_\i$};
        \path (B\i)+(0,-0.5) node {$y_\i$};
    }
}

\begin{example}
According to \cref{eq:psi,eq:sigma},
\begin{align}
  \bra{x_1, \dotsc, x_{5}}
  \, \psi^d_{3,2} \of[\Big]{\diagRand{}} \,
  \ket{y_1, \dotsc, y_{5}}
  &= \!\!\!
    \diagRand{\Labels} \\
 &= \delta_{x_1,y_1}
    \delta_{x_2,y_3}
    \delta_{x_3,x_5}
    \delta_{x_4,y_5}
    \delta_{y_2,y_4}
\end{align}
for any choice of $x_1, \dotsc, x_5 \in [d]$ and $y_1, \dotsc, y_5 \in [d]$.
\end{example}

A crucial fact about the matrix $\psi^d_{p,q}(\sigma)$ representing a diagram $\sigma$ is that its partial traces are related to those of the diagram.
Namely, for any $\sigma \in \B^d_{p,q}$ and $S \subseteq [p+q]$,
\begin{align}
  \Tr(\psi^d_{p,q}(\sigma)) &= \Tr(\sigma), &
  \Tr_S(\psi^d_{p,q}(\sigma)) &= \psi^d_{p',q'}(\Tr_S(\sigma)),
\end{align}
where the diagrammatic traces $\Tr(\sigma)$ and $\Tr_S(\sigma)$ are defined in \cref{eq:full trace,eq:partial trace}, respectively.
We formally establish these two identities in \cref{apx:traces}.

Similar to \cref{eq:Ap}, we denote the image of $\B^d_{p,q}$ under $\psi^d_{p,q}$ by
\begin{equation}
  \A^d_{p,q} \defeq \psi^d_{p,q}(\B^d_{p,q}).
  \label{eq:Apq}
\end{equation}
This is known as \emph{matrix algebra of partially transposed permutations} as it is generated by permutation matrices on $p+q$ qudit registers, partially transposed on the last $q$ registers.
Recall that the \emph{partial transpose} $\mathsf{\Gamma}\colon \End(V^{p+q}) \to \End(V^{p,q})$ is defined as
$(M \x N)\ptp \defeq M \x N\tp$,
for all $M \in \End(V^p)$ and $N \in \End(V^q)$.
This operation can be used to relate the maps
$\psi^d_{p,q}$ and $\psi^d_{p+q}$
defined in \cref{eq:psi basic,eq:psi}, respectively:
\begin{align}
  \psi^d_{p,q}(\sigma)
  &= \of[\Big]{\psi^d_{p+q}\of[\big]{\sigma\ptp}}\ptp,
\end{align}
where $\sigma\ptp$ denotes the partial transpose of the diagram $\sigma \in \B^d_{p,q}$,
see \cref{eq:diagram transpose}.
Hence, as vector spaces, the algebras $\A^d_{p,q}$ and $\A^d_{p+q}$ are related as follows:
\begin{equation}
  \A^d_{p,q} = \of[\big]{\A^d_{p+q}}\ptp.
\end{equation}
However, because of different product operations, the two algebras are generally not isomorphic.

\newcommand{\diagSWAP}{\,
  \tikz[scale = 0.5, baseline = (O)]{
    \BrauerTikZStyle
    \tikzset{looseness = 1}
    \path (1.5,0.35) coordinate (O);
    \foreach \i in {1,2} {
      \fill (\i,1) circle [dot] coordinate (A\i);
      \fill (\i,0) circle [dot] coordinate (B\i);
    }
    \draw (A1) to [l] (B2);
    \draw (A2) to [l] (B1);
  }\,
}

\newcommand{\diagContract}{\,
  \tikz[scale = 0.5, baseline = (O)]{
    \BrauerTikZStyle
    \tikzset{looseness = 1}
    \path (1.5,0.35) coordinate (O);
    \foreach \i in {1,2} {
      \fill (\i,1) circle [dot] coordinate (A\i);
      \fill (\i,0) circle [dot] coordinate (B\i);
    }
    \draw (A1) to [u] (A2);
    \draw (B1) to [n] (B2);
    \draw [dashed] (1.5,1.2) -- (1.5,-0.2);
  }\,
}

\begin{example}[Transposition versus contraction]
When $p = 2$ and $q = 0$, the only non-trivial element of $\B^d_{2,0}$ is the transposition $(12)$.
Its matrix version acts as
\begin{equation}
  \psi^d_{2,0}\of[\Big]{\diagSWAP}\colon \ket{i} \ket{j} \mapsto \ket{j} \ket{i},
\end{equation}
for all $i,j \in \set{1,\dotsc,d}$,
since the diagram is encoded by $\sigma^{i,j}_{k,l} = \delta_{i,l} \delta_{j,k}$.
More generally, any diagram that represents a permutation (i.e., has no edges across the wall)
simply translates into the corresponding permutation of the tensor factors.

When $p = q = 1$, the only non-trivial element of $\B^d_{1,1}$ is the contraction of~$1$ and~$2$.
The corresponding matrix acts as
\begin{equation}
  \psi^d_{1,1}\of[\Big]{\diagContract}\colon \ket{i} \ket{j} \mapsto \delta_{i,j} \sum_{k=1}^{d} \ket{k} \ket{k},
\end{equation}
for all $i,j \in \set{1,\dotsc,d}$,
since $\sigma^{i,j}_{k,l} = \delta_{i,j} \delta_{k,l}$ in this case.
In particular, when $d = 2$,
\begin{align}
  \psi^2_{2,0}\of[\Big]{\diagSWAP}
  &= \mx{
      1 & 0 & 0 & 0 \\
      0 & 0 & 1 & 0 \\
      0 & 1 & 0 & 0 \\
      0 & 0 & 0 & 1
    }, &
  \psi^2_{1,1}\of[\Big]{\diagContract}
  = \mx{
      1 & 0 & 0 & 1 \\
      0 & 0 & 0 & 0 \\
      0 & 0 & 0 & 0 \\
      1 & 0 & 0 & 1
    }
  = \psi^2_{2,0}\of[\Big]{\diagSWAP}\ptp,
  \label{eq:SWAP and proj}
\end{align}
which are known as the $\SWAP$ operator and the un-normalized projector onto the canonical maximally entangled state.
\end{example}

\subsection{Mixed Schur--Weyl duality}

Mixed Schur--Weyl duality is concerned with the action of $\U(d)$ and $\B^d_{p,q}$ on the
mixed tensor product space $V^{p,q} = V^{\otimes p} \x V^{* \otimes q}$ where $p,q \geq 0$ and $V = \C^d$.
The $q = 0$ case is equivalent to the usual Schur--Weyl duality discussed in \cref{sec:Schur-Weyl}
while the $p = 0$ case is isomorphic to it.

As a generalization of \cref{eq:phi},
consider the natural representation $\phi^d_{p,q} \colon \U(d) \to \End(V^{p,q})$ of $\U(d)$ defined as
\begin{equation}
  \phi^d_{p,q}(U) \defeq U\xp{p} \x \bar{U}\xp{q}
\end{equation}
for all $U \in \U(d)$, where the entry-wise complex conjugate $\bar{U}$ is also known as the \emph{dual}\footnote{More generally, the dual of the defining representation is $M \mapsto (M^{-1})\tp$ where $M \in \GL(d,\C)$. However, when $M$ is unitary, $(M^{-1})\tp = (M\ct)\tp = \bar{M}$.}
of the \emph{defining representation}.
Similar to \cref{eq:Up}, let
\begin{equation}\label{eq:Upq}
  \CA^d_{p,q} \defeq \spn_\C \set{\phi^d_{p,q}(U) : U \in \U(d)}.
\end{equation}

We are particularly interested in the matrix algebra $\End_{\CA^d_{p,q}}(V^{p,q})$,
i.e., the centralizer of the natural $\U(d)$ action on $V^{p,q}$,
which captures the unitary equivariance condition.
Indeed, recall from \cref{propl:Choi equivariance} that a $p \to q$ channel $\Phi$ is unitary-equivariant if and only if $X^\Phi \in \smash{\End_{\CA^d_{p,q}}(V^{p,q})}$.
The following result generalizes \cref{thm:schur-weyl} to the mixed tensor product space $V^{p,q}$ and says that
$\smash{\End_{\CA^d_{p,q}}(V^{p,q})}$
is equal to the matrix algebra $\A^d_{p,q}$ of partially transposed permutations (see \cref{sec:Apq}) and,
when $d \geq p + q$, isomorphic to the walled Brauer algebra $\B^d_{p,q}$.

\begin{theorem}[Mixed Schur--Weyl duality \cite{koike1989decomposition,bchlls}]\label{thm:gen-schur-weyl}
The algebra $\CA^d_{p,q}$ is the centraliser algebra of $\A^d_{p,q}$ in $\End(V^{p,q})$ and vice versa, i.e.,
\begin{align}
    \CA^d_{p,q} &= \End_{\A^d_{p,q}}(V^{p,q}), &
    \A^d_{p,q} &= \End_{\CA^d_{p,q}}(V^{p,q}).
\end{align}
Moreover, when $d \geq p + q$ the representation $\psi^d_{p,q}$ is faithful, i.e., $\A^d_{p,q} \cong \B^d_{p,q}$.
\end{theorem}

The simplest non-trivial instance of this duality is illustrated in \cref{ex:pq1d2} below.
It is similar to \cref{ex:p2d2}, except here the second system of
$\CA^2_{1,1}$ and $\A^2_{1,1}$
is subject to the dual action of $\U(d)$ and the partial transpose, respectively.

\begin{example}[$p = q = 1$ and $d = 2$]\label{ex:pq1d2}
The algebra $\A^2_{1,1}$ is generated by the identity matrix and the matrix
$\psi^d_{1,1}\of[\Big]{\diagContract}$
given in \cref{eq:SWAP and proj}:
\begin{equation}
  \A^2_{1,1} \defeq \set*{
    x \mx{
        1 & 0 & 0 & 0 \\
        0 & 1 & 0 & 0 \\
        0 & 0 & 1 & 0 \\
        0 & 0 & 0 & 1
    } +
    y \mx{
        1 & 0 & 0 & 1 \\
        0 & 0 & 0 & 0 \\
        0 & 0 & 0 & 0 \\
        1 & 0 & 0 & 1
    } :
    x,y \in \C
  }.
\end{equation}
Note from \cref{eq:A22} that
$\A^2_{1,1} = \of{\A^2_{2,0}}\ptp$.
We can describe $\CA^2_{1,1}$ by introducing the \emph{dual representation}:
for any invertible matrix $M$, let
$M^* \defeq (M^{-1})\tp$.
Note that $M^* = \bar{M}$ when $M$ is unitary and,
for a general $M$, the entries of $M^*$ are rational functions in the entries of $M$.
In particular,
\begin{equation}
  \mx{a & b \\ c & d}^*
  = \frac{1}{a d - b c} \mx{d & -c \\ -b & a}
\end{equation}
for any $a,b,c,d \in \C$ such that $a d - b c \neq 0$.
The algebra $\CA^2_{1,1}$ is then given by
\begin{equation}
  \CA^2_{1,1} \defeq \spn_\C \set*{
    \mx{a & b \\ c & d} \x
    \mx{a & b \\ c & d}^* :
    \mx{a & b \\ c & d} \in \U(2)
  }.
\end{equation}
Using the dual two-qubit Schur transform \cite{Majority}
\begin{equation}
  U \defeq
  \frac{1}{\sqrt{2}}
  \mx{
    1 & 0 & 0 & 1 \\
    0 & -\sqrt{2} & 0 & 0 \\
    1 & 0 & 0 & -1 \\
    0 & 0 & \sqrt{2} & 0
  },
\end{equation}
we can simultaneously block-diagonalize both algebras:
\begin{align}
  U \A^2_{1,1} U\tp
 ={}& \set*{
      \left(
      \begin{array}{@{}c|ccc@{}}
        x+2y & 0 & 0 & 0 \\ \hline
        0 & x & 0 & 0 \\
        0 & 0 & x & 0 \\
        0 & 0 & 0 & x
      \end{array}
      \right) : x,y \in \C
    }, \\
  U \CA^2_{1,1} U\tp
  = \spn_\C
  & \set*{
      \frac{1}{ad-bc}
      \left(
      \begin{array}{@{}c|ccc@{}}
        ad-bc & 0 & 0 & 0 \\ \hline
        0 & a^2 & \sqrt{2} a b & b^2 \\
        0 & \sqrt{2} a c & a d+b c & \sqrt{2} b d \\
        0 & c^2 & \sqrt{2} c d & d^2
      \end{array}
      \right) : \mx{a & b \\ c & d} \in \U(2)
    }.
\end{align}
These algebras centralize each other since
\begin{align}
  U \A^2_{1,1} U\tp &= \C \+ \C I_3, &
  U \CA^2_{1,1} U\tp &= \C \+ \End(\C^3).
\end{align}
While this appears identical to \cref{eq:A22 and U22}, the decompositions of
$\A^d_{p,q}$ and $\CA^d_{p,q}$
are generally very different from those of
$\A^d_{p+q}$ and $\CA^d_{p+q}$,
see \cref{sec:Bratteli for A and B}.
\end{example}

\begin{remark}\label{rm:Schur}
\Cref{thm:gen-schur-weyl} can be seen as an instance of the Double Centralizer Theorem, see \cite[Theorem~4.54]{etingof2011introduction}. Indeed, we can rephrase it as follows: as a representation of $\A^d_{p,q} \times \CA^d_{p,q}$, the space $V^{p,q}$ decomposes as
\begin{equation}
    V^{p,q} \cong \bigoplus_{\lambda \in \Irr(\A^d_{p,q})} V^\lambda \otimes U^\lambda,
    \label{eq:schur_weyl_alt}
\end{equation}
where $V^\lambda$ are simple modules of $\A^d_{p,q}$ and $U^\lambda$ are simple modules of $\CA^d_{p,q}$.
It follows from \cref{propl:Choi equivariance} that
there exists a \emph{mixed Schur transform} $U_{\textnormal{Sch}(p,q)} \in \U(V^{p,q})$ that block-diagonalizes any unitary-equivariant Choi matrix $X^\Phi$ as follows:
\begin{equation}
    U_{\textnormal{Sch}(p,q)} \, X^\Phi \, U_{\textnormal{Sch}(p,q)}\ct
    = \bigoplus_{\lambda \in \Irr(\A^d_{p,q})}
      \sof*{
        X^\Phi_\lambda \x I_{m_\lambda}
      },
      \label{rmk:eq_blocks}
\end{equation}
where the operator $X^\Phi_\lambda$ acts on a $\A^d_{p,q}$-register of dimension $d_\lambda$,
and the identity matrix $I_{m_\lambda}$ acts on a $\CA^d_{p,q}$-register of dimension $m_\lambda$.
When $q = 0$, $U_{\textnormal{Sch}(p,q)}$
reduces to the usual Schur transform considered in \cite{HarrowThesis}.
\end{remark}

The distinction between $\A^d_{p,q}$ and $\B^d_{p,q}$ is crucial if one is interested in small dimensions $d < p + q$ since the two algebras are not isomorphic in this case.
For example, the algebra $\A^d_{p,q}$ is always semisimple because $\CA^d_{p,q}$ is known to be semisimple from the representation theory of Lie groups \cite[Theorem 4.66]{etingof2011introduction}, so its commutant $\A^d_{p,q}$ must also be semisimple by the Double Centralizer Theorem \cite[Theorem~4.54]{etingof2011introduction}. However, $\B^d_{p,q}$ is not semisimple for integer $d < p + q - 1$ \cite{cox2008blocks}.

Previously the algebra $\A^d_{p,q}$ was studied in \cite{koike1989decomposition,bchlls} in the context of mixed Schur--Weyl duality. For $q = 1$, it was explicitly studied in \cite{zhang2007permutation,studzinski2013commutant,mozrzymas2014structure,mozrzymas2018simplified} motivated by applications in quantum information. Some aspects of it were also studied for general $q$ in \cite{studzinski2020efficient,mozrzymas2021optimal}.

One of our main technical results is the determination of primitive central idempotents of the matrix algebras $\A^d_{p,q}$, see \cref{sec:A idempotents}, which will play a central role in our results in \cref{sec:SDP} on linear programming with unitary-equivariant constraints.
In the following section we summarize a general method for constructing primitive central idempotents of multiplicity-free families of algebras due to \cite{doty2019canonical}.

\section{Review of the \texorpdfstring{\cite{doty2019canonical}}{[DLS18]} construction}\label{sec:DLS}

Inspired by the Okounkov--Vershik approach \cite{OV1996,VO2005} to the representation theory of the symmetric groups and their algebras,
the authors of \cite{doty2019canonical} present a general algorithm for finding the primitive central idempotents of any multiplicity-free family of algebras.
In this section, we introduce the necessary background on algebras and their representations, and then summarize the \cite{doty2019canonical} algorithm.
We will apply this algorithm in \cref{sec:A idempotents} to the algebra $\A^d_{p,q}$ of partially transposed permutation matrices defined in \cref{eq:Apq}.
The idempotents of $\smash{\A^d_{p,q}}$ are the main ingredient of our optimization framework in \cref{sec:SDP}.

\subsection{Background on algebras and their modules}\label{sec:algebras}

An \emph{algebra} over a field is a vector space over this field, equipped with a bilinear product.
A canonical example of an algebra is the full matrix algebra $\End(V)$, i.e., the set of all linear operators acting on some vector space $V$.
As part of our definition, we assume that all algebras are
\emph{complex}, \emph{finite-dimensional}, \emph{associative}, and \emph{unital}.
Namely, the vector space underlying an algebra $\A$ is complex and finite-dimensional,
the product operation in $\A$ is associative,
and there is a \emph{unit} element $1 \in \A$ such that
$1 a = a 1 = a$ for any $a \in \A$.
Similar to Cayley's theorem for groups, any algebra is isomorphic to a subalgebra of the full matrix algebra $\End(\C^{\dim \A})$.
Indeed, the action of $\A$ on any basis of $\A$ produces a matrix algebra that is analogous to the left-regular representation of a group.
Due to the Artin--Wedderburn Theorem, an algebra over $\C$ is \emph{semisimple} if it is isomorphic to a direct sum of full matrix algebras over $\C$.
If $\A$ and $\B$ are algebras,
$\varphi\colon \A \to \B$ is an \emph{algebra embedding}
if $\varphi$ is an injective homomorphism.
The embedding is \emph{unity-preserving} if $\varphi(1_\A) = 1_\B$.
We write $\A \hookrightarrow \B$ to mean that such embedding exists (in such case one can intuitively think of $\A$ as a subalgebra of $\B$). If $\B \subseteq \A$ is a subalgebra then we denote by $\Z_{\B}(\A)$ the \emph{centralizer} of $\B$ in $\A$:
\begin{equation}
    \Z_{\B}(\A) \defeq \set{a \in \A : ab = ba \text{ for every } b \in \B}.
    \label{eq:centralizer_abstract}
\end{equation}
If $\B = \A$ in the above definition then $\Z(\A) \defeq \Z_{\A}(\A)$ is known as the \emph{center} of the algebra $\A$.
Note that $\End_\B(V)$ defined in \cref{eq:centralizer} is a special case of $\Z_\B(\A)$ where $\A = \End(V)$.

If an algebra $\A$ acts on a complex vector space $V$, we call $V$ an \emph{$\A$-module}\footnote{Module is generally defined as a ring acting on an abelian group. Note that any unital, associative algebra is a ring and any vector space under addition is an abelian group.}
(in the context of groups it is analogous to the notion of a group representation).
A \emph{submodule} of an $\A$-module $V$ is a subspace $W$ of $V$ such that $a w \in W$
for all $w \in W$ and $a \in \A$. (Note that $W$ is an $\A$-module in its own right.)
If an $\A$-module $V$ has submodules $W_1$ and $W_2$ such that $V = W_1 \+ W_2$ as a vector space then we say that $V$ is the direct sum of $W_1$ and $W_2$.
A module $V$ is \emph{indecomposable} if it is not the direct sum of two non-zero submodules, and is \emph{decomposable} otherwise.
A module $V$ is \emph{simple} if $V$ has no submodules except $V$ and $0$ (in the context of groups, simple modules are known as irreducible representations).
We denote the set of all non-isomorphic simple modules of $\A$ by $\Irr(\A)$. Given a subalgebra $\A$ of an algebra $\B$ and an $\B$-module $V$ we can consider its \emph{restriction} to the subalgebra $\A$. We denote such $\A$-module $V$ by $\Res^\B_\A V$.
See \cite{FinitedimAlgebras,Cox} for more background on finite-dimensional algebras and their modules.

\subsection{Multiplicity-free families of algebras}

A particularly nice situation is when all restrictions of an algebra are multiplicity-free. This is formalized in the following definition which is motivated by the canonical example of the sequence $\CS_0 \hookrightarrow \CS_1 \hookrightarrow \dots \hookrightarrow \CS_n$ of symmetric group algebras considered in \cite{OV1996,VO2005}.

\begin{definition}[Definition~1.1 in \cite{doty2019canonical}]\label{def:mult-free family}
A family $\A_0, \dotsc, \A_n$ of finite-dimen\-sional semisimple\footnote{In \cite{doty2019canonical} the algebras are also required to be \emph{split}. However, since we consider only algebras over $\C$, which is an algebraically closed field, in our setting all algebras are automatically split.} algebras over $\C$ is \emph{multiplicity-free} if the following axioms hold:
\begin{enumerate}[(a)]
  \item $\A_{0} \cong \C$.
  \item For each $k$, there is a unity-preserving algebra embedding $\A_k \hookrightarrow \A_{k+1}$.
  \item The restriction of a simple $\A_k$-module to $\A_{k-1}$ is isomorphic to a direct sum of pairwise non-isomorphic simple $\A_{k-1}$-modules. We say that in that case the restriction from $\A_k$ to $\A_{k-1}$ is \emph{multiplicity-free}.
\end{enumerate}
\end{definition}

\subsection{Bratteli diagram}

Given a multiplicity-free family of algebras, one can create a graph that shows how different simple modules of these algebras restrict to their subalgebras.
We denote by $V^\lambda$ a simple $\A$-module corresponding to $\lambda \in \Irr(\A)$. The dimension of this module is denoted by $d_\lambda \defeq \dim V^\lambda$.
We can represent the multiplicity-free restrictions among $\A_k$ by a directed acyclic graph known as Bratteli diagram \cite{Bratteli}.

\begin{definition}\label{def:Bratteli}
Let $\A_0, \dotsc, \A_n$ be a multiplicity-free family of algebras.
Its \emph{Bratteli diagram} is a directed acyclic graph whose vertices are the isomorphism classes $\bigsqcup_{k=0}^n \Irr(\A_k)$ of simple $\A_k$-modules.
There is an edge $\lambda \rightarrow \mu$ from vertex $\lambda \in \Irr(\A_k)$ to vertex $\mu \in \Irr(\A_{k+1})$ if and only if $V^{\lambda}$ is isomorphic to a direct summand of
$\Res^{\A_{k+1}}_{\A_k}V^{\mu}$,
the restriction of $V^{\mu}$ to the subalgebra $\A_k$.
We call $\Irr(\A_k)$ the $k$-th \emph{level} of the Bratteli diagram.
We denote the unique vertex in $\Irr(\A_0)$ by $\0$\footnote{We will later use $(\0,\0)$ for the root of the Bratteli diagram of the walled Brauer algebra because the irreducible representations of this algebra are labeled by pairs of Young diagrams.} and call it the \emph{root}, while $\Irr(\A_n)$ are called \emph{leaves}.
\end{definition}

\begin{figure}
\begin{tikzpicture}[> = latex,
  cut/.style = {thin, dashed, red, rounded corners = 12pt, fill = red, fill opacity = 0.12},
  every node/.style = {inner sep = 3pt, anchor = west}]
  \def\W{2.0cm}
  \def\H{1cm}
  \foreach \i in {0,1,2,3,4} {
    \node at (\i*\W,2.9*\H) {$\CS_\i$};
  }
  \node (0)    at (0*\W, 0*\H) {$\0$};
  \node (1)    at (1*\W, 0*\H) {\yd{1}};
  \node (2)    at (2*\W, 1*\H) {\yd{2}};
  \node (11)   at (2*\W,-1*\H) {\yd{1,1}};
  \node (3)    at (3*\W, 2*\H) {\yd{3}};
  \node (21)   at (3*\W, 0*\H) {\yd{2,1}};
  \node (111)  at (3*\W,-2*\H) {\yd{1,1,1}};
  \node (4)    at (4*\W, 2.0*\H) {\yd{4}};
  \node (31)   at (4*\W, 1.0*\H) {\yd{3,1}};
  \node (22)   at (4*\W, 0.0*\H) {\yd{2,2}};
  \node (211)  at (4*\W,-1.0*\H) {\yd{2,1,1}};
  \node (1111) at (4*\W,-2.0*\H) {\yd{1,1,1,1}};
  \draw[->] (0) -- (1);
  \draw[->] (1) -- (2.190);
  \draw[->] (1) -- (11.170);
  \draw[->] (2) -- (3.190);
  \draw[->] (2) -- (21.165);
  \draw[->] (11) -- (21.195);
  \draw[->] (11) -- (111.170);
  \draw[->] (3) -- (4);
  \draw[->] (3) -- (31);
  \draw[->] (21) -- (31);
  \draw[->] (21) -- (22);
  \draw[->] (21) -- (211);
  \draw[->] (111) -- (211);
  \draw[->] (111) -- (1111);
  \begin{scope}[on background layer]
    \path (211) coordinate (v);
    \draw[cut, rounded corners = 18pt] (v)++(0.7,0.5) -- ++(-3.4,0) -- ++(0,-2.2) -- ++(3.4,0);
    \path (v)+(1.0,0) node {$d=2$};
    \path (v)+(0,-\H) coordinate (w);
    \draw[cut] (w)++(0.7,0.5) -- ++(-1.5,0) -- ++(0,-1) -- ++(1.5,0);
    \path (w)+(1.0,0) node {$d=3$};
  \end{scope}
\end{tikzpicture}
\caption{\label{fig:S4}Bratteli diagram for the symmetric group algebras
$\CS_0 \hookrightarrow
 \CS_1 \hookrightarrow
 \CS_2 \hookrightarrow
 \CS_3 \hookrightarrow
 \CS_4$,
also known as Young's lattice.
The Bratteli diagram for the permutation matrix algebras
$\A^d_0 \hookrightarrow
 \A^d_1 \hookrightarrow
 \A^d_2 \hookrightarrow
 \A^d_3 \hookrightarrow
 \A^d_4$
defined in \cref{eq:Ap} is the same when $d \geq 4$.
When $d = 2$ or $d = 3$, vertices with Young diagrams containing more than $d$ rows are removed.}
\end{figure}
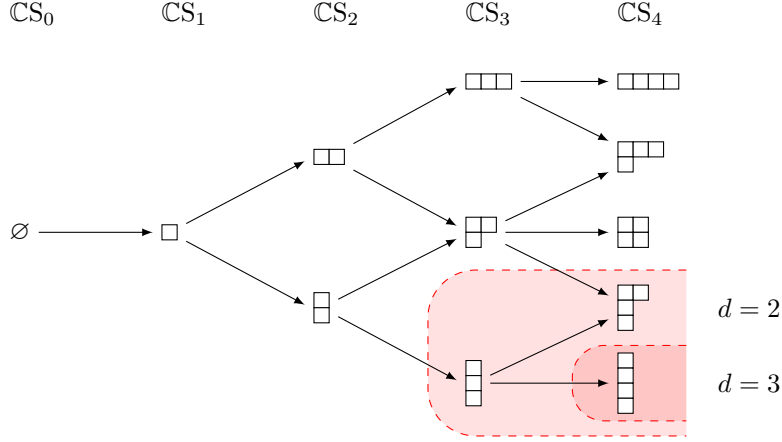

An example of a Bratteli diagram for a multiplicity-free family of semisimple symmetric group algebras is shown in \cref{fig:S4}.


\subsection{Primitive central idempotents}

Idempotents of an algebra play an important role in its structure.

\begin{definition}
An \emph{idempotent} $e$ in the algebra $\A$ is an element with the property $e^2=e$. Two idempotents $a,b \in \A$ are said to be \emph{orthogonal} if $ab=ba=0$. A \emph{central} idempotent $e \in \A$, is an idempotent that commutes with every element $a \in \A$, i.e., $ea=ae$.
\end{definition}

Certain types of idempotents are extremely useful for studying the representation theory of a given algebra. We define them as follows.

\begin{definition}
A \emph{primitive idempotent} is an idempotent that cannot be written as a sum of two nonzero orthogonal idempotents. A \emph{primitive central idempotent} is a central idempotent that cannot be written as a sum of two nonzero orthogonal central idempotents.
\end{definition}

If algebra $\A$ is semisimple, we have the isomorphism
\begin{equation}
  \A = \bigoplus_{\lambda \in \Irr(\A)} \varepsilon(\lambda) \A
  \cong \bigoplus_{\lambda \in \Irr(\A)} \End(V^\lambda)
  \label{eq:A direct sum}
\end{equation}
where $\varepsilon(\lambda)$ are the \emph{primitive central  idempotents} of $\A$ and the first direct sum should be understood as a decomposition of the left-regular representation of $\A$.
This isomorphism is very useful since it allows to think of an abstract semisimple algebra $\A$ as an algebra of block-diagonal matrices, a perspective that we will repeatedly use.
Note from \cref{eq:A direct sum} that
$\dim \A = \sum_{\lambda \in \Irr(\A)} d_\lambda^2$,
which is analogous to the dimension formula for irreducible representations of groups.
The primitive central idempotents $\varepsilon(\lambda)$ are in one-to-one correspondence with simple $\A$-modules labeled by $\lambda \in \Irr(\A)$ and provide a resolution of the unit element of the algebra $\A$: $\sum_{\lambda \in \Irr(\A)} \varepsilon(\lambda) = 1$.

\subsection{Gelfand--Tsetlin basis and subalgebra}\label{sec:GT subalgebra}

Let $\A_0, \dotsc, \A_n$ be a multiplicity-free family of algebras (see \cref{def:mult-free family}).
We denote by $\Paths(\lambda)$ the set of all paths in the Bratteli diagram starting from the root $\0 \in \Irr(\A_0)$ and terminating at the leaf $\lambda \in \Irr(\A_n)$.
Any $\T \in \Paths(\lambda)$ has the form
\begin{equation}
  \T = \lambda_0 \to \lambda_1 \to \cdots \to \lambda_n
  \label{eq:T}
\end{equation}
where $\lambda_{0}=\0$ and $\lambda_{n}=\lambda$.
Let $\T[k] \defeq \lambda_k$ denote the $k$-th vertex on the path $\T$.
We also define $\Paths(n) \defeq \bigcup_{\lambda \in \Irr(\A_n)} \Paths(\lambda)$ and say that
$\T \in \Paths(n)$ is a path of \emph{length} $n$.

We can now define a certain basis for the direct sum
$\bigoplus_{\lambda \in \Irr(\A_n)} V^\lambda$
of all simple $\A_n$-modules, where each element of the basis corresponds to a path in the Bratteli diagram.
This basis can be obtained by choosing any leaf $\lambda \in \Irr(\A_n)$ and considering the restriction $\smash{\Res^{\A_n}_{\A_{n-1}} V^\lambda}$ of the corresponding simple $\A_n$-module $V^\lambda$ to $\A_{n-1}$, which according to \cref{def:mult-free family} is multiplicity-free.
This restriction can then be iterated further along any path in the Bratteli diagram towards the root $\0$ that corresponds to the one-dimensional algebra $\A_0 \cong \C$.
Doing this along all $\Paths(\lambda)$ between $\0$ and $\lambda$ results in a decomposition of the chosen simple $\A_n$-module $V^\lambda$ into one-dimensional simple $\A_0$-modules.
Repeating this procedure for all leaves $\lambda \in \Irr(\A_n)$ produces the \emph{Gelfand--Tsetlin basis} of $\bigoplus_{\lambda \in \Irr(\A_n)} V^\lambda$:
\begin{equation}
    \set[\big]{\ket{\T} : \T \in \Paths(n)}.
\end{equation}
These vectors are labeled by elements of $\Paths(n)$ since each sequence of restrictions corresponds to some leaf-root path in the Bratteli diagram.

To find the Gelfand--Tsetlin basis explicitly, we can look at the maximal commutative subalgebras of $\A_k$. Due to \cref{eq:A direct sum} one can think of them as subalgebras of diagonal matrices, carrying the information about the projectors onto the Gelfand--Tsetlin basis. This motivates the following definition.

\begin{definition}\label{def:GT subalgebra}
For each $k \in [n]$, the corresponding \emph{Gelfand--Tsetlin subalgebra} is
\begin{equation}
  \X_k \defeq \ip{\Z(\A_1), \dotsc, \Z(\A_k)} \subseteq \A_k
\end{equation}
where $\Z(\A_i)$ denotes the center of the subalgebra $\A_i$.
\end{definition}

Note that $\X_1 \subseteq \dotsb \subseteq \X_n$.
The Gelfand--Tsetlin subalgebra $\X_k$ is a maximal commutative subalgebra of $\A_k$, see Proposition~1.1 of \cite{OV1996,VO2005}.
We will later find a particular set of generators for $\X_k$ that act nicely on the Gelfand--Tsetlin basis, which will help us to construct the primitive central idempotents of $\A_n$.

For each path
$\T = \lambda_0 \to \lambda_1 \to \cdots \to \lambda_n \in \Paths(n)$
in the Bratteli diagram, set
\begin{equation}
  \varepsilon_\T \defeq \varepsilon(\lambda_{1}) \varepsilon(\lambda_{2}) \cdots \varepsilon(\lambda_{n})
  \label{eq:epsT}
\end{equation}
where $\varepsilon(\lambda_i)$ are the primitive central idempotents of $\A_i$, see \cref{eq:A direct sum}.
Note that $\varepsilon_\T$ is an element of the Gelfand--Tsetlin subalgebra $\X_{n}$ since $\varepsilon(\lambda_i) \in \Z(\A_i)$ for each $i$.

\begin{proposition}[Proposition~1.6 and Corollary~1.7 \cite{doty2019canonical}]\label{prop:GT basis}
The collection $\set{\varepsilon_\T : \T \in \Paths(n)}$ is a family of orthogonal primitive idempotents in $\mathcal{A}_{n}$ that sums to the identity $1$ and is a basis for the Gelfand--Tsetlin subalgebra $\X_{n}$. Moreover, the primitive central idempotents of $\A_n$ are given by
\begin{equation}
  \varepsilon(\lambda) = \sum_{\T \in \Paths(\lambda)} \varepsilon_\T.
\end{equation}
\end{proposition}

Using the isomorphism in \cref{eq:A direct sum}, the primitive idempotents $\varepsilon_\T$ correspond to the projectors $\proj{\T}$ onto the Gelfand--Tsetlin basis vectors $\ket{\T} \in \bigoplus_{\lambda \in \Irr(\A_n)} V^\lambda$.

\subsection{Jucys--Murphy elements}

In this section, we define a certain nice set of elements of the algebra $\A_k$ that generate the Gelfand--Tsetlin subalgebra $\X_k$.
They are commonly known as Jucys--Murphy elements.

\begin{definition}[Definition~3.1 in \cite{doty2019canonical}]\label{def:add central and separating}
Let $\A_0, \dotsc, \A_n$ be a multiplicity-free family of algebras
and let $\X_1, \dotsc, \X_n$ be their Gelfand--Tsetlin subalgebras.
Let $J_1, \dotsc, J_n$ be a sequence of elements in $\A_n$ such that $J_{k} \in \X_{k}$ for each $k \in [n]$. This sequence is
\begin{enumerate}[(a)]
  \item \emph{additively central} if $J_1 + \dotsb + J_k \in \Z(\A_k)$ for all $k \in [n]$,
  \item \emph{separating} if $\X_k = \ip{J_1, \dotsc, J_k}$ for all $k \in [n]$.
\end{enumerate}
It is a \emph{Jucys--Murphy sequence} if it is both additively central and separating.
\end{definition}

Since $J_1, \dotsc, J_n \in \X_n$ and
$\set{\varepsilon_\T : \T \in \Paths(n)}$
is a basis of $\X_n$ due to \cref{prop:GT basis},
we can expand each $J_k$ as a linear combination of $\varepsilon_\T$.

\begin{definition}\label{def:J and c}
For a given sequence $J_1, \dotsc, J_n$ with $J_k \in \X_k$,
we define scalars $c_\T(1), \dotsc, c_\T(n) \in \C$ such that for all $k \in [n]$:
\begin{equation}
  J_k = \sum_{\T \in \Paths(n)} c_\T(k) \varepsilon_\T.
\end{equation}
\end{definition}

Note that under the isomorphism in \cref{eq:A direct sum} $\{ c_\T(k) : \T \in \Paths(n)\}$ are the eigenvalues of $J_k$.
An important observation regarding the $c_\T(k)$ is that the value of $c_\T(k)$ does not depend on the whole path $\T$ but only on the vertices $\T[k]$ and $\T[k-1]$, see Lemma~3.9 in \cite{doty2019canonical} which is a consequence of the property (a) in \cref{def:add central and separating}.
This means that the number $c_\T(k)$ can be assigned to the edge $\T[k-1] \to \T[k]$ in the Bratteli diagram and we can equivalently write
\begin{equation}
  c_{\T[k-1] \to \T[k]} \defeq c_\T(k).
  \label{def:c edge}
\end{equation}

The concepts introduced above are well-established for the group algebra $\CS_n$, as summarized in the following example.

\begin{example}[$\CS_n$]\label{ex:content}
The Bratteli diagram for $\CS_0 \hookrightarrow \CS_1 \hookrightarrow \dots \hookrightarrow \CS_n$ is known as \emph{Young lattice}, see \cref{fig:S4}. A path $\T$ in this Bratteli diagram can also be viewed as a \emph{standard Young tableau}. Jucys--Murphy elements of $\CS_n$ \cite{jucys1974symmetric,murphy1981new} are
\begin{equation}
  J_k \defeq
  \begin{cases}
    0 & \text {if } k = 1, \\
    \sum_{i=1}^{k-1} \sigma_{i,k} & \text{if } 2 \leq k \leq n,
  \end{cases}
  \label{eq:JM for Sn}
\end{equation}
where $\sigma_{i,k}$ is the transposition of elements $i$ and $k$. Moreover,
\begin{equation}
    c_\T(k) \defeq j - i,
    \label{eq:content}
\end{equation}
where $i$ and $j$ are the coordinates of the cell $(i, j)$ occupied by $k$ in the standard Young tableau $\T$. The number $j-i$ is also known as the \emph{content} of cell $(i,j)$ in a Young diagram.
The \emph{content} of a Young diagram $\lambda$ is defined as the total content of all its cells:
\begin{equation}
  \cont(\lambda) \defeq \sum_{(i,j) \in \lambda} (j-i)
  \label{eq:content of lambda}
\end{equation}
where the sum runs over all cells in the diagram $\lambda$.
\end{example}

\subsection{DLS algorithm for computing primitive idempotents}\label{sec:dls algorithm}

We have all ingredients to state the \cite{doty2019canonical} algorithm for computing primitive central and canonical primitive pairwise orthogonal idempotents of any multiplicity-free family $\A_0, \dotsc, \A_n$ of semisimple finite-dimensional algebras.

Following \cite{doty2019canonical}, we assign to each edge
$\lambda \to \mu$ between levels $k-1$ and $k$ of the Bratteli diagram an \emph{interpolating polynomial} $P_{\lambda \to \mu}$ of $x \in \A_k$ defined as
\begin{equation}
  P_{\lambda \to \mu} (x) \defeq
    \prod_{\tilde{\mu} : \lambda \to \tilde{\mu} \neq \mu}
    \frac{x-c_{\lambda \to \tilde{\mu}}}{c_{\lambda \to \mu}-c_{\lambda \to \tilde{\mu}}},
  \label{eq:P}
\end{equation}
where the product is over all edges $\lambda \to \tilde{\mu}$ (other than $\lambda \to \mu$) outgoing from the vertex $\lambda$.
According to their main result \cite[Theorem~3.11]{doty2019canonical},
the primitive central idempotents of $\A_k$ can be computed recursively
for any $k \in [n]$ and $\mu \in \Irr(\A_k)$
as follows:
\begin{equation}
  \varepsilon(\mu) =
    \sum_{\lambda : \lambda \to \mu}
    P_{\lambda \to \mu} (J_k) \varepsilon(\lambda),
    \label{eq:primitive_central_main}
\end{equation}
where the sum is over all edges $\lambda \to \mu$ incoming into $\mu$ and
$J_1, \dotsc, J_n$ is a Jucys--Murphy sequence for the algebras $\A_0, \dotsc, \A_n$. The base case of the recursion is $\varepsilon(\0) = 1$.
According to \cite[Theorem~3.8]{doty2019canonical},
the canonical primitive idempotents corresponding to the Gelfand--Tsetlin basis can be found by substituting \cref{eq:P} into \cref{eq:epsT}:
\begin{equation}
  \varepsilon_\T
  = \prod_{k=1}^n P_{\lambda_{k-1} \to \lambda_{k}} (J_k)
  = \prod_{k=1}^n
    \prod_{\mu : \lambda_{k-1} \to \mu \neq \lambda_{k}}
    \frac{J_k - c_{\lambda_{k-1} \to \mu}}{c_{\lambda_{k-1} \to \lambda_{k}}-c_{\lambda_{k-1} \to \mu}}
    \label{eq:primitive_main}
\end{equation}
where $\T = \lambda_0 \to \lambda_1 \to \cdots \to \lambda_n$ is a path in the Bratteli diagram.

Using these formulas requires the following data about the family $\A_0, \dotsc, \A_n$:
\begin{enumerate}
  \item the Bratteli diagram of $\A_0, \dotsc, \A_n$,
  \item a Jucys--Murphy sequence $J_1, \dotsc, J_n$ for $\A_0, \dotsc, \A_n$,
  \item the scalars $c_\T(k)$ for all $k \in [n]$ and paths $\T \in \Paths(n)$ in the Bratteli diagram.
  \end{enumerate}
In the following section we discuss how this information can be obtained for the family of partially transposed permutation matrix algebras $\A^d_{p,q}$ using the same known data for walled Brauer algebras $\B^d_{p,q}$.

\section{Adapting \texorpdfstring{\cite{doty2019canonical}}{[DLS18]} to the matrix algebras \texorpdfstring{$\A^d_{p,q}$}{Apq}}\label{sec:A idempotents}

In this section, we provide the necessary ingredients for applying the \cite{doty2019canonical} framework to the partially transposed permutation matrix algebras $\A^d_{p,q}$.
Our main technical contribution is \cref{thm:main technical} which shows
that Jucys--Murphy elements of partially transposed permutation matrix algebras can be obtained from Jucys--Murphy elements of walled Brauer algebras, even when the corresponding walled Brauer algebras are not semisimple.

Consider the following multiplicity-free family of walled Brauer algebras:
\begin{equation}
  \C \cong
  \B^d_{0,0} \hookrightarrow
  \B^d_{1,0} \hookrightarrow \cdots \hookrightarrow
  \B^d_{p,0} \hookrightarrow
  \B^d_{p,1} \hookrightarrow \cdots \hookrightarrow
  \B^d_{p,q},
\end{equation}
where the embeddings correspond to adding on the right of the diagram an extra pair of nodes that are connected with a vertical line.
For the sake of brevity, let us denote this family by $\B \defeq (\B_0, \dotsc, \B_{p+q})$ where
\begin{equation}\label{eq:Bk}
  \B_k \defeq
  \begin{cases}
    \C & \text{if } k = 0, \\
    \B^d_{k,0} & \text{if } 1 \leq k \leq p, \\
    \B^d_{p,k-p} & \text{if } p+1 \leq k \leq p+q.
  \end{cases}
\end{equation}

Similarly, let $\A \defeq (\A_0, \dotsc, \A_{p+q})$ where,
for every $k \in \set{0,\dotsc,p+q}$,
\begin{equation}
  \A_k \defeq \psi^d_{p,q}(\B_k) \subseteq \End(V^{p,q})
\end{equation}
is the corresponding partially transposed permutation matrix algebra
and $\psi^d_{p,q}$ is the map from \cref{eq:psi}.
Note that $\A_{k-1}$ is the subalgebra of $\A_k$ that consists of all matrices of the form $M \x I_d$ for some $M$.

For every $k \in [p+q]$, let
\begin{align}
  \X^\B_k &\defeq \ip{\Z(\B_1), \dotsc, \Z(\B_k)}, &
  \X^\A_k &\defeq \ip{\Z(\A_1), \dotsc, \Z(\A_k)}
  \label{eq:GT subalgebras}
\end{align}
denote the Gelfand--Tsetlin subalgebras of $\B$ and $\A$, respectively, see \cref{def:GT subalgebra}. Sometimes we abuse the notation and refer to $\A$ (or $\B$) as the Bratteli diagram of the corresponding algebra family.

\subsection{Bratteli diagram for walled Brauer algebras}\label{sec:Bratteli for B}
The simple modules of the walled Brauer algebra $\B^\delta_{p,q}$ are labeled by pairs of Young diagrams $(\lambda^l,\lambda^r)$ where $\lambda^l$ and $\lambda^r$ are partitions of $p-k$ and $q-k$ for some $0 \leq k \leq \min(p,q)$:
\begin{equation}
    \Irr(\B^\delta_{p,q}) = \set[\big]{ \lambda = (\lambda^l,\lambda^r) : 0 \leq k \leq \min(p,q), \lambda^l \pt p - k, \lambda^r \pt q - k}.
    \label{eq:IrrB}
\end{equation}
The Bratteli diagram (see \cref{def:Bratteli}) for the family $\B$ of walled Brauer algebras is defined as follows \cite{FusionProcedure}.
The only vertex at level $k=0$ is the root $(\0, \0)$.
For any $k \in [p+q]$, the vertices at level $k$ are given by $\Irr(\B_k)$, see \cref{eq:Bk,eq:IrrB}.
For any pair of adjacent levels $k-1$ and $k$ where $k \in [p+q]$, an edge $\lambda \to \mu$ between $\lambda \in \Irr(\B_{k-1})$ and $\mu \in \Irr(\B_k)$ is present if and only if
\begin{enumerate}
    \item $k \leq p$ and the diagram $\mu$ is obtained from $\lambda$ by adding one cell to the diagram $\lambda^l$,
    \item $k > p$ and $\mu$ is obtained from $\lambda$ by either adding a cell to $\lambda^r$ or removing a cell from $\lambda^l$.
\end{enumerate}
For example, the Bratteli diagram for the multiplicity-free family ending with $\B^\delta_{2,2}$ is given in \cref{fig:B22}.

By construction, the number of paths from the root vertex to any leaf $\lambda$ in the Bratteli diagram is equal to the dimension $d_\lambda = \dim(V^\lambda)$ of the corresponding simple module of $\B^\delta_{p,q}$.


\newcommand{\br}[1]{$\of*{#1}$}

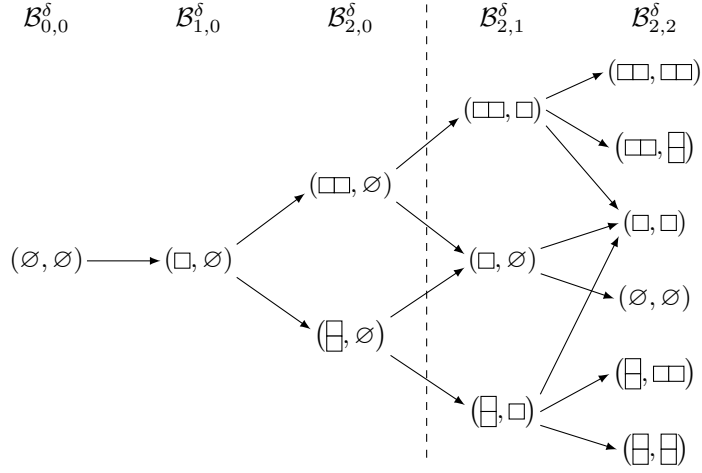
\begin{figure}
\begin{tikzpicture}[> = latex,
  every node/.style = {inner sep = 1pt}]
  \def\W{2.0cm}
  \def\H{1cm}
  \foreach \i/\p/\q in {0/0/0, 1/1/0, 2/2/0, 3/2/1, 4/2/2} {
    \node at (\i*\W,3.2*\H) {$\B^\delta_{\p,\q}$};
  }
  \draw[dashed] (2.5*\W,3.4*\H) -- (2.5*\W,-2.7*\H);
  \node (0!0)   at (0*\W, 0*\H) {\br{\0,\0}};
  \node (1!0)   at (1*\W, 0*\H) {\br{\yd{1},\0}};
  \node (2!0)   at (2*\W, 1*\H) {\br{\yd{2},\0}};
  \node (11!0)  at (2*\W,-1*\H) {\br{\yd{1,1},\0}};
  \node (2!1)   at (3*\W, 2*\H) {\br{\yd{2},\yd{1}}};
  \node (1!0')  at (3*\W, 0*\H) {\br{\yd{1},\0}};
  \node (11!1)  at (3*\W,-2*\H) {\br{\yd{1,1},\yd{1}}};
  \node (2!2)   at (4*\W, 2.5*\H) {\br{\yd{2},\yd{2}}};
  \node (2!11)  at (4*\W, 1.5*\H) {\br{\yd{2},\yd{1,1}}};
  \node (1!1)   at (4*\W, 0.5*\H) {\br{\yd{1},\yd{1}}};
  \node (0!0')  at (4*\W,-0.5*\H) {\br{\0,\0}};
  \node (11!2)  at (4*\W,-1.5*\H) {\br{\yd{1,1},\yd{2}}};
  \node (11!11) at (4*\W,-2.5*\H) {\br{\yd{1,1},\yd{1,1}}};
  \draw[->] (0!0) -- (1!0);
  \draw[->] (1!0.north east) -- (2!0.190);
  \draw[->] (1!0.south east) -- (11!0.170);
  \draw[->] (2!0.north east) -- (2!1.190);
  \draw[->] (2!0.south east) -- (1!0'.170);
  \draw[->] (11!0.north east) -- (1!0'.190);
  \draw[->] (11!0.south east) -- (11!1.170);
  \draw[->] (2!1.15) -- (2!2.180);
  \draw[->] (2!1.0) -- (2!11.180);
  \draw[->] (2!1.-20) -- (1!1.160);
  \draw[->] (1!0') -- (1!1.180);
  \draw[->] (1!0') -- (0!0'.180);
  \draw[->] (11!1.25) -- (1!1.200);
  \draw[->] (11!1.0) -- (11!2.180);
  \draw[->] (11!1.-20) -- (11!11.180);
\end{tikzpicture}
\caption{Bratteli diagram associated to the multiplicity-free family $\C \cong \B^\delta_{0,0} \hookrightarrow \B^\delta_{1,0} \hookrightarrow \B^\delta_{2,0} \hookrightarrow
\B^\delta_{2,1} \hookrightarrow
\B^\delta_{2,2}$ of walled Brauer algebras
when they are semisimple.}
\label{fig:B22}
\end{figure}

\subsection{Jucys--Murphy elements for walled Brauer algebras}

If $\delta \in \C$ is such that the walled Brauer algebras $\B^\delta_{p,q}$ are semisimple, their Jucys--Murphy elements are given by
\cite{brundan2012gradings,sartori2015walled,jung2020supersymmetric}
(cf.~\cref{ex:content})
\begin{equation}
  J^{\B}_k \defeq
  \begin{cases}
    0 & \text {if } k = 1, \\
    \sum_{i=1}^{k-1} \sigma_{i,k} & \text{if } 2 \leq k \leq p, \\
    \sum_{i=p+1}^{k-1} \sigma_{i,k} - \sum_{i=1}^p \bar{\sigma}_{i,k} + \delta & \text{if } p+1 \leq k \leq p+q,
  \end{cases}
  \label{eq:JM for B}
\end{equation}
where $\sigma_{i,k}$ is the transposition of elements $i$ and $k$, and $\bar{\sigma}_{i,k}$ is the corresponding contraction. When the walled Brauer algebra is not semisimple, we still define $J^{\B}_{k}$ via the above formula.

\subsection{Content vectors for walled Brauer algebras}\label{sec:WBA content}

Let $\T \in \Paths(p+q)$ be an arbitrary root-leaf path in the Bratteli diagram of $\B^\delta_{p,q}$ and let $\T[k-1] \to \T[k]$ where $k \in [p+q]$ denote an edge on this path.
Recall from \cref{eq:IrrB} that each vertex $\T[k]$ of $\T$ is labeled by some bipartition $(\lambda^l,\lambda^r)$.
The number $c_{\T[k-1] \to \T[k]}$ introduced in \cref{def:J and c,def:c edge} that corresponds to this edge is calculated via the following rule \cite{FusionProcedure,jung2020supersymmetric}:
\begin{enumerate}
    \item if $1 \leq k \leq p$ and $\T[k]$ is obtained from $\T[k-1]$ by adding a cell $(i, j)$ to the first diagram in the bipartition $\T[k-1]$ then $$c_{\T[k-1] \to \T[k]} = j-i,$$
    \item if $p+1 \leq k \leq p+q$ and $\T[k]$ is obtained from $\T[k-1]$ by removing a cell $(i, j)$ from the first diagram in the bipartition $\T[k-1]$ then $$c_{\T[k-1] \to \T[k]} = i-j,$$
    \item if $p+1 \leq k \leq p+q$ and $\T[k]$ is obtained from $\T[k-1]$ by adding a cell $(i, j)$ to the second diagram in the bipartition $\T[k-1]$ then $$c_{\T[k-1] \to \T[k]} = j-i+\delta.$$
\end{enumerate}
Since these formulas are very similar to the definition of the \emph{content} $j-i$ of a cell $(i,j)$ in a Young tableau, see \cref{eq:content}, we refer to $(c_\T(1), \dotsc, c_\T(p+q))$ as the \emph{content vector} of the path $\T \in \Paths(p+q)$ where $c_\T(k) = c_{\T[k-1] \to \T[k]}$.

\subsection{Adapting the Bratteli diagram from \texorpdfstring{$\B^\delta_{p,q}$}{Bpq} to \texorpdfstring{$\A^d_{p,q}$}{Apq}}\label{sec:Bratteli for A and B}


\begin{figure}
\begin{tikzpicture}[> = latex,
  cut/.style = {thin, dashed, red, rounded corners = 12pt, fill = red, fill opacity = 0.12},
  bad/.style = {draw = red!40, line width = 3pt},
  every node/.style = {inner sep = 1pt}]
  \def\W{2.0cm}
  \def\H{1cm}
  \foreach \i/\p/\q in {0/0/0, 1/1/0, 2/2/0, 3/2/1, 4/2/2} {
    \node at (\i*\W,3.2*\H) {$\A^d_{\p,\q}$};
  }
  \draw[dashed] (2.5*\W,3.4*\H) -- (2.5*\W,-2.7*\H);
  \node (0!0)   at (0*\W, 0*\H) {\br{\0,\0}};
  \node (1!0)   at (1*\W, 0*\H) {\br{\yd{1},\0}};
  \node (2!0)   at (2*\W, 1*\H) {\br{\yd{2},\0}};
  \node (11!0)  at (2*\W,-1*\H) {\br{\yd{1,1},\0}};
  \node (2!1)   at (3*\W, 2*\H) {\br{\yd{2},\yd{1}}};
  \node (1!0')  at (3*\W, 0*\H) {\br{\yd{1},\0}};
  \node (11!1)  at (3*\W,-2*\H) {\br{\yd{1,1},\yd{1}}};
  \node (2!2)   at (4*\W, 2.5*\H) {\br{\yd{2},\yd{2}}};
  \node (2!11)  at (4*\W, 1.5*\H) {\br{\yd{2},\yd{1,1}}};
  \node (1!1)   at (4*\W, 0.5*\H) {\br{\yd{1},\yd{1}}};
  \node (0!0')  at (4*\W,-0.5*\H) {\br{\0,\0}};
  \node (11!2)  at (4*\W,-1.5*\H) {\br{\yd{1,1},\yd{2}}};
  \node (11!11) at (4*\W,-2.5*\H) {\br{\yd{1,1},\yd{1,1}}};
  \draw[->] (0!0) -- (1!0);
  \draw[->] (1!0.north east) -- (2!0.190);
  \draw[->] (1!0.south east) -- (11!0.170);
  \draw[->] (2!0.north east) -- (2!1.190);
  \draw[->] (2!0.south east) -- (1!0'.170);
  \draw[->] (11!0.north east) -- (1!0'.190);
  \draw[->] (11!0.south east) -- (11!1.170);
  \draw[->] (2!1.15) -- (2!2.180);
  \draw[->] (2!1.0) -- (2!11.180);
  \draw[->] (2!1.-20) -- (1!1.160);
  \draw[->] (1!0') -- (1!1.180);
  \draw[->] (1!0') -- (0!0'.180);
  \draw[->] (11!1.25) -- (1!1.200);
  \draw[->] (11!1.0) -- (11!2.180);
  \draw[->] (11!1.-20) -- (11!11.180);
  \begin{scope}[on background layer]
    \draw[cut, rounded corners = 18pt] (11!2)++(0.9,0.5) -- ++(-3.6,0) -- ++(0,-2.3) -- ++(3.6,0);
    \path (11!11)+(1.5,0) node {$d=3$};
    \draw[cut] (2!11)++(0.9,0.5) -- ++(-1.7,0) -- ++(0,-1) -- ++(1.7,0);
    \draw[cut] (11!11)++(0.9,0.5) -- ++(-1.7,0) -- ++(0,-1) -- ++(1.7,0);
    \path (2!11)+(1.5,0) node {$d=2$};
    \path (11!2)+(1.5,0) node {$d=2$};
    \draw[bad] (0!0) -- (1!0);
    \draw[bad] (1!0.south east) -- (11!0.170);
    \draw[bad] (11!0.south east) -- (11!1.170);
    \draw[bad] (11!1.25) -- (1!1.200);
  \end{scope}
\end{tikzpicture}
\caption{Bratteli diagram associated to the multiplicity-free family $\C \cong \A^d_{0,0} \hookrightarrow \A^d_{1,0} \hookrightarrow \A^d_{2,0} \hookrightarrow
\A^d_{2,1} \hookrightarrow
\A^d_{2,2}$ of partially transposed permutation matrix algebras for different values of the local dimension $d$.
When $d \geq 4$, this diagram coincides with that of the walled Brauer algebras (see \cref{fig:B22}).
However, for small values of $d$ (i.e., $d = 2$ and $d = 3$) the diagram has to be modified by removing the designated vertices.
Note that removing the vertex \br{\yd{1,1},\yd{1}} when $d = 2$ eliminates the highlighted path from the root \br{\0,\0} to the leaf \br{\yd{1},\yd{1}}, which decreases the dimension of the corresponding simple $\A^d_{2,2}$-module $V^{\of*{\yd{1},\yd{1}}}$ by one, i.e.,
$\dim\of{V^{\of*{\yd{1},\yd{1}}}} = 4$ if $d > 2$ while
$\dim\of{V^{\of*{\yd{1},\yd{1}}}} = 3$ if $d = 2$.
}
\label{fig:A22}
\end{figure}

According to \cite[Theorem~1.11]{bchlls},
the Bratteli diagram for the family of partially transposed permutation matrix algebras $\A$
can be obtained from the Bratteli diagram for semisimple walled Brauer algebras $\B$
by removing all vertices $(\lambda^l, \lambda^r)$ that violate the condition $\len(\lambda^l) + \len(\lambda^r) \leq d$, where $\len(\mu)$ denotes the length of the first column of the Young diagram $\mu$.
For example, \cref{fig:A22} shows how \cref{fig:B22} should be adapted for small values of $d$.

Note that along with the removed vertices we also remove their incident edges.
Depending on the local dimension $d$ we may need to remove some vertices that are not leaves of the diagram, which in turn can decrease the number of root-leave paths, thus affecting the dimension of $V^\lambda$ (see \cref{fig:A22}). In particular, $d_\lambda = \dim(V^\lambda)$ of the simple $\A_n$-module $V^\lambda$ generally depends on the local dimension $d$.

\subsection{Adapting JM elements and content vectors from \texorpdfstring{$\B^d_{p,q}$}{Bpq} to \texorpdfstring{$\A^d_{p,q}$}{Apq}}

The Jucys--Murphy elements for semisimple walled Brauer algebras $\B$,
given in \cref{eq:JM for B},
can be used in the DLS algorithm in \cref{sec:dls algorithm} to find the primitive central idempotents $\varepsilon^\B(\lambda)$ and canonical primitive idempotents $\varepsilon^\B_\T$ of $\B^\delta_{p,q}$.
In this section, we show how this procedure can be adapted to the partially transposed permutation matrix algebras $\smash{\A^d_{p,q}}$.
To construct the primitive central idempotents $\varepsilon^\A(\lambda)$ and canonical primitive idempotents $\varepsilon^\A_\T$ of $\smash{\A^d_{p,q}}$, we can use the modified Bratteli diagram from \cref{sec:Bratteli for A and B}, and it only remains to adapt the Jucys--Murphy elements and content vectors from $\smash{\B^\delta_{p,q}}$ to $\smash{\A^d_{p,q}}$.
Below we show that we can use lifted versions of Jucys--Murphy elements of $\B^\delta_{p,q}$ and their content vectors for that purpose.

Throughout this section we set $\delta \defeq d$ and,
for every $k \in [p+q]$, let
\begin{equation}
  J^\A_k \defeq \psi^d_{p,q}(J^\B_k) \in \A^d_{p,q},
  \label{eq:JM for A}
\end{equation}
where $J^\B_k$ are the Jucys--Murphy elements of $\B$ given in \cref{eq:JM for B}
and $\psi^d_{p,q}$ is the map from \cref{eq:psi}.
To establish that
$J^{\A}_1, \dotsc, J^{\A}_{p+q}$ is a Jucys--Murphy sequence for $\A^d_{p,q}$,
we need to show that it is both additively central and separating
(see \cref{def:add central and separating}).

\begin{lemma}\label{lem:additively central}
The sequence $J^{\A}_1, \dotsc, J^{\A}_{p+q}$ is additively central in $\A^d_{p,q}$.
\end{lemma}

\begin{proof}
For any $k \in [p+q]$, one can verify that the Jucys--Murphy elements
$J^\B_k$ defined in \cref{eq:JM for B} satisfy
$J^\B_k \in \X^\B_k$ and $J^\B_1 + \dotsb + J^\B_k \in \Z(\B_k)$ \cite{jung2020supersymmetric}.
Since $\psi^d_{p,q}$ is a homomorphism,
$\psi^d_{p,q}\of[\big]{\Z(\B_k)} \subseteq \Z(\A_k)$
and hence
$\psi^d_{p,q}(\X^\B_k) \subseteq \X^\A_k$.
Therefore
$J^\A_k = \psi^d_{p,q}(J^\B_k) \in \psi^d_{p,q}(\X^\B_k) \subseteq \X^\A_k$
and the sequence $J^\A_1, \dotsc, J^\A_{p+q}$ is additively central in $\A^d_{p,q}$.
\end{proof}

\begin{restatable}{lemma}{technical}\label{lem:main_lem}
For any $\T \in \Paths(p+q)$ in the Bratteli diagram of $\A^d_{p,q}$,
\begin{equation}
  \of{J^\A_1 + \dotsb + J^\A_{p+q}} \, \varepsilon^\A_\T
  = \of[\big]{\cont(\lambda^l) + \cont(\lambda^r) + d \cdot \size(\lambda^r)} \, \varepsilon^\A_\T
\end{equation}
where $\varepsilon^\A_\T$ is the corresponding canonical primitive idempotent of $\A^d_{p,q}$, $\lambda = (\lambda^l,\lambda^r) = \T[p+q]$ is the last vertex of the path $\T$, $\cont(\lambda)$ is the total content of all cells of the Young diagram $\lambda$, see \cref{eq:content of lambda}, and $\size(\lambda)$ is the number of cells in $\lambda$.
\end{restatable}

\begin{proof}
See Appendix~\ref{apx:proof of main lemma} for proof. Our proof is reminiscent of \cite[Lemma~2.3]{brundan2012gradings}, which is a similar statement for the walled Brauer algebras.
\end{proof}

\begin{corollary}\label{cor:jm content}
For any $k \in [p+q]$ and $\T \in \Paths(p+q)$, $J^{\A}_k \varepsilon^\A_\T = c_\T(k) \varepsilon^\A_\T$ where $c_\T(k) = c_{\T[k-1] \to \T[k]}$ is the notion of content for the walled Brauer algebra $\B^d_{p,q}$, see \cref{sec:WBA content}.
\end{corollary}

\begin{proof}
Let $\T^k \defeq \lambda_0 \to \lambda_1 \to \cdots \to \lambda_k$ denote the first $k$ edges of the path $\T$.
Recall from \cref{eq:epsT} that the canonical primitive idempotents of $\A$ are given by
\begin{equation}
    \varepsilon^\A_{\T^{k}} = \varepsilon^\A(\lambda_{1}) \varepsilon^\A(\lambda_{2}) \cdots \varepsilon^\A(\lambda_{k}).
\end{equation}
Each value of $k$ effectively corresponds to truncating the Bratteli diagram to a certain $p$ and $q$.
Thus \cref{lem:main_lem} with appropriate $p$ and $q$ allows us to compute the eigenvalue of $J^{\A}_{1}+\cdots+J^{\A}_{k}$ for two consecutive $\varepsilon^\A_{\T^{k}}$:
\begin{align}
    \of{J^\A_1 + \dotsb + J^\A_{k}} \, \varepsilon^\A_{\T^{k}}
  &= \of[\big]{\cont(\lambda_{k}^l) + \cont(\lambda_{k}^r) + d \cdot \size(\lambda_{k}^r)} \, \varepsilon^\A_{\T^{k}}, \\
  \of{J^\A_1 + \dotsb + J^\A_{k-1}} \, \varepsilon^\A_{\T^{k-1}}
  &= \of[\big]{\cont(\lambda_{k-1}^l) + \cont(\lambda_{k-1}^r) + d \cdot \size(\lambda_{k-1}^r)} \, \varepsilon^\A_{\T^{k-1}}.
  \label{eq:cor1main}
\end{align}
Multiplying both equations with the primitive central idempotents of $\A$ ranging from
$\varepsilon^\A(\lambda_{k})$ to $\varepsilon^\A(\lambda_{p+q})$
transforms the subscripts $\T^{k}$ and $\T^{k-1}$ into $\T^{p+q} = \T$:
\begin{align}
    \of{J^\A_1 + \dotsb + J^\A_{k}} \, \varepsilon^\A_{\T}
  &= \of[\big]{\cont(\lambda_{k}^l) + \cont(\lambda_{k}^r) + d \cdot \size(\lambda_{k}^r)} \, \varepsilon^\A_{\T}, \\
  \of{J^\A_1 + \dotsb + J^\A_{k-1}} \, \varepsilon^\A_{\T}
  &= \of[\big]{\cont(\lambda_{k-1}^l) + \cont(\lambda_{k-1}^r) + d \cdot \size(\lambda_{k-1}^r)} \, \varepsilon^\A_{\T}.
  \label{eq:cor1main2}
\end{align}
Subtracting these two equations we get
\begin{align}
    J^{\A}_k \varepsilon^\A_{\T}
    = \Bigl(
       \cont(\lambda_{k}^l) - \cont(\lambda_{k-1}^l)
    &+ \cont(\lambda_{k}^r) - \cont(\lambda_{k-1}^r) \\
    &+ d \cdot \of[\big]{\size(\lambda_{k}^r) - \size(\lambda_{k-1}^r)}
    \Bigr) \, \varepsilon^\A_{\T}. \nonumber
\end{align}

If $k \leq p$ then $\lambda_{k}^r = \lambda_{k-1}^r = \0$ and $\cont(\lambda_{k}^l) - \cont(\lambda_{k-1}^l) = j - i$, where $(i,j)$ is the location where adding a cell to the Young diagram $\lambda_{k-1}^l$ transforms it into $\lambda_{k}^l$.
If $k > p$ then there are two cases.
If $\lambda_{k}^r = \lambda_{k-1}^r$ then $\cont(\lambda_{k}^l) - \cont(\lambda_{k-1}^l) = i - j$ because the cell $(i,j)$ is removed from $\lambda_{k-1}^l$.
If $\lambda_{k}^l = \lambda_{k-1}^l$ then $\cont(\lambda_{k}^l) - \cont(\lambda_{k-1}^l) = j - i$ and $\size(\lambda_{k}^r) - \size(\lambda_{k-1}^r) = 1$, where $(i,j)$ is the location where adding a cell to the Young diagram $\lambda_{k-1}^r$ transforms it into $\lambda_{k}^r$.
In either case,
\begin{equation}
    J^{\A}_k \varepsilon^\A_{\T}  = c_\T(k) \varepsilon^\A_{\T},
\end{equation}
where $c_\T(k) = c_{\T[k-1] \to \T[k]}$ is exactly the notion of content for the walled Brauer algebra $\B^d_{p,q}$ as defined in \cref{sec:WBA content}.
\end{proof}

{

\renewcommand{\S}{\mathrm{S}}

\begin{lemma}\label{lem:content}
If $\S,\T \in \Paths(p+q)$ are two paths in the Bratteli diagram of $\A^d_{p,q}$ then
$\S = \T$ if and only if $c_\S = c_\T$.
\end{lemma}

\begin{proof}
The forward direction is obvious. For the reverse implication, assume that $\S \neq \T$ and $c_\S = c_\T$. We consider two cases depending on the first location $k \in [p+q]$ where the two paths differ,
i.e., $\S[k] \neq \T[k]$ while $\S[k-1] = \T[k-1] \eqdef (\lambda^l,\lambda^r)$.

If $k \leq p$, we can only add cells to the Young diagram $\lambda^l$,
so let $(i,j)$ and $(i',j')$ denote the two possible locations.
It follows from \cref{cor:jm content} that $c_\T(k) = j - i$ and $c_\S(k) = j' - i'$,
and hence $j - i = j' - i'$.
Since any Young diagram has at most one location on any diagonal where a new cell can be added, $(i,j) = (i',j')$ and therefore $\S[k] = \T[k]$.

If $k > p$, the condition $c_\T(k) = c_\S(k)$ can be satisfied only if we assume (without loss of generality) that $c_\T(k) = i^l - j^l$ and $c_\S(k) = j^r - i^r + d$, implying that $i^l + i^r = j^l + j^r + d$ for the removed cell $(i^l,j^l)$ of $\lambda^l$ and the added cell $(i^r,j^r)$ of $\lambda^r$. In particular, $i^l + i^r = j^l + j^r + d \geq d + 2$ since $j^l > 0$ and $j^r > 0$. On the other hand, the total length of the two diagrams satisfies
$\len(\lambda^l) + \len(\lambda^r) \leq d$, which implies that $i^l + i^r \leq d $,
a contradiction.
\end{proof}
}

\begin{corollary}\label{cor:sep}
The sequence $J^{\A}_1, \dotsc, J^{\A}_{p+q}$ is separating in $\A$.
\end{corollary}

\begin{proof}
This follows from \cref{lem:content} and Proposition~3.5 of~\cite{doty2019canonical}.
\end{proof}

\begin{theorem}\label{thm:main technical}
$J^\A_1, \dotsc, J^\A_{p+q}$ is a Jucys--Murphy sequence for $\A$.
\end{theorem}

\begin{proof}
This follows from \cref{lem:additively central,cor:sep}.
\end{proof}

\Cref{thm:main technical} establishes that the operators $J^\A_k$, which were obtained in \cref{eq:JM for A} by lifting the Jucys--Murphy sequence $J^\B_k$ of the walled Brauer algebras $\B^d_{p,q}$ to the matrix algebras $\A^d_{p,q}$, are indeed Jucys--Murphy elements of $\A^d_{p,q}$. Furthermore, \cref{cor:jm content} shows that the content vectors of the two families of algebras agree.

\subsection{Primitive central and canonical primitive idempotents of \texorpdfstring{$\A^d_{p,q}$}{Apq}}\label{sec:copmute_preimages_idem_A}

We have now established the three ingredients required to apply the \cite{doty2019canonical} algorithm to the partially transposed matrix algebras $\A^d_{p,q}$:
\begin{enumerate}
    \item the Bratteli diagram for $\A$ is obtained by truncating the Bratteli diagram of $\B$ as discussed in \cref{sec:Bratteli for A and B},
    \item the Jucys--Murphy elements of $\A$ are obtained by applying $\psi^d_{p,q}$ to the Jucys--Murphy elements of $\B$ given in \cref{eq:JM for B},
    \item the content vectors of $\A$ agree with those of $\B$ and are given in \cref{sec:WBA content}.
\end{enumerate}
This allows us to use the algorithm described in \cref{sec:dls algorithm} to compute the primitive central idempotents and canonical primitive idempotents of $\A$.

A major advantage of our approach is that the entire computation can be performed by employing linear combinations of diagrams instead of actual matrices
(i.e., staying within the diagrammatic walled Brauer algebra $\B^d_{p,q}$ rather than working in the matrix algebra $\smash{\A^d_{p,q}}$).
This results in a diagrammatic representation of an idempotent of $\smash{\A^d_{p,q}}$ as a preimage of the actual idempotent under $\psi^d_{p,q}$.
More explicitly, the primitive central idempotents of $\smash{\A^d_{p,q}}$ can be computed iteratively as
\begin{equation}
\label{eq:primitive_central_A}
  \varepsilon^\A(\mu) \defeq
    \sum_{\lambda : \lambda \to \mu} \psi^d_{p,q} \left( P_{\lambda \to \mu} (J_k^\B) \right) \varepsilon^\A(\lambda),
\end{equation}
where the polynomials $P_{\lambda \to \mu}$ from \cref{eq:P} are evaluated for the Bratteli diagram of the family $\A$ described in \cref{sec:Bratteli for A and B}. Similarly to \cref{eq:primitive_main}, canonical primitive idempotents $\varepsilon_\T$ are adapted to the matrix algebras $\A$ as follows:
\begin{equation}
    \varepsilon^\A_\T
  = \psi^d_{p,q} \left( \prod_{k=1}^{p+q}
    \prod_{\mu : \lambda_{k-1} \to \mu \neq \lambda_{k}}
    \frac{J_k^\B - c_{\lambda_{k-1} \to \mu}}{c_{\lambda_{k-1} \to \lambda_{k}}-c_{\lambda_{k-1} \to \mu}} \right),
    \label{eq:primitive_main_A_d}
\end{equation}
where the second product runs over edges in the Bratteli diagram of the family $\A$.

This representation of idempotents is more compact compared to the naive one when $d$ is large, and easily amenable to further fast diagrammatic calculations.
This allows to significantly lower the computational complexity of various tasks within the partially transposed permutation matrix algebra, as illustrated in the next section for a certain class of optimization problems.

\section{Reducing unitary-equivariant SDPs to LPs}\label{sec:SDP}

In this section we derive our main result---a pre-processing algorithm for LP solvers which accepts a sparse SDP with a $\U(d)$-equivariant constraint as input.
Our algorithm also requests one of several additional symmetries (see \cref{sec:symmetries}) that guarantee that the provided SDP reduces to an LP.
While the input problem has a compact representation due to all involved matrices being sparse, naively solving it might be impossible in practice due to a prohibitively large $d$ (the SDP matrix variable has dimension $d^{p+q}$).
Our algorithm converts the implicit input LP to an explicit smaller LP whose naive representation has size that no longer depends on $d$.
Although it may generally not be sparse, this LP is much smaller and can thus be further supplied as input to any standard LP solver.

In \cref{sec:symmetries} we list the additional symmetries our algorithm requires,
in \cref{sec:input} we specify the input format of our algorithm,
and in \cref{sec:main} we state and prove our main result.

\subsection{Types of symmetries}\label{sec:symmetries}

To achieve a reduction from SDP to LP, the SDP matrix variable $X$ needs some further symmetry in addition to unitary equivariance. For example, one option is the $\S_p \times \S_q$ permutational symmetry which is natural in the context of $p \to q$ quantum channels, see \cref{lem:choi_perm}. We show in \cref{sec:Sp x Sq symmetry} that an SDP with such symmetry reduces to an LP when $\min(p,q) \leq 2$.
The full list of possible symmetries we consider for the SDP variable $X$ is as follows.

\begin{definition}\label{def:symmetries}
A matrix $X \in \End(V^{p,q})$ possesses
\begin{itemize}
    \item the \emph{$\S_p \times \S_q$ permutational symmetry} if for every  $\sigma \in \C(\S_p \times \S_q) \subset \B^d_{p,q}$
    \begin{equation}
        \sof[\big]{X, \psi^d_{p,q}(\sigma)} = 0,
    \end{equation}
    or equivalently $X \in \Z_{\psi^d_{p,q}(\C(\S_p \times \S_q))}(\A^d_{p,q})$;
    \item the \emph{walled Brauer algebra symmetry} if for every $\sigma \in \B^d_{p,q}$, see \cref{sec:wba},
    \begin{equation}
        \sof[\big]{X, \psi^d_{p,q}(\sigma)} = 0,
    \end{equation}
    or equivalently $X \in \Z(\A^d_{p,q})$;
    \item the \emph{Gelfand--Tsetlin symmetry} if for every $A \in \X^\A_{p+q}$, see \cref{eq:GT subalgebras},
    \begin{equation}
        \sof{X, A} = 0,
    \end{equation}
    or equivalently $X \in \X^\A_{p+q}$ since $\X^\A_{p+q}$ is the maximal commutative subalgebra of $\A^d_{p,q}$ (see \cref{sec:GT subalgebra}).
\end{itemize}
\end{definition}

The last two symmetries have an intuitive interpretation in Schur basis.
Recall from \cref{rm:Schur} that there exists a mixed Schur transform $U_{\textnormal{Sch}(p,q)} \in \U(V^{p,q})$ that block-diagonalizes any unitary-equivariant $X$:
\begin{align}
    U_{\textnormal{Sch}(p,q)} \, X \, U_{\textnormal{Sch}(p,q)}\ct
    = \bigoplus_{\lambda \in \Irr(\A^d_{p,q})}
      \sof*{
        X_\lambda \x I_{m_\lambda}
      }.
    \label{eq:X in Schur basis}
\end{align}
We will assume that $U_{\textnormal{Sch}(p,q)}$ is adapted to the Gelfand--Tsetlin basis, meaning that each $X_\lambda$ is expressed in this basis.
Each symmetry from \cref{def:symmetries} results in some simplification of the matrix variable $X$ since each block $X_\lambda$ assumes a special form.
We discuss this in more detail in the following sections.

\subsubsection{Full walled Brauer algebra symmetry}

The maximal possible symmetry that can be assumed is the full \emph{walled Brauer algebra symmetry}, which means that each $X_\lambda$ in \cref{eq:X in Schur basis} is proportional to the identity matrix, i.e.,
\begin{equation}
    X_\lambda = v_\lambda I_{d_\lambda}
\end{equation}
for some scalar $v_\lambda \in \R$.
In this case, the semidefinite constraint $X_\lambda \succeq 0$ reduces to $v_\lambda \geq 0$, thus simplifying the problem from an SDP to an LP with variables $\set{v_\lambda : \lambda \in \Irr(\A^d_{p,q})}$. The total number of variables in the LP is
\begin{equation}
    n_{p,q}^d(\text{Brauer})
    \defeq \abs[\big]{\Irr(\A^d_{p,q})}
    = \sum_{k = 0}^{\min(p,q)} f_{p-k,q-k}^d,
    \label{eq:n full Brauer}
\end{equation}
where $f_{p-k,q-k}^d$ is the number of pairs $(\lambda^l,\lambda^r)$ of Young diagrams  such that $\lambda^l \pt p-k$, $\lambda^r \pt q-k$ and $\len(\lambda^l) + \len(\lambda^r) \leq d$.
In this case, we can write $X$ as a linear combination of the primitive central idempotents from \cref{eq:primitive_central_A}:
\begin{equation}
    X = \sum_{\lambda \in \Irr(\A^d_{p,q})} v_\lambda \varepsilon^\A(\lambda).
    \label{eq:X central}
\end{equation}

\subsubsection{Gelfand--Tsetlin symmetry}

A minimal symmetry that can be assumed is the \emph{Gelfand--Tsetlin symmetry}, which means that each $X_\lambda$ is diagonal in the Gelfand--Tsetlin basis, i.e.,
\begin{equation}
    X_\lambda = \sum_{i=1}^{d_\lambda} v_{\lambda,i} \proj{i}
    \label{eq:X primitive lambda}
\end{equation}
for some scalars $v_{\lambda,i} \in \R$ that become the variables of the LP.
The $X \succeq 0$ constraint then reduces to $v_{\lambda,i} \geq 0$ for all $\lambda \in \Irr(\A^d_{p,q})$ and $i \in [d_\lambda]$.
The total number of LP variables in this case is
\begin{equation}
    n_{p,q}^d(\text{Gelfand--Tsetlin})
    \defeq \sum_{\lambda \in \Irr(\A^d_{p,q})} d_\lambda,
    \label{eq:n GT}
\end{equation}
where $d_\lambda = \dim(V^\lambda)$ is the dimension of the corresponding simple module of $\A^d_{p,q}$.
In this case, we can write $X$ as a linear combination of the canonical primitive idempotents from \cref{eq:primitive_main_A_d}:
\begin{equation}
    X = \sum_{\T \in \Paths(\A^d_{p,q})} v_\T \varepsilon^\A_\T.
    \label{eq:X primitive}
\end{equation}

\subsubsection{\texorpdfstring{$\S_p \times \S_q$}{Sp x Sq} permutational symmetry}\label{sec:Sp x Sq symmetry}

A somewhat intermediate symmetry that can be assumed is the \emph{$\S_p \times \S_q$ permutational symmetry}. This symmetry allows us to simplify $X_\lambda$ to
\begin{align}
    X_\lambda
    &\cong \bigoplus_{\substack{
        \mu \pt p \text{, } \len(\mu) \leq d \\
        \nu \pt q \text{, } \len(\nu) \leq d}}
    (I_{\mu} \otimes I_{\nu})
    \otimes \tilde{X}^{\lambda}_{\mu,\nu}
    \label{eq:X with three registers}
\end{align}
where $\tilde{X}^{\lambda}_{\mu,\nu}$ is a Hermitian matrix of dimension $m_{\mu, \nu}^{\lambda}(d)$, see \cref{eq:mult_main} in \cref{apx:restriction to Spq}, and a priori we do not know the basis on the right-hand side.

There are two cases when this symmetry leads to a reduction from SDP to LP. We show in \cref{lem:multiplicity m,lem:multiplicity m d=2} that $m_{\mu, \nu}^{\lambda}(d) \in \set{0,1}$ when $\min(p,q) \leq 2$ or $d=2$, meaning that the corresponding term in \cref{eq:X with three registers} either drops out or $\smash{\tilde{X}^{\lambda}_{\mu,\nu}}$ becomes a scalar $\smash{\tilde{x}^{\lambda}_{\mu,\nu}} \in \R$:
\begin{align}
    X_\lambda
    &\cong \bigoplus_{\substack{
    \mu \pt p \text{, } \len(\mu) \leq d  \\
    \nu \pt q \text{, } \len(\nu) \leq d \\
    m_{\mu, \nu}^{\lambda}(d) = 1}}
    \tilde{x}^{\lambda}_{\mu,\nu}
    ( I_{\mu} \otimes I_{\nu}).
\end{align}
We show in \cref{prop:Sp1_and_unitary} that when $\min(p,q) = 1$ each block $X_\lambda$ becomes diagonal specifically in the Gelfand--Tsetlin basis.
In contrast to \cref{eq:X primitive lambda}, some of the diagonal entries of $X_\lambda$ must be equal in this case. In the following proposition we assume $ q = 1$. The argument is completely analogous when $p = 1$ since one just has to use a different sequence $\A$ of algebras $\C \hookrightarrow \A^d_{0,1} \hookrightarrow \cdots \hookrightarrow \A^d_{0,q} \hookrightarrow \A^d_{1,q}$ when constructing the Bratteli diagram and the corresponding Gelfand--Tsetlin basis.

\begin{proposition}\label{prop:Sp1_and_unitary}
Fix $d \geq 2$ and $p \geq 1$, and let $X \in \End(V^{p,1})$ be a Hermitian matrix with unitary equivariance and $\S_p \times \S_1$ symmetry:
\begin{align}
    \sof[\big]{X,U\xp{p} \x \bar{U}} &= 0, \qquad \forall U \in \U(V), \label{eq:XU} \\
    \sof[\big]{X, \psi^d_p(\pi) \x I_{d}} &= 0, \qquad \, \forall \pi \in \S_p. \label{eq:XP}
\end{align}
Then $X$ can be written as
\begin{equation}
    X = \sum_{\T \in \Paths(\A^d_{p,1})} v_{\T[p],\T[p+1]} \varepsilon^\A_\T,
    \label{eq:X when q=1}
\end{equation}
where $v_{\T[p],\T[p+1]} \in \R$ depends only on the last edge of path $\T$.
\end{proposition}

\begin{proof}
Let us first understand in what way the $\S_p \times \S_1$ symmetry is special among general $\S_p \times \S_q$ symmetries.
Since the group $\S_1$ is trivial, $\C(\S_p \x \S_1) \cong \CS_p$.
Moreover,
$\psi^d_{p,1}(\pi \times e) = \psi^d_p(\pi) \x I_d$
for any $\pi \in \S_p$, where $e \in \S_1$ denotes the identity permutation.
Hence
\begin{equation}
  \psi^d_{p,1}\of[\big]{\C(\S_p \x \S_1)}
  = \psi^d_p(\CS_p) \x I_d
  = \A^d_{p,0} \hookrightarrow \A^d_{p,1},
  \label{eq:psi SpS1}
\end{equation}
so the algebra $\A^d_{p,0}$ generated by the $\S_p \times \S_1$ symmetry appears in the multiplicity-free family
$\C \hookrightarrow
\A^d_{1,0} \hookrightarrow \cdots \hookrightarrow
\A^d_{p,0} \hookrightarrow
\A^d_{p,1}$.

We can use this observation to get a better grip on the matrix $X$.
Thanks to the mixed Schur--Weyl duality (\cref{thm:gen-schur-weyl}), \cref{eq:XU} implies that $X \in \A^d_{p,1}$.
Also, notice from \cref{eq:psi SpS1} that \cref{eq:XP} is equivalent to $[X, \A^d_{p,0}] = 0$.
By combining these two observations we see that $X \in \Z_{\A^d_{p,0}}(\A^d_{p,1})$.

Writing any $A \in \A^d_{p,1}$ in Schur basis we get
\begin{equation}
    U_{\textnormal{Sch}(p,1)} \, A \, U_{\textnormal{Sch}(p,1)}\ct
    = \bigoplus_{\lambda \in \Irr(\A^d_{p,1})}
      \sof*{
        A_\lambda \x I_{m_\lambda}
      },
    \label{eq:A blocks}
\end{equation}
where the blocks $A_\lambda$ are expressed in the Gelfand--Tsetlin basis.
If we instead take $A \in \A^d_{p,0}$ then, thanks to how the Gelfand--Tsetlin basis is recursively constructed from paths in the Bratteli diagram,
the $A_\lambda$ in \cref{eq:A blocks} are block-diagonal.
That is, $A_\lambda = \bigoplus_\mu A_{\lambda,\mu}$
(there are no multiplicities here since the embedding $\A^d_{p,0} \hookrightarrow \A^d_{p,1}$ is part of a multiplicity-free family).

Since $[X,A] = 0$ for all $A \in \A^d_{p,0}$, the blocks $X_\lambda$ of $X$ are of the form $X_\lambda = \bigoplus_\mu c_{\lambda,\mu} I_{\lambda,\mu}$ for some $c_{\lambda,\mu} \in \R$.
In particular, they are diagonal in the Gelfand--Tsetlin basis, so
$X = \sum_{\T \in \Paths(\A^d_{p,1})} v_\T \varepsilon^\A_\T$
for some $v_\T \in \R$ for each path $\T$.
Moreover, all variables $v_\T$ that correspond to paths $\T$ that go through a fixed vertex at level $p$ of the Bratteli diagram have the same value (this vertex labels a simple $\A^d_{p,0}$-module).
Formally this means that $v_{\T} = v_{\S}$ for every $\T, \S \in \Paths(\A^d_{p,1})$ such that the paths $\T$ and $\S$ in the Bratteli diagram of $\A^d_{p,1}$ share the same last edge, i.e., $(\T[p],\T[p+1]) = (\S[p],\S[p+1])$. In particular, if $(\T[p],\T[p+1]) = (\mu,\lambda)$ then $v_\T = c_{\lambda,\mu}$.
\end{proof}

The number of variables in this case is
\begin{equation}
    n_{p,q}^d(\S_p \times \S_q)
    \defeq \sum_{
    \substack{
    \lambda \in \Irr(\A^d_{p,q}) \\
    \mu \pt p \text{, } \len(\mu) \leq d \\
    \nu \pt q \text{, } \len(\nu) \leq d }} \of[\big]{m_{\mu, \nu}^{\lambda}(d)}^2.
    \label{eq:n Spq}
\end{equation}
This number can be easily calculated numerically using the results of \cref{apx:restriction to Spq}. The results of this calculation for small $d,p,q$ can be found in \cref{apx:numeric tables}.

\subsubsection{Summary}

The chosen symmetry, together with unitary equivariance, guarantees that $X$ can be expressed as a linear combination of idempotents of the algebra $\A^d_{p,q}$, see \cref{eq:X primitive,eq:X central,eq:X when q=1}:
\begin{equation}
    X = \sum_{i=1}^n v_i \varepsilon^\A_i,
    \label{eq:X expansion}
\end{equation}
where the number of terms $n$ is given by \cref{eq:n full Brauer,eq:n GT,eq:n Spq}, respectively (see \cref{table:symmetries} for a summary).
Since we effectively know the basis in which $X$ is diagonal, the SDP reduces to an LP.
In particular, the scalars $v_1, \dotsc, v_n \in \R$ in \cref{eq:X expansion} will be the variables of the output LP produced by our algorithm. The number of variables $n$ varies dramatically depending on the chosen symmetry type, see \cref{apx:numeric tables}.

\begin{table}
\begin{tabular}{l|c|c|c}
    Symmetry & Ansatz for $X$ & Formula for $n_{p,q}^d$ & Values of $n_{p,q}^d$ \\ \hline
    Full walled Brauer algebra & \cref{eq:X central} & \cref{eq:n full Brauer} & \cref{table:FwBa} \\
    $\S_p \times \S_q$ & \cref{eq:X when q=1} & \cref{eq:n Spq} & \cref{table:SpSq} \\
    Gelfand--Tsetlin & \cref{eq:X primitive} & \cref{eq:n GT} & \cref{table:GT} \\
    Only unitary equivariance & \cref{eq:X in Schur basis} & $\dim(\A^d_{p,q})$ & \cref{table:SDP}
\end{tabular}
\caption{\label{table:symmetries}Summary of symmetries from \cref{def:symmetries}.
For each symmetry, we provide pointers to the corresponding ansatz for $X$, the number of variables $n_{p,q}^d$ in this ansatz, and tables for numerical values of $n_{p,q}^d$. Note that the ansatz~\eqref{eq:X when q=1} of $X$ in case of the $\S_p \times \S_q$ symmetry is valid only when $q=1$; a similar formula can also be obtained for $p=1$ as explained just before \cref{prop:Sp1_and_unitary}.}
\end{table}

\subsection{Input specification}\label{sec:input}

Our algorithm accepts a sparse SDP composed of the following objects:
\begin{enumerate}
    \item $p,q \geq 0$ -- the number of input and output systems,
    \item $d \geq 2$ -- the local dimension of $V^{p,q} = (\C^d)\xp{p} \x (\C^d)^{*\x q}$,
    \item $X \succeq 0$ -- a Hermitian matrix variable acting on $V^{p,q}$ that has to obey the $\U(d)$-equivariance condition in \cref{eq:Choi equivariance},
    \item (in)equality constraints that involve constant sparse matrices,
    \item a desired additional type of symmetry (see \cref{sec:symmetries} for possible options) which guarantees that the problem reduces to an LP.
\end{enumerate}
Our algorithm outputs an explicit LP, equivalent to the input one, whose size does not depend on $d$ and which can be further fed as an input to a standard LP solver.
We are concerned only with this pre-processing step and its complexity.

The input to our algorithm is an SDP in the following form:
\begin{equation}
    \begin{aligned}
        \max_X \quad & \Tr(CX) \\
        \textrm{s.t.} \quad
        & \Tr(A_k X) \leq b_k, & \forall k &\in [\ftr], \\
        & \Tr_{S_k} (X) = D_k, & \forall k &\in [\ptr], \\
        & \sof[\big]{X, U\xp{p} \x \bar{U}\xp{q}} = 0, & \forall U &\in \U(\C^d), \\
        & X \succeq 0,
    \end{aligned}
    \label{eq:input SDP}
\end{equation}
where $\ftr$ and $\ptr$ denote the number of constraints that involve full trace and partial trace, respectively.
Recall from \cref{eq:Choi equivariance} that the penultimate condition is equivalent to unitary-equivariance of  the superoperator associated to $X$.

The Hermitian matrices $C$ and $A_k$ in \cref{eq:input SDP} are constant and \emph{$s$-sparse}, meaning that they can be written as a linear combination of at most $s$ terms, where each term is either $\psi^d_{p,q}(\sigma)$ for some diagram $\sigma$ of the walled Brauer algebra $\B^d_{p,q}$ or an elementary standard basis matrix whose all entries are $0$ and only one entry is $1$, i.e., $\ketbra{i}{j}$ for some $i,j \in [d]^{p+q}$.
Each $D_k$ is a Hermitian linear combination of at most $s$ diagrams from the walled Brauer algebra $\B^d_{p_k,q_k}$ obtained by removing the nodes $S_k \subseteq [p+q]$ from $\B^d_{p,q}$.
The remaining number of nodes on each side of the wall of $\B^d_{p_k,q_k}$ is
$p_k \defeq p - |S_k \cap [p]|$ and
$q_k \defeq q - |(S_k-p) \cap [q]|$.
The total size of the input SDP in terms of the number of scalars needed to specify the matrices $C, A_k, D_k$ is\footnote{For simplicity, we do not count the $\ptr$ additional parameters needed to specify the scalars $b_k$ and the additional information needed to specify the subsets $S_k$ in \cref{eq:input SDP}.}
\begin{equation}
    (1 + \ftr + \ptr) s.
\end{equation}
The SDP may contain additional scalar variables that need not obey the unitary equivariance condition.
Such variables do not require any pre-processing by our algorithm, so they do not incur additional costs in our setting.

If the input SDP contains partial trace constraints,
our algorithm has the following technical restriction:
we require the local dimension $d$ to be sufficiently large, namely
\begin{equation}
    d \geq p + q - \min_k |S_k|.
    \label{eq:d big}
\end{equation}
In particular, if $X$ is a Choi matrix of a $p \to q$ channel and we include the partial trace constraint $\Tr_{V^q}(X) = I_{V^p}$ to capture trace preservation, we require that $d \geq p$.
Due to this we cannot apply our formalism to the setting of \cite{Majority} where $d = 2$ and $p$ is large.
To remove this restriction, one would have to know the kernel of the map $\psi^d_{p,q}$ in order to correctly process the partial trace constraints in the SDP (see \cref{ex:faithfulness} for an instance where the kernel is non-trivial).

\begin{remark}\label{rem:kernel}
As outlined in \cref{apx:blocks of Apq}, one way to obtain the kernel of $\psi^d_{p,q}$ is by using the primitive idempotents $\varepsilon^\A_\T$ to compute the blocks of $\psi^d_{p,q}\of[\big]{\sum_{j=1}^{(p+q)!} b_j \sigma_j}$ where $b_j$ are symbolic variables.
Equating these blocks to zero produces linear equations in $b_j$ that reveal the linear dependencies among the matrices $\psi^d_{p,q}(\sigma_j)$.
We can store this information in a database and use it to reduce the complexity of multiplication in the diagrammatic algebra $(\psi^d_{p,q})^{-1}(\A^d_{p,q})$, i.e., the preimage of $\smash{\A^{d}_{p,q}}$ under $\smash{\psi^d_{p,q}}$.
This can lead to an improved complexity in our main result \cref{thm:main} since the complexity parameter $N$ can be lowered from $(p+q)!$ to $\dim(\A^d_{p,q})$.
\end{remark}

\subsection{Main result}\label{sec:main}

The input to our algorithm is an SDP of the form \eqref{eq:input SDP} which involves a unitary-equivariant constraint and has the following parameters:
\begin{itemize}
    \item $p$ and $q$ -- number of input and output systems,
    \item $d$ -- local dimension of each system,
    \item $s$ -- sparsity of matrices $C, A_k, D_k$,
    \item $\ftr$ -- number of inequality constraints,
    \item $\ptr$ -- number of equality constraints with partial trace.
\end{itemize}
We assume that $p$ and $q$ are small constants while $d$ may generally be large.
The complexity of our algorithm will scale in
\begin{equation}
    N \defeq (p+q)!
\end{equation}
but not $d$.
This is in contrast to the naive approach of solving an SDP with a matrix variable $X$ of dimension $d^{p+q}$.
While the naive approach quickly becomes impractical as $d$ grows, our method does not suffer from this problem.
In fact, it even offers performance improvements for small $d$ such as $d = 2$.
Our algorithm requires assuming that $X$ has one of the symmetries listed in \cref{sec:symmetries}, which guarantees that the SDP reduces to an LP.

The following is a formal statement of our main result.

\newcommand{\sym}{sym}

\begin{theorem}\label{thm:main}
Any SDP of the form \eqref{eq:input SDP}, where $X$ has one of the symmetries listed in \cref{def:symmetries}, can be converted to an equivalent LP with $n$ variables and $\ftr + \ptr N + n$ constraints.
The number of variables $n \defeq n_{p,q}^d(\sym)$, which depends on $p,q,d$ and the chosen symmetry $\sym$ (see \cref{table:symmetries}), can always be bounded as
\begin{equation}
    n_{p,q}^d(\sym) \leq \dim(\A_{p,q}^d) \leq (p+q)! = N.
\end{equation}
The algorithm consists of two parts:
\begin{enumerate}
    \item an input-independent pre-computation that needs to be done only once for each set of parameters $p,q,d$, and whose complexity does not scale in $d$,
    \item and SDP-to-LP conversion that takes time
    \begin{equation}
        (1 + \ftr + \ptr) s \cdot n N
    \end{equation}
    where $(1 + \ftr + \ptr) s$ is the size of the input SDP.
\end{enumerate}
If $d \geq p+q$, the run-time of the pre-computation does not scale in $d$, while for small $d < p+q$ additional speedup is gained.
If the SDP contains partial trace constraints, i.e., $\ptr > 0$, we require that $d \geq p + q - \min_k |S_k|$.
\end{theorem}

\begin{remark}
For the sake of simplicity we will ignore various details in our analysis.
In particular, we will assume that the following operations take constant time:
multiplying two walled Brauer algebra diagrams,
contracting a diagram with a rank-$1$ matrix,
or computing the (partial) trace of a diagram.
In reality the complexity of these operations scales with $p+q$, which is small compared to our yardstick $N = (p+q)!$.
Similarly, we will ignore the fact that the input size scales as $2 (p+q) \log_2 d$ when the SDP contains rank-$1$ matrices.
Finally, we will also ignore the fact that storing the value $d^{\loops(\sigma)}$ requires $(p+q) \log_2 d$ bits.
\end{remark}

\begin{proof}
Our algorithm consists of two parts:
(1)~pre-computation of a database of $\A^d_{p,q}$ idempotents and
(2)~processing the input SDP to an LP.

The pre-computation of a database of $\A^d_{p,q}$ idempotents can be done upfront since it depends only on the parameters $p,q,d$ but not the input SDP.
The type of idempotents needed depends on the specified symmetry type $\sym$ (see \cref{sec:symmetries}).
In either case, they can be computed diagrammatically using the DLS algorithm from \cref{sec:copmute_preimages_idem_A}.
It produces a list of preimages $\varepsilon_1, \dotsc, \varepsilon_n \in \B^{d}_{p,q}$ of $\A^d_{p,q}$ idempotents, with each $\varepsilon_i$ expressed as a linear combination of walled Brauer algebra diagrams $\sigma_j$\footnote{By knowing the kernel of $\psi^d_{p,q}$, we can express each $\varepsilon_i$ more economically as $\varepsilon_i = \sum_{j=1}^m \alpha_{ij} \sigma_j$ where $m = \dim(\A^d_{p,q}) \leq (p+q)!$.}:
\begin{equation}
    \varepsilon_i = \sum_{j=1}^{(p+q)!} \alpha_{ij} \sigma_j.
    \label{eq:idempotent expansion}
\end{equation}
The resulting $n \times (p+q)!$ coefficient matrix $\alpha$ is the output of the pre-computation step.
By construction, its entries are rational.

The second part of the algorithm requires a $\U(d)$-equivariant SDP as input and reduces it to an explicit LP whose size is $d$-independent.
The main idea of this algorithm is that we can evaluate all traces appearing in \cref{eq:input SDP} diagrammatically without ever explicitly computing any of the $d^{p+q} \times d^{p+q}$ matrices involved (see \cref{apx:traces}).
Let us discuss this step in more detail.

Due to unitary equivariance and the additional symmetry $\sym$, we can express the matrix variable $X$ as a linear combination of idempotents $\psi^{d}_{p,q}(\varepsilon_i)$ with unknown coefficients $v_i \in \R$ as in \cref{eq:X expansion}:
\begin{equation}
    X = \sum_{i=1}^n v_i \psi^{d}_{p,q}(\varepsilon_i).
    \label{eq:X expansion with psi}
\end{equation}
These coefficients will be the variables of the output LP.
The number $n = n^d_{p,q}(\sym)$ of variables $v_i$ and idempotent preimages $\varepsilon_i$ depends on the type of symmetry (see \cref{table:symmetries} in \cref{sec:symmetries}).
Since $\psi^{d}_{p,q}(\varepsilon_i) \succeq 0$ and these idempotents are mutually orthogonal for different $i$, the positive semidefinite constraint $X \succeq 0$ reduces to
\begin{equation}
    v_i \geq 0, \quad \forall i = 1, \dotsc, n.
\end{equation}

For the target function and each of the constraints in \cref{eq:input SDP}, we can evaluate the corresponding trace via diagram contraction.
The main idea is to expand $X$ as a linear combination of $\psi^d_{p,q}(\sigma_j)$ using \cref{eq:idempotent expansion,eq:X expansion with psi}:
\begin{equation}
    X = \sum_{i=1}^n v_i \sum_{j=1}^{(p+q)!} \alpha_{ij} \psi^{d}_{p,q}(\sigma_j).
    \label{eq:X full expansion}
\end{equation}
Since the constant matrices $C$ and $A_k$ in \cref{eq:input SDP} are $s$-sparse, they are already provided to us as linear combinations of diagrams and elementary rank-$1$ matrices.
Using these expansions together with \cref{eq:X full expansion}, we can diagrammatically evaluate all traces in \cref{eq:input SDP} by linearity (see \cref{apx:traces} for more details).

In particular, the objective function can be written in terms of the LP variables $v_i$ as follows:
\begin{align}
    \Tr(CX)
    &= \sum_{i=1}^n v_i \sum_{j=1}^{(p+q)!} \alpha_{ij} \Tr\of[\big]{C \psi^{d}_{p,q}(\sigma_j)}
     = \sum_{i=1}^n v_i c_i
     = c\tp v
\end{align}
where $c \in \R^n$ is a vector with entries
\begin{equation}
    c_i \defeq \sum_{j=1}^{(p+q)!} \alpha_{ij} \Tr\of[\big]{C \psi^{d}_{p,q}(\sigma_j)}.
\end{equation}
Since $C$ is $s$-sparse, we can use \cref{prop:full trace} in \cref{apx:traces} to evaluate the trace.
The total time it takes to compute the vector $c$ is
\begin{equation}
    \#i \cdot \#j \cdot s
    = n N s.
    \label{eq:c complexity}
\end{equation}

Similarly, the $k$-th inequality constraint can be expressed as
\begin{equation}
    \Tr(A_k X) = a_k\tp v \leq b_k
\end{equation}
where each $a_k \in \R^n$ is a vector with entries
\begin{equation}
    (a_k)_i \defeq \sum_{j=1}^{(p+q)!} \alpha_{ij} \Tr\of[\big]{A_k \psi^{d}_{p,q}(\sigma_j)}.
\end{equation}
The total time it takes to compute the tensor $a$ is
\begin{equation}
    \#k \cdot \#i \cdot \#j \cdot s
    = \ftr n N s.
    \label{eq:a complexity}
\end{equation}

Next, let us fix $k$ and deal with the $k$-th partial trace equality constraint.
First, we expand the partial trace $\Tr_{S_k} (X)$ by linearity using \cref{eq:X full expansion}:
\begin{equation}
    \Tr_{S_k}(X)
    = \sum_{i=1}^n v_i \sum_{j=1}^{(p+q)!} \alpha_{ij} \Tr_{S_k}\of[\big]{\psi^{d}_{p,q}(\sigma_j)}.
    \label{eq:partial trace by linearity}
\end{equation}
If $\sigma_{j}^{S_k}$ denotes the diagram $\sigma_j$ with pairs of nodes in the set $S_k$ that are opposite to each other contracted, then the matrix corresponding to $\sigma_j$ has partial trace
\begin{equation}
    \Tr_{S_k}\of[\big]{\psi^{d}_{p,q}(\sigma_j)}
    = d^{\loops_{S_k}(\sigma_j)} \psi^{d}_{p_k,q_k}(\sigma_{j}^{S_k})
    \label{eq:partial trace of gamma}
\end{equation}
where $\loops_{S_k}(\sigma_j)$ is the number of loops formed and $(p_k,q_k)$ is the remaining number of systems on each side of the wall, see \cref{prop:partial trace} in \cref{apx:traces}.
Substituting this into \cref{eq:partial trace by linearity},
\begin{align}
    \Tr_{S_k}(X)
    = \sum_{i=1}^n v_i \sum_{j=1}^{(p+q)!} \alpha_{ij}
    d^{\loops_{S_k}(\sigma_j)} \psi^{d}_{p_k,q_k}(\sigma_{j}^{S_k}).
    \label{eq:X long}
\end{align}
We need to compare this to $D_k$ and derive a set of linear constraints.

Since $D_k$ is a linear combination of diagrams,
\begin{equation}
    D_k = \sum_{l=1}^{(p_k+q_k)!} e_{kl} \, \psi^{d}_{p_k,q_k}(\rho_l)
    \label{eq:D expansion}
\end{equation}
for some coefficients $e_{kl} \in \R$, where $\rho_l \in \B^d_{p_k,q_k}$ ranges over all walled Brauer algebra diagrams on $p_k + q_k = p + q - |S_k|$ nodes.
Since $\sigma_j^{S_k}$ is a diagram in $\B^d_{p_k,q_k}$,
\begin{equation}
    \psi^{d}_{p_k,q_k}(\sigma_j^{S_k})
    = \sum_{l=1}^{(p_k+q_k)!} \delta\of{\sigma_j^{S_k},\rho_l} \, \psi^{d}_{p_k,q_k}(\rho_l)
\end{equation}
where $\delta$ denotes the Kronecker delta function.
Substituting this in \cref{eq:X long},
\begin{equation}
    \Tr_{S_k}(X)
    = \sum_{l=1}^{(p_k+q_k)!}
    \sum_{i=1}^n v_i
    \sum_{j=1}^{(p+q)!} \alpha_{ij}
    d^{\loops_{S_k}(\sigma_j)}
    \delta\of{\sigma_j^{S_k},\rho_l} \,
    \psi^{d}_{p_k,q_k}(\rho_l).
    \label{eq:X extra long}
\end{equation}
Because of the assumption \eqref{eq:d big} that $d \geq p + q - \min_k |S_k|$, the representation $\psi^d_{p_k,q_k}$ is faithful due to \cref{thm:gen-schur-weyl}, and hence the matrices
$\psi^{d}_{p_k,q_k}(\rho_l)$
are linearly independent.
Comparing the coefficients at $\psi^{d}_{p_k,q_k}(\rho_l)$ in \cref{eq:D expansion,eq:X extra long},
we conclude that
\begin{equation}
    \sum_{i=1}^n v_i
    \sum_{j=1}^{(p+q)!} \alpha_{ij}
    d^{\loops_{S_k}(\sigma_j)}
    \delta\of{\sigma_j^{S_k},\rho_l}
    = e_{kl}.
\end{equation}
In other words, for every $1 \leq l \leq (p_k+q_k)!$ we get a linear constraint
\begin{equation}
    \sum_{i=1}^n v_i (d_{kl})_i = d_{kl}\tp v = e_{kl}
\end{equation}
where each $d_{kl} \in \R^n$ is a vector with entries
\begin{equation}
    (d_{kl})_i
    \defeq
    \sum_{j=1}^{(p+q)!} \alpha_{ij}
    d^{\loops_{S_k}(\sigma_j)}
    \delta\of{\sigma_j^{S_k},\rho_l}.
    \label{eq:dkl}
\end{equation}

To compute all entries $d_{kli}$ of the above tensor,
we can fix $k$ and $i$ and then evaluate the sum over $j$.
For each $j$, we determine $l$ such that
$\sigma_j^{S_k} = \rho_l$
and add the contribution
$\smash{\alpha_{ij} d^{\loops_{S_k}(\sigma_j)}}$
to the corresponding entry $d_{kli}$.
The total time it takes to perform this computation is
\begin{equation}
    \#k \cdot \#i \cdot \#j
    = \ptr \cdot
    n \cdot
    (p + q)!
    = \ptr n N.
    \label{eq:d complexity}
\end{equation}

Combining everything together, the output of our algorithm is the following LP:
\begin{equation}
    \begin{aligned}
        \max_{v} \quad & c\tp v \\
        \textrm{s.t.} \quad
        & a_k\tp v \leq b_k, & \forall k &\in [\ftr], \\
        & d_{kl}\tp v = e_{kl}, & \forall k &\in [\ptr], l \in \sof[\big]{(p_k+q_k)!}, \\
        & v_i \geq 0, & \forall i &\in [n].
    \end{aligned}
\end{equation}
It has $n \leq N$ variables $v_i$ and
\begin{equation}
    \ftr + \sum_{k=1}^{\ptr}(p_k+q_k)! + n
    \leq \ftr + \ptr N + n
\end{equation}
constraints.
The total number of scalar constants needed to specify the tensors $c,a,d$ appearing\footnote{The tensor $e$ defined in \cref{eq:D expansion} is already given to us as part of the input.}  in this LP is
\begin{equation}
    (1 + \ftr + \ptr N) n.
\end{equation}
The total amount of time it takes to compute these tensors is obtained by adding together \cref{eq:c complexity,eq:a complexity,eq:d complexity}:
\begin{equation}
    n N s + \ftr n N s + \ptr n N \leq (1 + \ftr + \ptr) n N s.
\end{equation}
This completes the description and complexity analysis of our algorithm.
\end{proof}

\begin{remark}
Our proof did not use the assumption that the input matrices $D_k$ are $s$-sparse.
This assumption only helps to keep the input SDP more compact and makes it easier to compare its size to that of the output LP.
The size of the problem description grows by a factor of $nN$ during the conversion.
\end{remark}

\begin{remark}
Our framework can be straightforwardly generalized from SDPs of the form \eqref{eq:input SDP} to the following slightly more general form:
\begin{equation}
    \begin{aligned}
        \max_{X,x_1,\dotsc,x_M} \quad & c_1 x_1 + \dotsb + c_M x_M \\
        \textrm{s.t.} \quad
        & \Tr(A_k X) \leq x_1 a_{k1} + \dotsb + x_M a_{kM} + b_k, & \forall k &\in [\ftr], \\
        & \Tr_{S_k} (\tilde{A}_k X) = D_k, & \forall k &\in [\ptr], \\
        &   \sof[\big]{X, U\xp{p} \x \bar{U}\xp{q}} = 0, & \forall U &\in \U(\C^d), \\
        & X \succeq 0,
    \end{aligned}
    \label{eq:input SDP2}
\end{equation}
where $A_k$ are constant $s$-sparse matrices that are provided as a linear combination of diagrams and elementary rank-$1$ matrices,
and $\tilde{A}_k,D_k$ are linear combinations of at most $s$ walled Brauer diagrams on registers that are left after tracing out systems $S_k$.
We use this more general form of SDPs in \cref{sec:cov maj vote}.
\end{remark}

\section{Applications}\label{sec:applications}

In this section we discuss several applications of our framework, focusing on four natural unitary-equivariant problems in quantum information theory: deciding the principal eigenvalue of a quantum state, quantum majority vote, asymmetric cloning, and transformation of a black-box unitary operation
(they are inspired by
\cite{SpectrumEstimation},
\cite{Majority},
\cite{AsymmetricCloning}, and
\cite{quintino2021deterministic}, respectively).
These are only meant as toy examples that illustrate how our framework can be easily applied to a variety of problems, hence we do not attempt to derive full analytical solutions.
Similarly, this list of applications is by no means exhaustive. Other potential applications (within quantum information) include
entanglement witness construction \cite{balanzójuandó2021positive},
and quantum error-correction \cite{kong2021nearoptimal}
and machine learning \cite{zheng2021speeding}.
We leave exploring these other applications in more depth to future work.

The four main applications mentioned above are discussed in separate sections below.
For each application we provide a separate \emph{Wolfram Mathematica} notebook \cite{github} that performs the required calculations and produces the resulting plots.

\subsection{Deciding the principal eigenvalue}

This application is inspired by the problem of estimating the spectrum of a given quantum state \cite{SpectrumEstimation}.
Let $\rho$ be a $d$-dimensional quantum state
that is picked from some unitary-invariant measure.
Given $\rho\xp{n}$ and a threshold value $c \in [1/d,1]$,
the problem is to decide whether
$\lambda_{\max} < c$ or
$\lambda_{\max} \geq c$,
where $\lambda_{\max}$ is the principal eigenvalue of $\rho$.

For concreteness, we assume that $\rho$ is produced by choosing a uniformly random pure state in $\C^d \x \C^k$ and discarding the ancillary $k$-dimensional system.
This guarantees that $\rho$ has a unitary-invariant measure.
Moreover, the eigenvalues $\lambda = (\lambda_1, \dotsc, \lambda_d)$ of such $\rho$ have the same probability density as those of a normalized Wishart matrix \cite[Proposition~4]{Nechita2007}:
\begin{equation}
  \mu_{d,k}(\lambda_1, \dotsc, \lambda_d)
  \defeq
  \frac{1}{\sqrt{d}}
  C_{d,k}
  V(\lambda)^2
  \prod_{i=1}^d \lambda_i^{k-d}
  \label{eq:mu}
\end{equation}
where $C_{d,k}$ is a normalization constant and $V(\lambda)$ is the Vandermonde determinant:
\begin{align}
  C_{d,k} &\defeq \frac{\Gamma(dk)}{\prod_{j=0}^{d-1} \Gamma(d+1-j) \Gamma(k-j)}, &
  V(\lambda) &\defeq \prod_{1 \leq i < j \leq d} (\lambda_i - \lambda_j).
\end{align}
The $1/\sqrt{d}$ factor in \cref{eq:mu} accounts for the fact that (unlike in \cite{Nechita2007}) we treat all $\lambda_1, \dotsc, \lambda_d$ as independent variables.
The density $\mu_{d,k}$ is normalized to $1$ on the standard probability simplex
\begin{equation}
  \Delta_{d-1} \defeq \set{(\lambda_1,\dotsc,\lambda_d) : \lambda_1 + \dotsb + \lambda_d = 1, \lambda_1, \dotsc, \lambda_d \geq 0}.
\end{equation}

Any strategy for this problem can be described by a two-outcome measurement with operators $P,Q \succeq 0$ such that $P + Q = I_{d^n}$, where $P$ and $Q$ correspond to outcomes
$\lambda_{\max} < c$ and
$\lambda_{\max} \geq c$, respectively.
The optimal probability of distinguishing the two cases correctly is
\begin{align}
    & p^n_{d,k}(c) =
    \min_{U \in \U(d)}
    \max_{\substack{P, Q \succeq 0 \\ P + Q = I}} \label{eq:pndk} \\
    & \of*{
        \int_{\lambda \in \Delta_{d-1}}
        \mu_{d,k}(\lambda)
        \Tr \sof[\Big]{
            \rho(\lambda,U)\xp{n}
            \of[\Big]{
                \delta\of*{\tfrac{1}{d} \leq \lambda_{\max} < c} P
                + \delta(c \leq \lambda_{\max} \leq 1) Q
            }
        }
    }, \nonumber
\end{align}
where $\rho(\lambda,U) \defeq U \diag(\lambda) U\ct$ and $\delta$ denotes the indicator function for the corresponding subregion of $\Delta_{d-1}$.

To simplify this expression, we focus on the trace.
Using the cyclic property, we can move the unitary dependence from $\rho$ onto $P$ and $Q$:
\begin{equation}
    \Tr \sof[\Big]{
        \diag(\lambda)\xp{n}
        \of[\Big]{
            \delta\of*{\tfrac{1}{d} \leq \lambda_{\max} < c} U\ctxp{n} P U\xp{n}
            + \delta(c \leq \lambda_{\max} \leq 1) U\ctxp{n} Q U\xp{n}
        }
    }.
    \label{eq:simplified trace}
\end{equation}
Then, by twirling over $\U(d)$, we can turn the worst case probability into the average case and thus remove the minimization over $U$ in \cref{eq:pndk} altogether.
Hence, without loss of generality $P,Q \in \A^d_{n,0}$ in an optimal strategy, i.e., they can be written as linear combinations of $n$-qudit permutations, see \cref{eq:Ap}.
Moreover, since $\rho(\lambda,U)\xp{n}$ is invariant under qudit permutations, we can also twirl $P$ and $Q$ over $\S_n$.
Hence, we can write $P$ as a non-negative linear combination of primitive central idempotents of $\A^d_{n,0}$, see \cref{eq:X expansion}, and set $Q = I_{d^n} - P$.

With these simplifications, we can evaluate the trace in \cref{eq:simplified trace} diagrammatically, giving us a polynomial in the eigenvalues $\lambda_i$.
Plugging this back into \cref{eq:pndk} allows us to evaluate the integral over $\lambda$.
The resulting expression depends only on the decomposition of $P$ into idempotents.
This reduces the problem from an SDP to an LP, where we only need to optimize the weights in the decomposition of $P$.
The following example provides an explicit formula for $p^n_{d,k}(c)$, for a specific combination of parameters, obtained using this procedure.

\begin{figure}[ht]
  \includegraphics[width = 0.75\textwidth]{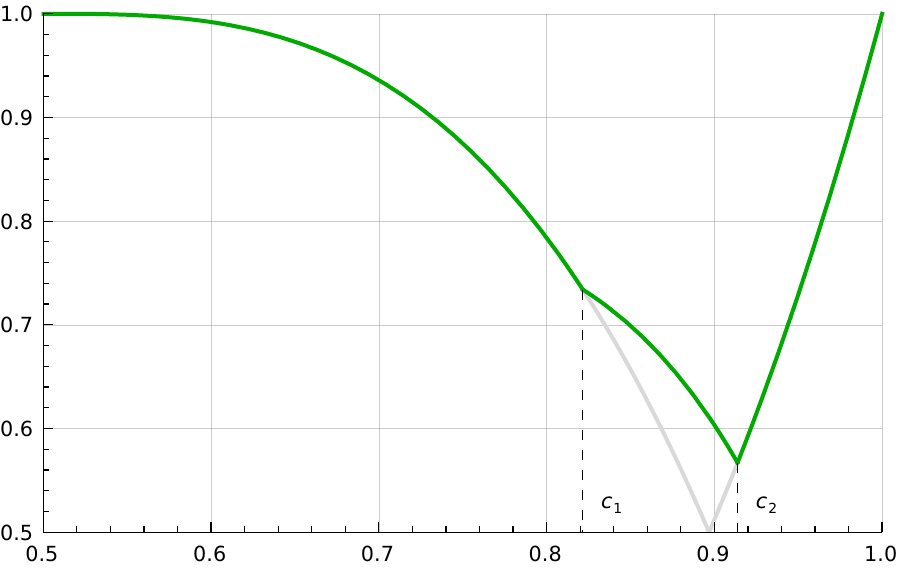}
  \caption{\label{fig:p222}Plot of the success probability $p^3_{2,2}(c)$ from \cref{eq:p222} as a function of $c \in [1/2,1]$.
  The gray curves represent the trivial lower bound from \cref{eq:trivial bound} obtained by setting $n = 1$.}
\end{figure}

\begin{figure}[ht]
  \includegraphics[width = 0.75\textwidth]{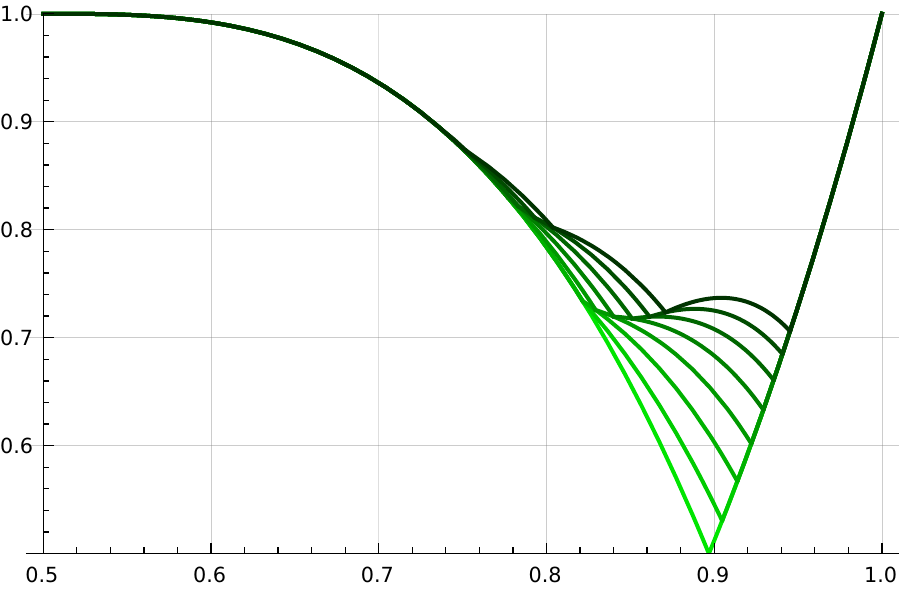}
  \caption{\label{fig:pn22}Plots of $p^n_{2,2}(c)$ for $n = 1, \dotsc, 8$
  (darker lines correspond to larger values of $n$).
  As $n$ gets larger, the curves move upwards and the number of their segments increases.}
\end{figure}

\begin{example}[$n = 3$, $d = 2$, $k = 2$]
An exact formula for the success probability as a function of the threshold value $c$ in this case is given by
\begin{equation}
  p^3_{2,2}(c) \defeq
  \begin{cases}
    2 (1-c) (4 c^2 - 2 c + 1) & \text{if $c \in [1/2, c_1]$}, \\
    \frac{7}{5} - \frac{6}{5} c (16 c^4 - 40 c^3 + 40 c^2 - 20 c + 5) & \text{if $c \in [c_1, c_2]$}, \\
    (2 c-1)^3 & \text{if $c \in [c_2, 1]$},
  \end{cases}
  \label{eq:p222}
\end{equation}
where
$c_1 \approx 0.821569391$ and
$c_2 \approx 0.913830846$
are roots of the polynomials
$96 x^5-240 x^4+200 x^3-60 x^2+3$ and
$24 x^5-60 x^4+70 x^3-45 x^2+15 x-3$, respectively.
A plot of the function $p^3_{2,2}(c)$ is shown in \cref{fig:p222}.
Note that $p^3_{2,2}(1/2) = p^3_{2,2}(1) = 1$ since the problem of deciding the largest eigenvalue becomes trivial for extreme values of $c$.
The success probability $p^3_{2,2}(c)$ is always at least $p^3_{2,2}(c_2) \approx 0.566968020$
and never below the trivial $n = 1$ lower bound
\begin{equation}
  p^1_{2,2}(c) \defeq \max \set{2 (1-c) (4 c^2 - 2 c + 1), (2 c-1)^3}
  \label{eq:trivial bound}
\end{equation}
whose minimum is $1/2$ at $c = \frac{1}{2} + \frac{1}{2^{4/3}}$.
Using the same procedure we also obtained explicit expressions for $p^n_{2,2}(c)$ with $n = 1, \dotsc, 8$. Their plots are shown in \cref{fig:pn22}.
\end{example}

\subsection{Quantum majority vote} \label{sec:cov maj vote}

This application of our framework is inspired by the work of \cite{Majority} on optimal unitary-equivariant quantum channels for evaluating permutation-equivariant and symmetric Boolean functions.

As usual, let $p,q \geq 0$ denote the number of inputs and outputs, and let $d \geq 2$ denote their dimension.
We are interested in functions of the form $f\colon [d]^p \to [d]^q$, or more generally in multi-valued functions or relations $f \subseteq [d]^p \times [d]^q$.

We call $f$ \emph{equivariant} with respect to the symmetric group $\S_d$ on the alphabet $[d]$ if $f(\pi \cdot x)=\pi \cdot f(x)$ for every $\pi \in \S_d$, where
$\pi \cdot (x_1,\dotsc,x_n) \defeq (\pi(x_1),\dotsc,\pi(x_n))$ for all $x_1, \dotsc, x_n \in [d]$ (this is a classical analogue of \cref{def:equivariance}).
We say that $f$ is \emph{symmetric} if its inputs and outputs have $\S_p \times \S_q$ permutation symmetry, i.e., if $f(\pi \circ x) = \sigma \circ f(x)$ for every $(\pi, \sigma) \in \S_p \times \S_q$, where $\pi \circ x \defeq (x_{\pi^{-1}(1)},\dotsc,x_{\pi^{-1}(p)})$ and $\sigma \circ f(x)$ is defined similarly (this is a classical analogue of \cref{def:sym channel}).

The most important example of an equivariant and symmetric function is the \emph{majority vote} function $f\colon [d]^p \to [d]$ that outputs the most frequently occurring input symbol. Since generally this symbol may not be unique, we prefer to think of $f \subseteq [d]^p \times [d]$ as a relation.
One can also consider generalizations with $q \geq 1$ where the $q$ most popular symbols must be output in any order.

A natural quantum generalization of an equivariant function $f$ is a unitary-equivariant map $\Psi_f\colon \End(V^p) \to \End(V^q)$ such that $\Psi_f(\proj{x}) = \proj{f(x)}$ for all $x \in [d]^p$. In case of a multi-valued function $f$, the ideal quantum map can be taken as $\Psi_f(\proj{x}) = \Pi_{f(x)}$, where $\Pi_{f(x)} \defeq \sum_{y \in f(x)} \proj{y}$ is the rank-$\abs{f(x)}$ orthogonal standard basis projector on all valid output states.
Note that the ideal map $\Psi_f$ may not be a quantum channel in general.

Given a classical $\S_d$-equivariant and $\S_p \times \S_q$ symmetric relation $f \subseteq [d]^p \times [d]^q$, we would like to find a unitary-equivariant $p \to q$ quantum channel $\Phi_f$ that approximates the ideal functionality.
Namely, one that maximizes the worst-case fidelity
\begin{equation}
  \min_{x \in [d]^p} \Tr\of[\big]{\Phi_f(\proj{x}) \, \Pi_{f(x)}}.
\end{equation}
We can formulate this as an SDP for computing the worst-case fidelity $F \in \R$ of an optimal quantum channel represented by its Choi matrix $X \in \End(V^{p,q})$:
\begin{equation}
  \begin{aligned}
    \max_{X,F} \quad & F \\
    \textrm{s.t.} \quad & \Tr \of[\Big]{X\of[\big]{\proj{x} \x \Pi_{f(x)}}} \geq F,
    & \forall x &\in [d]^p, \\
    & \Tr_{V^q} (X) = I_{V^p}, \\
    & \sof[\big]{X, U\xp{p} \x \bar{U}\xp{q}} = 0, & \forall U &\in \U(\C^d), \\
    & X \succeq 0.
  \end{aligned}
  \label{eq:function SDP}
\end{equation}

As a generalization of \cite{Majority}, we consider the majority relation on any alphabet $[d]$ with $d \geq 3$. For simplicity, we restrict to the case of $p = 3$ inputs (and $q = 1$ outputs). In this case the majority relation for any $d \geq 3$ is fully defined by
\begin{align}
  111 &\mapsto 1, \nonumber \\
  112 &\mapsto 1, \label{eq:maj} \\
  123 &\mapsto 1,2,3, \nonumber
\end{align}
which are extended to the whole domain using symmetry and equivariance.
These rules cover three distinct cases: when all three inputs are equal,
when one of them is different,
and when all three are different.
Since there is no clear majority in the last case, the relation can output any of the three symbols.

For quantum majority vote with $p = 3$, $q = 1$, and $d \geq 3$ the symmetries of the Choi matrix $X$ of the optimal quantum channel $\Phi_f$ allow us to reduce the SDP \eqref{eq:function SDP} to an LP using the ansatz \eqref{eq:X expansion} for $X$:
\begin{equation}
    X = \sum_{i=1}^{n(d)} v_i \varepsilon^\A_{\T_i},
    \label{eq:X expansion maj vote}
\end{equation}
where $\varepsilon_{\T_i}$ are primitive idempotents that correspond to distinct root-leaf paths $\T_i$ in the Bratteli diagram of $\A^d_{3,1}$, and $n(d) \defeq n_{3,1}^d(\text{Gelfand--Tsetlin})$ is the total number of such paths, see \cref{eq:n GT}.
Based on \cref{eq:maj}, which defines the majority relation on three symbols, the resulting LP for any $d \geq 3$ has the following form:
\begin{equation}
\begin{aligned}
  \max_{F, v_1, \dotsc, v_{n(d)}} \quad & F \\
  \textrm{s.t.} \quad
  & \sum_{i=1}^{n(d)} v_i \bra{111,1} \varepsilon^\A_{\T_i} \ket{111,1} \geq F, \\
  & \sum_{i=1}^{n(d)} v_i \bra{112,1} \varepsilon^\A_{\T_i} \ket{112,1} \geq F, \\
  & \sum_{i=1}^{n(d)} v_i \sum_{y=1}^3 \bra{123,y} \varepsilon^\A_{\T_i} \ket{123,y} \geq F, \\
  & \sum_{i=1}^{n(d)} v_i \Tr_{\Vout} \varepsilon^\A_{\T_i} = I_{\Vin}, \\
  & v_i \geq 0, \quad \forall i \in [n(d)], \\
  & v_i = v_j \text{ whenever $\T_i$ and $\T_j$ share the same last edge}.
  \label{eq:lp_maj_vote}
\end{aligned}
\end{equation}
The last condition is a consequence of \cref{prop:Sp1_and_unitary} and ensures $\S_p \times \S_1$ symmetry of the majority relation. Enforcing this symmetry effectively decreases the number of variables in the LP \eqref{eq:lp_maj_vote} from $n(d)$, which corresponds to the Gelfand--Tsetlin symmetry, to the smaller value of $n_{3,1}^d(\S_3 \times \S_1)$ (see \cref{eq:n Spq}) that corresponds to the $\S_3 \times \S_1$ symmetry.
Explicit values of these numbers can be found in \cref{table:GT,table:SpSq} in \cref{apx:numeric tables}.

The LP (\ref{eq:lp_maj_vote}) can be solved exactly and the optimal fidelity turns out to be $F = 8/9$ for all $d \geq 2$, thus extending the $d = 2$ result of \cite{Majority}.

\subsection{Asymmetric cloning} \label{sec:asymmetric cloning}

In this section, we provide an example of how our approach based on primitive idempotents can be used to solve a general unitary-equivariant SDP that does not reduce to an LP. This example is based on the problem of asymmetric cloning from \cite{AsymmetricCloning}. The problem is to find a $1 \to q$ quantum channel $\Phi\colon \End(V) \rightarrow \End(V^q)$ whose marginals $\Phi_i \defeq {\Tr_{[q] \backslash \set{i}}} \circ \Phi$ satisfy
\begin{equation}
  \Phi_i (\rho) = p_i \rho + (1 - p_i) \frac{I}{d}
  \label{eq:marginals}
\end{equation}
for all $i \in [q]$ and states $\rho \in \D(V)$ where $V = \C^d$.
We consider the case $q=3$ with $d=2$ and $d=3$, and plot the set of triples $(p_1, p_2, p_3)$ in \cref{eq:marginals} that are physically realizable.
Note that this set is invariant under permutations of $p_i$.

According to \cite{AsymmetricCloning}, the Choi matrix $X^\Phi$ of the channel $\Phi$ is a linear combination of partially transposed permutation matrices, i.e., $X^\Phi \in \A^d_{1,q}$. We can use the primitive idempotents from \cref{eq:primitive_main_A_d} to construct the blocks $X^\Phi_\lambda$ in \cref{rmk:eq_blocks} without explicitly computing the Schur transform $U_{\textnormal{Sch}(1,q)}$, see \cref{apx:blocks of Apq} for more details. The resulting blocks are then subject to positive semidefinite and trace constraints. In this way we can formulate the question of physical realizability of the channel $\Phi$ as a semidefinite feasibility problem.

Two examples of SDPs resulting from this procedure are given below.
They characterize asymmetric $1 \to 3$ cloning in dimensions $d = 2$ and $d = 3$.
Here we denote for brevity $X^i \defeq X^\Phi_{\lambda_i}$ for every $\lambda_i \in \Irr(\A^d_{p,q})$.

\begin{figure}[ht]
\includegraphics[width = 0.5\textwidth]{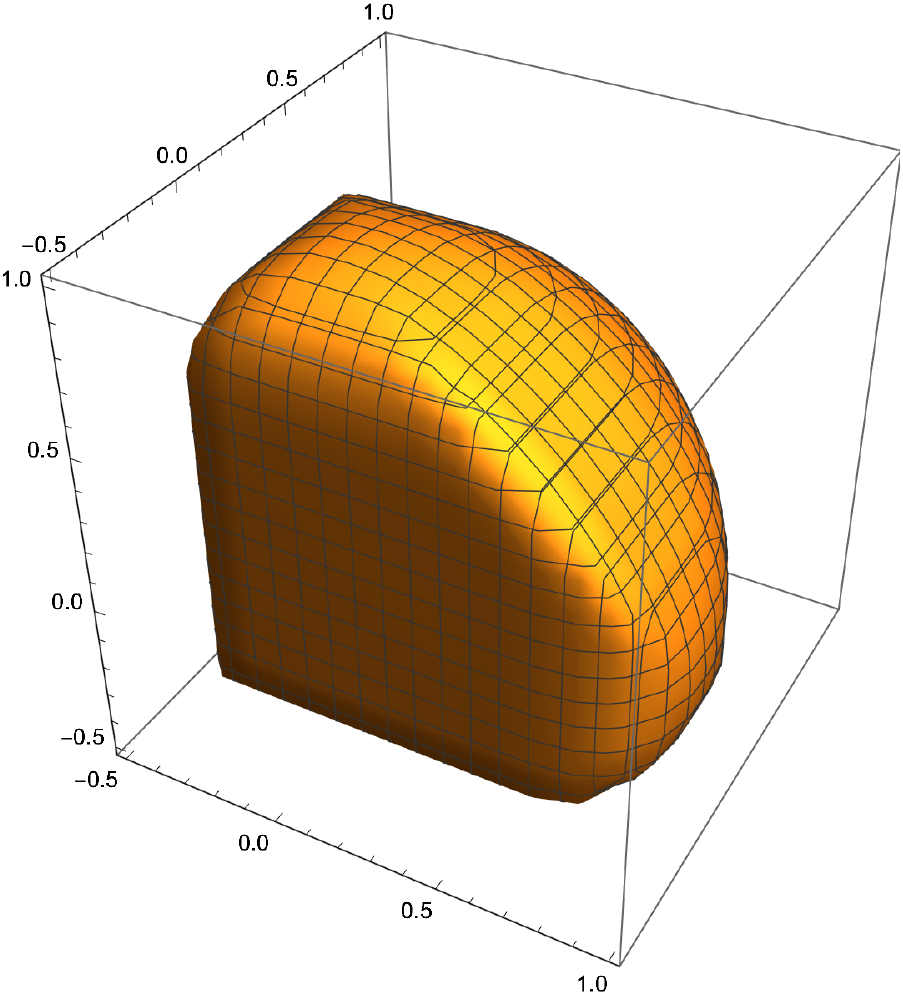}
\caption{\label{fig:app3_d2}Plot of the feasible region for $(p_1,p_2,p_3)$ in the asymmetric cloning SDP for $q=3$ and $d=2$.}
\end{figure}

\begin{example}[$q = 3$ and $d = 2$]
Positive semidefinite constraints:
\begin{align}
    \mx{
        X_{1,1}^1
    } &\succeq 0, &
    \mx{
        X_{1,1}^2 & X_{1,2}^2 \\
        X_{2,1}^2 & X_{2,2}^2
    } &\succeq 0, &
    \mx{
        X_{1,1}^3 & X_{1,2}^3 & X_{1,3}^3 \\
        X_{2,1}^3 & X_{2,2}^3 & X_{2,3}^3 \\
        X_{3,1}^3 & X_{3,2}^3 & X_{3,3}^3
    } &\succeq 0.
\end{align}
Trace constraint:
\begin{equation}
    5 X_{1,1}^1 + X_{1,1}^2 + X_{2,2}^2 + 3 X_{1,1}^3 + 3 X_{2,2}^3 + 3 X_{3,3}^3 = 2.
\end{equation}
Expressions for realizable triples $(p_1, p_2, p_3)$:
\begin{align}
    p_1 &= \frac{1}{3} \left(
      3 X_{1,1}^3 + X_{1,1}^2 + 5 X_{1,1}^1 + 3 X_{2,2}^3 + 9 X_{3,3}^3 + 3 X_{2,2}^2 - 3
      \right), \\
    p_2 &= \frac{1}{6} \left(
      6 X_{1,1}^3 + 5 X_{1,1}^2 + 10 X_{1,1}^1 + 15 X_{2,2}^3 + 3 \sqrt{3} X_{2,3}^3 + 3 \sqrt{3} X_{3,2}^3
      \right. \\ & \nonumber \quad \left. {}
      + 9 X_{3,3}^3 + \sqrt{3} X_{1,2}^2 + \sqrt{3} X_{2,1}^2 + 3 X_{2,2}^2 - 6
      \right), \nonumber \\
    p_3 &= \frac{1}{6} \left(
      14 X_{1,1}^3 + 5 X_{1,1}^2 + 10 X_{1,1}^1 + 2 \sqrt{2} X_{1,2}^3 + 2 \sqrt{2} X_{2,1}^3 + 7 X_{2,2}^3
      \right. \\ & \nonumber \quad \left. {}
      + 9 X_{3,3}^3 + 2 \sqrt{6} X_{1,3}^3 + 2 \sqrt{6} X_{3,1}^3 + \sqrt{3} X_{2,3}^3 + \sqrt{3} X_{3,2}^3 - \sqrt{3} X_{1,2}^2
      \right. \\ & \nonumber \quad \left. {}
      - \sqrt{3} X_{2,1}^2 + 3 X_{2,2}^2 - 6
      \right).
\end{align}
The feasible region for $(p_1, p_2, p_3)$ is shown in \cref{fig:app3_d2}.
\end{example}

\begin{figure}[ht]
\includegraphics[width = 0.5\textwidth]{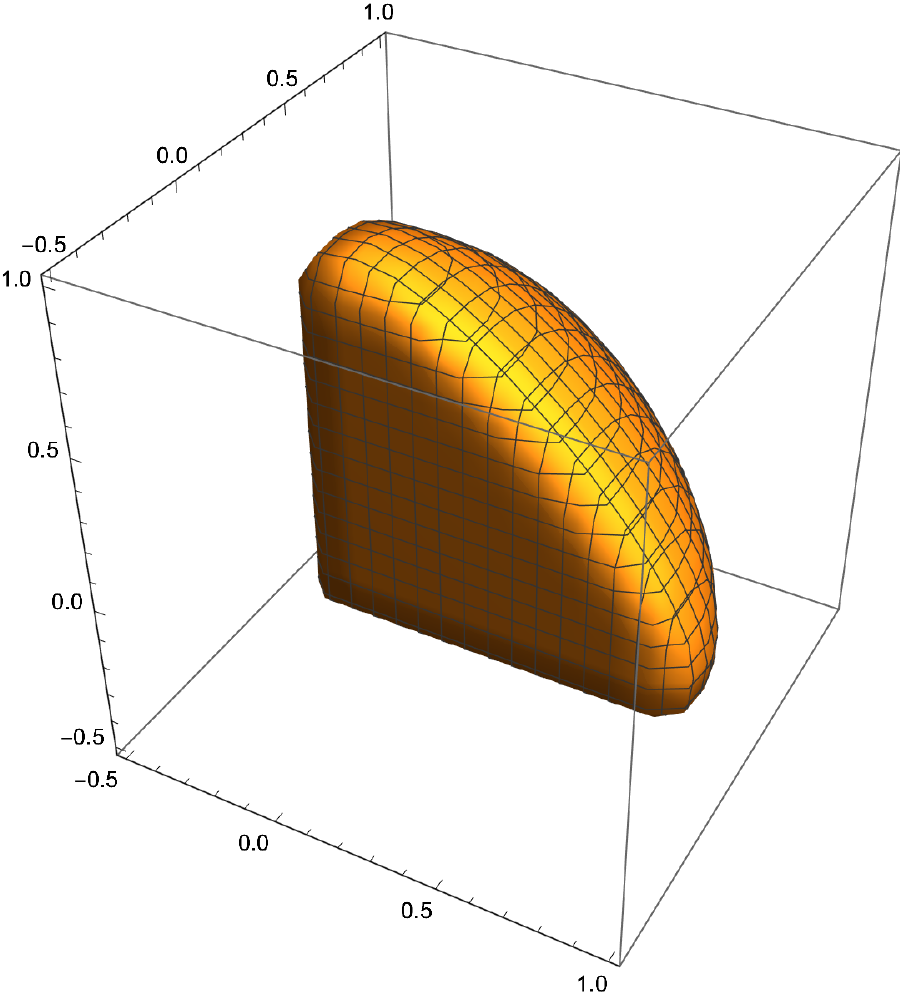}
\caption{\label{fig:app3_d3}Plot of the feasible region for $(p_1,p_2,p_3)$ in the asymmetric cloning SDP for $q=3$ and $d=3$.}
\end{figure}

\begin{example}[$q = 3$ and $d = 3$]
Positive semidefinite constraints:
$\mx{X_{1,1}^1} \succeq 0$,
\begin{align}
\mx{
    X_{1,1}^2 & X_{1,2}^2 \\
    X_{2,1}^2 & X_{2,2}^2
} &\succeq 0, &
\mx{
    X_{1,1}^3 & X_{1,2}^3 & X_{1,3}^3 \\
    X_{2,1}^3 & X_{2,2}^3 & X_{2,3}^3 \\
    X_{3,1}^3 & X_{3,2}^3 & X_{3,3}^3
} &\succeq 0, &
\mx{
    X_{1,1}^4 & X_{1,2}^4 & X_{1,3}^4 \\
    X_{2,1}^4 & X_{2,2}^4 & X_{2,3}^4 \\
    X_{3,1}^4 & X_{3,2}^4 & X_{3,3}^4
} &\succeq 0.
\end{align}
Trace constraint:
\begin{equation}
    X_{1,1}^4+2 X_{1,1}^3+5 X_{1,1}^2+8 X_{1,1}^1+X_{2,2}^4+X_{3,3}^4+2 X_{2,2}^3+2 X_{3,3}^3+5 X_{2,2}^2 = 1.
\end{equation}
Expressions for realizable triples $(p_1, p_2, p_3)$:
\begin{align}
    p_1 &= \frac{1}{8} \left(
      3 X_{1,1}^4 + 6 X_{1,1}^3 + 15 X_{1,1}^2 + 24 X_{1,1}^1 + 3 X_{2,2}^4 + 12 X_{3,3}^4 + 6 X_{2,2}^3
      \right. \\ & \nonumber \quad \left. {}
      + 24 X_{3,3}^3 + 15 X_{2,2}^2 - 4
      \right), \\
    p_2 &= \frac{1}{8} \left(
      3 X_{1,1}^4 + 6 X_{1,1}^3 + 15 X_{1,1}^2 + 24 X_{1,1}^1 + 11 X_{2,2}^4 + 2 \sqrt{2} X_{2,3}^4 + 2 \sqrt{2} X_{3,2}^4
      \right. \\ & \nonumber \quad \left. {}
      + 4 X_{3,3}^4 + 22 X_{2,2}^3 + 4 \sqrt{2} X_{2,3}^3 + 4 \sqrt{2} X_{3,2}^3 + 8 X_{3,3}^3 + 15 X_{2,2}^2 - 4
      \right), \\
    p_3 &= \frac{1}{8} \left(
      9 X_{1,1}^4 + 21 X_{1,1}^3 + 15 X_{1,1}^2 + 24 X_{1,1}^1 + 8 X_{3,3}^3 + 15 X_{2,2}^2 + 5 X_{2,2}^4
      \right. \\ & \nonumber \quad \left. {}
      + 7 X_{2,2}^3 + 4 X_{3,3}^4 + 2 \sqrt{3} X_{1,2}^4 + 2 \sqrt{3} X_{2,1}^4 - \sqrt{2} X_{2,3}^4 - \sqrt{2} X_{3,2}^4
      \right. \\ & \nonumber \quad \left. {}
      - \sqrt{6} X_{3,1}^4 - \sqrt{6} X_{1,3}^4 + \sqrt{15} X_{1,2}^3 + \sqrt{15} X_{2,1}^3 + \sqrt{30} X_{3,1}^3 + \sqrt{30} X_{1,3}^3
      \right. \\ & \nonumber \quad \left. {}
      + \sqrt{2} X_{2,3}^3 + \sqrt{2} X_{3,2}^3 - 4
      \right).
\end{align}
The feasible region for $(p_1, p_2, p_3)$ is shown in \cref{fig:app3_d3}.
\end{example}

\subsection{Transformation of a black-box unitary operation}\label{sec:quantum_comb}

Following the work of \cite{quintino2021deterministic} and others \cite{quintino2019reversing,quintino2019probabilistic,yoshida2021universal,yoshida2022reversing,ComplexConjugation}, we present another application of our method---transforming a black-box unitary operation. Consider the following general problem: given $n$ copies of an unknown $d$-dimensional unitary $U$, the task is to find a universal protocol that implements $f(U)$, where $f$ is some function of $U$. This protocol can be either \emph{deterministic} or \emph{probabilistic}, depending on whether it always succeeds or not, and either \emph{exact} or \emph{non-exact}, depending on the channel fidelity between the ideal channel and the one implemented by the protocol. In particular, we focus on the deterministic case where $f(U)=U\tp$. Our main result in this section is the existence of an exact and deterministic protocol which transforms $4$ copies of a black-box single-qubit unitary $U$ into $f(U) = U\tp$. This result is similar to the recent work \cite{yoshida2022reversing}, which proves the same claim for the function $f(U) = U^{-1}$. For more detailed background on this topic we refer the reader to  \cite{quintino2021deterministic,yoshida2022reversing}; here we only introduce the necessary ingredients needed to describe our approach.

Following previous work, we use the formalism of quantum superchannels. We search for a deterministic sequential protocol accomplishing our task by expressing it as a quantum sequential superchannel \cite{quintino2021deterministic}.
A \emph{quantum superchannel} is a linear map $\mathcal{C}\colon \bigotimes_{i=1}^{n} \of[\big]{\End(\I_i) \rightarrow \End(\O_i)} \rightarrow \of[\big]{\End(\P) \rightarrow \End(\F)}$ that transforms $n$ quantum channels into a new quantum channel.
Here the spaces $\I_i = \O_i = \P = \F = \C^d$ correspond to the inputs $\I_i$ and outputs $\O_i$ of the $i$-th copy of the channel $\mathcal{U}(\rho) \defeq U \rho U\ct$ associated with the unknown input unitary $U$, and $\P$ and $\F$ are the input and output spaces of the desired output channel $\mathcal{U}_f(\rho) := f(U) \rho f(U)\ct$ that represents the target unitary $f(U)$. Let $\I^n \defeq \bigotimes_{i=1}^n \I_i$ and $\O^n \defeq \bigotimes_{i=1}^n \O_i$.
A \emph{quantum sequential superchannel} $\mathcal{C}$ (also known as a \emph{quantum comb}) is a quantum superchannel with certain additional constraints on its Choi matrix $C \in \End(\P \otimes \I^n \otimes \O^n \otimes \F)$ \cite{chiribella2008qca}:
\begin{align}
    C &\succeq 0, \label{eq:comb_constraint_psd}\\
    \Tr C &= 1, \label{eq:comb_constraint_tr} \\
    \Tr_{\I_i} C_i &= C_{i-1} \otimes I_{\O_{i-1}}, \quad \forall i \in [n+1],
    \label{eq:comb_constraints_lin}
\end{align}
where $C_{n+1} \defeq C$, $\I_{n+1} \defeq \F$, $\O_{0} \defeq \P$ and $C_{i-1} \defeq \frac{1}{d}\Tr_{\I_i \O_{i-1}} C_i$.

Finding a deterministic sequential superchannel $\mathcal{C}$ which implements the operation $\mathcal{C} ( \mathcal{U}\xp{n}) = \mathcal{U}\tp$ with highest possible average channel fidelity is equivalent to solving the following SDP for the Choi matrix $C$ of $\mathcal{C}$ \cite{quintino2021deterministic}:
\begin{equation}
  \begin{aligned}
    \max_{C} \quad & \Tr \of{C \Omega_{n,d}} \\
    \textrm{s.t.} \quad & C \text{ satisfies \eqref{eq:comb_constraint_psd}--\eqref{eq:comb_constraints_lin}},
  \end{aligned}
  \label{eq:black-box unitary SDP}
\end{equation}
where $\Omega_{n,d}$ is given by
\begin{equation}
    \Omega_{n,d} \defeq \frac{1}{d^2} \sum_{\substack{\lambda \in \Irr(\A^d_{n,1}) \\ i,j \in [d_\lambda]}} \frac{ \of{E^{\lambda}_{ij}}_{\O^n \P} \otimes \of{E^{\lambda}_{ij}}_{\I^n \F} }{m_\lambda}
\end{equation}
and $E^{\lambda}_{ij} \defeq  \psi^d_{n,1}(\varepsilon^{\lambda}_{ij})$, where $\varepsilon^{\lambda}_{ij}$ is the same as $\varepsilon_{ij}$ for a given $\lambda \in \Irr(\A^d_{n,1})$ in \cref{apx:blocks of Apq} where $(p,q) = (n,1)$. Notice that $\Omega_{n,d}$ has the mixed unitary symmetry:
\begin{equation}
    \sof[\big]{\Omega_{n,d}, U\xp{n}_{\O^n} \x \bar{U}_{\P} \x V\xp{n}_{\I^n} \x \bar{V}_{\F}} = 0, \qquad \forall U, V \in \U(\C^d).
    \label{eq:app4_symmetry_Omega}
\end{equation}
Therefore without loss of generality the optimal solution of the SDP (\ref{eq:black-box unitary SDP}) also has the same symmetry:
\begin{equation}
    \sof[\big]{C, U\xp{n}_{\O^n} \x \bar{U}_{\P} \x V\xp{n}_{\I^n} \x \bar{V}_{\F}} = 0, \qquad \forall U, V \in \U(\C^d),
    \label{eq:app4_symmetry_C}
\end{equation}
which allows us to use the following ansatz for $C$:
\begin{equation}
    C = \sum_{\lambda, \mu \in \Irr(\A^d_{n,1})} \sum_{i,j = 1}^{d_\lambda} \sum_{k,l = 1}^{d_\mu} c^{\lambda \mu}_{ijkl} \of{E^{\lambda}_{ij}}_{\O^n \P} \otimes \of{E^{\mu}_{kl}}_{\I^n \F} .
    \label{eq:app4_ansatz_C}
\end{equation}

\begin{table}[!t]
\begin{tabular}{c|c||c|c|c|c}
    $f(U)$ & \diagbox{$d$}{$n$} &   1   &  2     &  3     &  4     \\ \hline \hline
    \multirow{2}{*}{$U\tp$}   &   2   &  0.500000  & 0.750000 & 0.933013 & 1.000000 \\
        &   3  & 0.222222 & 0.407407 & 0.626597 & 0.799250      \\ \hline
    \multirow{2}{*}{$U^{-1}$} &   2   &  0.500000  & 0.750000 & 0.933013 & 1.000000 \\
        &   3  & 0.222222 & 0.333333 & 0.444444 &  0.555556
\end{tabular}
\caption{\label{table:app4_results} Optimal values of the SDP (\ref{eq:black-box unitary SDP}). The column $f(U)$ indicates the task, for which we want to find a deterministic sequential superchannel $\mathcal{C}$. In both tasks we reproduce the results of \cite{quintino2021deterministic} for $d = 2$, $n \leq 3$ and $d = 3$, $n \leq 2$. We also reproduce the results for $f(U)=U^{-1}$ \cite{yoshida2022reversing} when $d = 2$, $n = 4$ and $d = 3$, $n = 3$ and $d = 3$, $n = 4$. Finally, we obtain new results for the unitary transposition task for $d = 2$, $n = 4$ and $d = 3$, $n = 3$ and $d = 3$, $n = 4$. }
\end{table}

Both matrices $C$ and $\Omega_{n,d}$ can be thought of as linear combinations of tensor products of two walled Brauer algebra diagrams, so our techniques from \cref{sec:SDP} can be used to rewrite the SDP \eqref{eq:black-box unitary SDP} explicitly in terms of the variables $c^{\lambda \mu}_{ijkl}$ and then numerically solve it.

The trace constraint \eqref{eq:comb_constraint_tr} involves a linear combination of diagrams and hence is easy to evaluate using \cref{eq:full trace}. The semidefinite constraint \eqref{eq:comb_constraint_psd} becomes
\begin{equation}
    C \succeq  0 \qquad \Leftrightarrow \qquad \sof[\big]{c^{\lambda \mu}_{ijkl}}_{(ik),(jl)} \succeq  0, \quad \forall \lambda, \mu \in \Irr(\A^d_{n,1}),
\end{equation}
where we think of $\sof[\big]{c^{\lambda \mu}_{ijkl}}_{(ik),(jl)} \in \End(\C^{d_\lambda} \otimes \C^{d_\mu})$ as matrices. The only tricky constraint is \eqref{eq:comb_constraints_lin}. The $i$-th constraint says that a certain linear combination of diagrams from the algebra $\B^d_{i-1,1} \otimes \B^d_{i-1,0}$ equals to $0$ under the map $\psi^d_{i-1,1} \otimes \psi^d_{i-1,0}$, i.e., it belongs to the kernel of this map. One way of dealing with this type of constraint is to map each diagram on both sides of the tensor product to all of its irreducible blocks. This can be done using the method outlined in \cref{apx:blocks of Apq}.

We can also solve a similar problem for the function $f(U)=U^{-1}$ and verify the results obtained in \cite{yoshida2022reversing}. In that case, the SDP \eqref{eq:black-box unitary SDP} has the same form, except that the matrices $C$ and $\Omega_{n,d}$ posses a different symmetry: they commute with $U\xp{n+1}_{\O^n \P} \x V\xp{n+1}_{\I^n\F}$ for every $U, V \in \U(\C^d)$. The elementary matrices $E^{\lambda}_{ij} \defeq  \psi^d_{n+1,0}(\varepsilon^{\lambda}_{ij})$ are labeled by irreducible representations $\lambda \in  \Irr(\A^d_{n+1,0})$ of the symmetric group algebra $\CS_{n+1}$. This situation corresponds to the $(p,q) = (n+1,0)$ case of our formalism.

Our numerical results are summarized in \cref{table:app4_results}, and our \emph{Wolfram Mathematica} code for obtaining these values and the corresponding optimal Choi matrices $C$ can be found on GitHub \cite{github}.

\printbibliography

\appendix
\newpage

\section{Lifting traces from \texorpdfstring{$\B^d_{p,q}$}{Bpq} to \texorpdfstring{$\A^d_{p,q}$}{Apq}}
\label{apx:traces}

Representing the elements of the matrix algebra $\A^d_{p,q}$ as preimages of diagrams under $\psi^d_{p,q}$ is particularly useful when computing traces and partial traces of $\psi^d_{p,q}(\sigma)$ for any $\sigma \in \B^d_{p,q}$.
The following two propositions relate the matrix traces of $\psi^d_{p,q}$ to the diagrammatic traces of $\sigma$ defined in \cref{eq:full trace,eq:partial trace}.

\begin{proposition}\label{prop:full trace}
For any $\sigma \in \B^d_{p,q}$,
\begin{equation}
  \Tr(\psi^d_{p,q}(\sigma)) = \Tr(\sigma).
  \label{eq:full trace relation}
\end{equation}
In particular, when $\sigma$ is a single diagram,
\begin{equation}
  \Tr(\psi^d_{p,q}(\sigma)) = d^{\loops(\sigma)}
\end{equation}
where $\loops(\sigma)$ is the number of loops created by connecting all pairs of opposite vertices of $\sigma$.
\end{proposition}

\begin{proof}
To establish \cref{eq:full trace relation}, it suffices to consider only the case when $\sigma$ is a single diagram since the general case follows by linearity.
Using \cref{eq:psi,eq:sigma},
\begin{align}
  \label{eq:full contract derivation}
  \Tr(\psi^d_{p,q}(\sigma))
 &= \sum_{i_1, \dotsc, i_{p+q} \in [d]}
    \of[\big]{ \bra{i_1} \x \dotsb \x \bra{i_{p+q}} }
    \, \psi^d_{p,q}(\sigma) \,
    \of[\big]{ \ket{i_1} \x \dotsb \x \ket{i_{p+q}} } \\
 &= \sum_{\substack{i_1, \dotsc, i_{p+q} \in [d] \\ i_{\p{1}}, \dotsc, i_{\p{p+q}} \in [d]}}
    \sigma^{i_1, \dotsc, i_{p+q}}_{i_{\p{1}}, \dotsc, i_{\p{p+q}}}
    \prod_{k \in [p+q]} \delta_{i_k,i_{\p{k}}} \\
 &= \sum_{\substack{i_1, \dotsc, i_{p+q} \in [d] \\ i_{\p{1}}, \dotsc, i_{\p{p+q}} \in [d]}}
    \prod_{(r,s) \in \sigma} \delta_{i_r,i_s}
    \prod_{k \in [p+q]}\delta_{i_k,i_{\p{k}}} \\
 &= d^{\loops(\sigma)}.
\end{align}
The last equality follows by partitioning the product of delta functions into closed loops and observing that all indices on a given loop must have the same value.
The final value agrees with the diagrammatic definition of $\Tr(\sigma)$ in \cref{eq:full trace}.
\end{proof}

The following generalization allows to graphically compute the partial trace $\Tr_{S}(\psi^d_{p,q}(\sigma))$ for any subset of systems $S \subseteq [p+q]$.

\begin{proposition}\label{prop:partial trace}
For any $\sigma \in \B^d_{p,q}$ and subset $S \subseteq [p+q]$,
\begin{equation}
    \Tr_{S}\of[\big]{\psi^{d}_{p,q}(\sigma)}
    = \psi^d_{p',q'}(\Tr_S(\sigma))
\end{equation}
where $\Tr_S(\sigma)$ is defined in \cref{eq:partial trace} and $p'$ and $q'$ denote the number of leftover systems on both sides of the wall.
In particular, when $\sigma$ is a single diagram,
\begin{equation}
    \Tr_{S}\of[\big]{\psi^{d}_{p,q}(\sigma)}
    = d^{\loops_S(\sigma)} \psi^{d}_{p',q'}(\sigma^S)
\end{equation}
where $\sigma^{S}$ denotes the diagram $\sigma$ with opposite pairs of nodes that belong to $S$ contracted, and $\loops_S(\sigma)$ is the number of loops formed in this process.
\end{proposition}

\begin{proof}
By linearity, it suffices to establish the result for any diagram $\sigma \in \B^d_{p,q}$.
Note from \cref{eq:psi} that
\begin{equation}
  \psi^d_{p,q}(\sigma)
  = \!\!\!
    \sum_{\substack{l_1, \dotsc, l_{p+q} \in [d] \\ l_{\p{1}}, \dotsc, l_{\p{p+q}} \in [d]}}
    \!\!\!
    \sigma^{l_1, \dotsc, l_{p+q}}_{l_{\p{1}}, \dotsc, l_{\p{p+q}}} \,
    \ketbra{l_{\p{1}}}{l_1} \x \dotsb \x \ketbra{l_{\p{p+q}}}{l_{p+q}}
  = \sum_{l,\p{l} \in [d]^{p+q}} \!\!
    \sigma^l_{\p{l}} \,
    \ketbra{\p{l}}{l}
  \label{eq:psi explicit}
\end{equation}
where $l = (l_1,\dotsc,l_{p+q})$ and $\p{l} = (l_{\p{1}},\dotsc,l_{\p{p+q}})$.
Letting $\bar{S} \defeq [p+q] \setminus S$, we can generalize \cref{eq:full contract derivation} as follows:
\begin{align}
  \Tr_S(\psi^d_{p,q}(\sigma))
 &= \sum_{i \in [d]^S}
    \of[\big]{
      \bra{i}_S
      \x I_{\bar{S}}
    }
    \, \psi^d_{p,q}(\sigma) \,
    \of[\big]{
      \ket{i}_S
      \x I_{\bar{S}}
    } \\
 &= \sum_{i \in [d]^S}
    \of[\big]{
      \bra{i}_S
      \x I_{\bar{S}}
    }
    \of[\Big]{
      \sum_{l,\p{l} \in [d]^{p+q}}
      \sigma^l_{\p{l}} \,
      \ketbra{\p{l}}{l}
    }
    \of[\big]{
      \ket{i}_S
      \x I_{\bar{S}}
    } \\
    &= \sum_{i \in [d]^S}
    \sum_{l,\p{l} \in [d]^{p+q}}
    \sigma^l_{\p{l}}
    \of[\Big]{
      \prod_{k \in S}
      \delta_{i_k,l_{\p{k}}}
      \delta_{l_{k},i_k}
    }
    \of[\Big]{
      \bigotimes_{k \in \bar{S}}
      \ketbra{l_{\p{k}}}{l_k}_k
    } \\
 &= \sum_{l,\p{l} \in [d]^{p+q}}
    \sigma^l_{\p{l}}
    \of[\Big]{\prod_{k \in S} \delta_{l_{\p{k}},l_k}}
    \of[\Big]{
      \bigotimes_{k \in \bar{S}}
      \ketbra{l_{\p{k}}}{l_k}_k
    },
\end{align}
where we eliminated the sum over $i$ in the last equality by noting that
\begin{equation}
  \sum_{i_k \in [d]} \delta_{i_k,l_{\p{k}}} \delta_{l_{k},i_k}
  = \delta_{l_{\p{k}},l_k}
  \label{eq:contraction}
\end{equation}
for all $k \in S$.
Substituting the definition of $\sigma^l_{\p{l}}$ from \cref{eq:sigma} we get
\begin{align}
  \Tr_S(\psi^d_{p,q}(\sigma))
 &= \sum_{l,\p{l} \in [d]^{p+q}}
    \of[\Big]{\prod_{(r,s) \in \sigma} \delta_{l_r,l_s}}
    \of[\Big]{\prod_{k \in S} \delta_{l_k,l_{\p{k}}}}
    \of[\Big]{
      \bigotimes_{k \in \bar{S}}
      \ketbra{l_{\p{k}}}{l_k}_k
    } \\
 &= d^{\loops_S(\sigma)}
    \sum_{l',\p{l}' \in [d]^{\bar{S}}}
    \of[\Big]{\prod_{(u,v) \in \sigma^S} \delta_{l'_u,l'_v}}
    \ketbra{\p{l}'}{l'},
\end{align}
where the second equality is obtained by using a generalization of \cref{eq:contraction} to contract the chains of delta functions along all loops and paths.
This collapses the sum from $[d]^{p+q}$ to $[d]^{\bar{S}}$ and reduces the product to run over edges $(u,v)$ in the remaining smaller diagram $\sigma^S$.
Using \cref{eq:sigma,eq:psi explicit} in reverse,
\begin{align}
  \Tr_S(\psi^d_{p,q}(\sigma))
 &= d^{\loops_S(\sigma)}
    \sum_{l',\p{l}' \in [d]^{\bar{S}}}
    (\sigma^S)^{l'}_{\p{l}'}
    \ketbra{\p{l}'}{l'} \\
 &= d^{\loops_S(\sigma)}
    \psi^d_{p',q'}(\sigma^S) \\
 &= \psi^d_{p',q'}(\Tr_S(\sigma)),
\end{align}
where we used \cref{eq:partial trace} that defines the diagrammatic partial trace $\Tr_S(\sigma)$.
\end{proof}

Finally, note from \cref{eq:psi explicit} that we can also easily evaluate the trace of $\psi^d_{p,q}(\sigma)$ with any elementary rank-$1$ standard basis matrix $\ketbra{i}{j}$ for any $i,j \in [d]^{p+q}$:
\begin{equation}
  \Tr \of[\big]{\psi^d_{p,q}(\sigma) \, \ketbra{i}{j}}
  = \sigma^i_j,
\end{equation}
where $\sigma^i_j$ is given in \cref{eq:sigma}.
The partial trace $\Tr_S\of[\big]{\psi^d_{p,q}(\sigma) \, \ketbra{i}{j}}$ can be evaluated similarly.

\section{Restriction to \texorpdfstring{$\S_p \times \S_q$}{Sp x Sq} permutational symmetry}
\label{apx:restriction to Spq}

In this appendix we describe how the simple module $V^\lambda$ of $\A^d_{p,q}$ restricts to the algebra $\psi^d_{p,q}(\C (\S_p \times \S_q) )$. This question was first answered in \cite{king1970generalized,king1971modification}. It also follows from \cite[Proposition 2.2, Corollary 2.3.1]{koike1989decomposition} and \cite[Theorem 1.7]{halverson1996characters} that a general formula for this restriction is
\begin{equation}
    \Res^{\A^d_{p,q}}_{\S_p \times \S_q} V^{\lambda}
    \cong
    \bigoplus_{\substack{
            \mu \pt p \text{, } \len(\mu) \leq d \\
            \nu \pt q \text{, } \len(\nu) \leq d
        }
    }
    (S^\mu \x S^\nu)^{\oplus m_{\mu,\nu}^\lambda(d)}
    \label{eq:sp_sq multiplicity}
\end{equation}
where $S^\mu$ and $S^\nu$ are simple modules of $\CS_p$ and $\CS_q$, respectively.
For $\lambda = (\lambda^l, \lambda^r) \in \Irr(\A^d_{p,q})$ with $\len(\lambda^l) + \len(\lambda^r) \leq d$ and $\mu \pt p, \nu \pt q$ with $\len(\mu), \len(\nu) \leq d$ the multiplicity $m_{\mu,\nu}^\lambda(d)$ is given by the following formula:
\begin{equation}
    m_{\mu,\nu}^\lambda(d) \defeq
    \sum_{\substack{\tilde{\lambda} : f(\tilde{\lambda},d)=\lambda \\ \len(\tilde{\lambda}^l) \leq d \\ \len(\tilde{\lambda}^r) \leq d}}
    g(\tilde{\lambda},d)
    \sum_{\gamma \pt k(\tilde{\lambda})} c^\mu_{\gamma \tilde{\lambda}^l} c^\nu_{\gamma \tilde{\lambda}^r}
    \label{eq:mult_main}
\end{equation}
where $\tilde{\lambda} = (\tilde{\lambda}^l, \tilde{\lambda}^r)$ is a pair of partitions $\tilde{\lambda}^l \pt p - k(\tilde{\lambda})$ and $\tilde{\lambda}^r \pt q - k(\tilde{\lambda})$ for some integer $k(\tilde{\lambda}) \geq 0$, and $c^\mu_{\gamma \tilde{\lambda}^r}$, $c^\nu_{\gamma \tilde{\lambda}^l}$ are the Littlewood--Richardson coefficients.
The numbers $g(\tilde{\lambda},d) \in \set{-1,0,1}$ and the bipartitions $f(\tilde{\lambda},d)$ are defined as follows.
If $\len(\tilde{\lambda}^l) + \len(\tilde{\lambda}^r) \leq d$ then $f(\tilde{\lambda},d) \defeq \tilde{\lambda}$
and $g(\tilde{\lambda},d) \defeq 1$.
If $\len(\tilde{\lambda}^l) + \len(\tilde{\lambda}^r) > d$ then $f(\tilde{\lambda},d)$ and $g(\tilde{\lambda},d)$ are obtained via the following procedure (here $\mu^\prime$ denotes the transpose of the Young diagram $\mu$):
\begin{itemize}
    \item If $d - \tilde{\lambda}^{r\prime}_i - \tilde{\lambda}^r_1 + i = \tilde{\lambda}^{l\prime}_j - \tilde{\lambda}^r_1 - j + 1$ for some $i \in [\tilde{\lambda}^r_1]$ and $j \in [\tilde{\lambda}^l_1]$, then $g(\tilde{\lambda},d) \defeq 0$ and $f(\tilde{\lambda},d)$ is left undefined.
    \item Otherwise, sort the (distinct) numbers
    \begin{equation}
        \of[\big]{d - \tilde{\lambda}^{r\prime}_i - \tilde{\lambda}^r_1 + i : i = \tilde{\lambda}^r_1, \dotsc, 1}
        \cup
        \of[\big]{\tilde{\lambda}^{l\prime}_j - \tilde{\lambda}^r_1 - j + 1 : j = 1, \dotsc, \tilde{\lambda}^l_1}
    \end{equation}
    in decreasing order and denote the resulting list by $\of{k_1, \dotsc, k_{\tilde{\lambda}^r_1 + \tilde{\lambda}^l_1}}$. Denote the permutation that achieves this sorting by $\pi$ and let $g(\tilde{\lambda},d) \defeq \mathrm{sgn}(\pi)$. The bipartition $f(\tilde{\lambda},d) = \of[\big]{f(\tilde{\lambda},d)^l, f(\tilde{\lambda},d)^r}$ is then defined via its transpose as follows:
    \begin{align}
        f(\tilde{\lambda},d)^{l\prime} &\defeq
        \of[\big]{d - k_i - i + 1 : i = \tilde{\lambda}^r_1, \dotsc, 1}, \\
        f(\tilde{\lambda},d)^{r\prime} &\defeq
        \of[\big]{k_{\tilde{\lambda}^r_1+j} + \tilde{\lambda}^r_1 + j - 1 : j = 1, \dotsc, \tilde{\lambda}^l_1}.
    \end{align}
\end{itemize}
Using a more geometric understanding of this procedure from \cite[eq.~(2.18)]{king1971modification} we arrive at the following multiplicity-free results.

\begin{lemma}\label{lem:multiplicity m}
If $\min(p,q) \leq 2$ then the multiplicity $m_{\mu, \nu}^\lambda(d)$ defined in \cref{eq:mult_main} is either $0$ or $1$ for any valid $\lambda, \mu, \nu, d$.
\end{lemma}

\begin{proof}
Fix a valid combination of $\lambda, \mu, \nu, d$, i.e., $\mu \pt p$, $\nu \pt q$ with $\len(\mu), \len(\nu) \leq d$ and $\lambda = (\lambda^l,\lambda^r)$ with $p - \size(\lambda^l) = q - \size(\lambda^r) \geq 0$ and $\len(\lambda^l) + \len(\lambda^r) \leq d$. Assume that $\tilde{\lambda} = (\tilde{\lambda}^l, \tilde{\lambda}^r)$ is a pair of partitions $\tilde{\lambda}^l \pt p - k(\tilde{\lambda})$ and $\tilde{\lambda}^r \pt q - k(\tilde{\lambda})$ for some integer $k(\tilde{\lambda}) \geq 0$. Without loss of generality it is enough to consider only the $q=1$ and $q=2$ cases.

\textit{Case $q = 1$}.
Assume that $\len(\tilde{\lambda}^l) \leq d$ and $\len(\tilde{\lambda}^r) = 1$. Then according to \cite{king1971modification} we have that
\begin{equation}
    g(\tilde{\lambda},d) =
    \begin{cases}
        1 & \text{if } \len(\tilde{\lambda}^l) + \len(\tilde{\lambda}^r) \leq d, \\
        0 & \text{if }  \len(\tilde{\lambda}^l) + \len(\tilde{\lambda}^r) = d + 1.
    \end{cases}
\end{equation}
The set $\{\tilde{\lambda} : f(\tilde{\lambda},d)=\lambda \}$ contains at most two elements corresponding to $\len(\tilde{\lambda}^l)+ \len(\tilde{\lambda}^r) \leq d$ and $\len(\tilde{\lambda}^l) + \len(\tilde{\lambda}^r) = d + 1$.
Moreover, there is only one term in the sum
$\sum_{\gamma \pt k(\tilde{\lambda})} c^\mu_{\gamma \tilde{\lambda}^l} c^\nu_{\gamma \tilde{\lambda}^r}$
in \cref{eq:mult_main} corresponding to either the empty partition $\gamma = \0$ or $\gamma = (1)$.
Each of these Littlewood--Richardson coefficients is either zero or one due to the Pieri rule \cite{macdonald1998symmetric}, which is a special case of the Littlewood--Richardson rule \cite{berenshtein1988involutions,remmel1998simple,gasharov1998short,stembridge2002concise}. Therefore $m_{\mu,\nu}^\lambda(d) \in \set{0,1}$.

\textit{Case $q = 2$}.
Assume that $\len(\tilde{\lambda}^l) \leq d$ and $\len(\tilde{\lambda}^r) \leq 2$. Analogously, from \cite{king1971modification} it follows that
\begin{equation}
    g(\tilde{\lambda},d) =
    \begin{cases}
        1 & \text{if } \len(\tilde{\lambda}^l) + \len(\tilde{\lambda}^r) \leq d, \\
        -1 & \text{if } \len(\tilde{\lambda}^l) = d \text{ and } \len(\tilde{\lambda}^r) = 2 \text{ and } \tilde{\lambda}^{l\prime}_1 > \tilde{\lambda}^{l\prime}_2, \\
        0 & \text{otherwise}.
    \end{cases}
\end{equation}
There are three possible cases depending on the value of $k(\tilde{\lambda}) \in \set{0,1,2}$:
\begin{itemize}
    \item If $k(\tilde{\lambda}) = 0$ then the only term in the sum $\sum_{ \gamma \pt k(\tilde{\lambda})} c^\mu_{\gamma \tilde{\lambda}^l} c^\nu_{\gamma \tilde{\lambda}^r}$ corresponds to $\gamma = \0$ and each Littlewood--Richardson coefficient is either zero or one, which implies that $ \sum_{ \gamma \pt k(\tilde{\lambda})} c^\mu_{\gamma \tilde{\lambda}^l} c^\nu_{\gamma \tilde{\lambda}^r} \in \set{0,1}$.
    \item If $k(\tilde{\lambda}) = 1$ then $\tilde{\lambda}^r = (1)$ and the only term in the sum $\sum_{ \gamma \pt k(\tilde{\lambda})} c^\mu_{\gamma \tilde{\lambda}^l} c^\nu_{\gamma \tilde{\lambda}^r}$ corresponds to $\gamma = (1)$, which means that $\sum_{ \gamma \pt k(\tilde{\lambda})} c^\mu_{\gamma \tilde{\lambda}^l} c^\nu_{\gamma \tilde{\lambda}^r} \in \set{0,1}$.
    \item If $k(\tilde{\lambda}) = 2$ then $\tilde{\lambda}^r = \0$ and therefore
    \begin{equation}
        \sum_{ \gamma \pt k(\tilde{\lambda})} c^\mu_{\gamma \tilde{\lambda}^l} c^\nu_{\gamma \tilde{\lambda}^r} = \sum_{ \gamma \pt k(\tilde{\lambda})} c^\mu_{\gamma \tilde{\lambda}^l} \delta_{\nu,\gamma} = c^\mu_{\nu \tilde{\lambda}^l} \in \set{0,1},
    \end{equation}
    where the conclusion follows from $\nu \pt 2$ and the Pieri rule.
\end{itemize}

Moreover, for any valid $\lambda$ there is either exactly one $\tilde{\lambda}$ for which $f(\tilde{\lambda},d) = \lambda$ (namely $\tilde{\lambda} = \lambda$) or exactly two different $\tilde{\lambda}_1, \tilde{\lambda}_2$ with $g(\tilde{\lambda}_1,d) = -1$ and $g(\tilde{\lambda}_2,d) = 1$. Therefore $m_{\mu,\nu}^\lambda(d) \in \set{0, 1}$.
\end{proof}

\begin{lemma}\label{lem:multiplicity m d=2}
If $d = 2$ then the multiplicity $m_{\mu, \nu}^\lambda(d)$ defined in \cref{eq:mult_main} is either $0$ or $1$ for any valid $\lambda, \mu, \nu$.
\end{lemma}

\begin{proof}
Fix a valid combination of $\lambda, \mu, \nu$, i.e., $\mu \pt p$, $ \nu \pt q$ with $\len(\mu), \len(\nu) \leq 2$ and $\lambda = (\lambda^l,\lambda^r)$ with $k \defeq p - \size(\lambda^l) = q - \size(\lambda^r) \geq 0$ and $\len(\lambda^l) + \len(\lambda^r) \leq d$. Assume that $\tilde{\lambda} = (\tilde{\lambda}^l, \tilde{\lambda}^r)$ is a pair of partitions $\tilde{\lambda}^l \pt p - k(\tilde{\lambda})$ and $\tilde{\lambda}^r \pt q - k(\tilde{\lambda})$ for some integer $k(\tilde{\lambda}) \geq 0$ with $\len(\tilde{\lambda}^l) \leq 2$ and $\len(\tilde{\lambda}^r) \leq 2$. Again, we use a result of \cite{king1971modification} stating that for $d=2$,
\begin{equation}
    \label{eq:Spq d=2 g}
    g(\tilde{\lambda},2) =
    \begin{cases}
        1 & \text{if } \len(\tilde{\lambda}^l) + \len(\tilde{\lambda}^r) \leq 2, \\
        -1 & \text{if } \len(\tilde{\lambda}^l) = \len(\tilde{\lambda}^r) = 2 \text{ and }  \tilde{\lambda}^{l\prime}_1 > \tilde{\lambda}^{l\prime}_2 \text{ and } \tilde{\lambda}^{r\prime}_1 > \tilde{\lambda}^{r\prime}_2, \\
        0 & \text{otherwise}.
    \end{cases}
\end{equation}
We need to consider only two different cases.

\textit{Case $k = \min(p,q)$}.
Without loss of generality assume $k=q$, i.e., $\lambda^r = \0$. Then
according to \cref{eq:Spq d=2 g} the first sum in \cref{eq:mult_main} contains only one term corresponding to $\tilde{\lambda} = \lambda$:
\begin{equation}
    m_{\mu,\nu}^\lambda(2) =
    \sum_{\gamma \pt k} c^\mu_{\gamma \lambda^l} c^\nu_{\gamma \0} = \sum_{\gamma \pt k} c^\mu_{\gamma \lambda^l} \delta_{\nu,\gamma} = c^\mu_{\nu \lambda^l}.
\end{equation}
Since $\len(\mu),\len(\nu),\len(\lambda^l) \leq 2$, it follows from the Littlewood--Richardson rule that $m_{\mu,\nu}^\lambda(2) = c^\mu_{\nu \lambda^l} \in \{0,1\}$.

\textit{Case $k < \min(p,q)$}. In this case $\lambda^l, \lambda^r$ are non-empty Young diagrams with only one row, i.e., $\len(\lambda^l) = \len(\lambda^r) = 1$. According to \cref{eq:Spq d=2 g} there are now two $\tilde{\lambda}$ such that $f(\tilde{\lambda},2) = \lambda$ and $g(\tilde{\lambda},2) \neq 0$, namely $\tilde{\lambda} = \lambda$ which corresponds to $g(\tilde{\lambda},2) = 1$ and $\tilde{\lambda} = (\tilde{\lambda}^l,\tilde{\lambda}^r) =  \of{(\lambda^l_1,1), (\lambda^r_1,1)}$ which corresponds to $g(\tilde{\lambda},2) = -1$. Therefore
\begin{equation}
    m_{\mu,\nu}^\lambda(2) = \sum_{\gamma \pt k} c^\mu_{\gamma \lambda^l} c^\nu_{\gamma \lambda^r} - \sum_{\gamma \pt k-1} c^\mu_{\gamma (\lambda^l_1,1)} c^\nu_{\gamma (\lambda^r_1,1)}.
    \label{eq:Spq d=2 m intermediate}
\end{equation}
Since $\len(\lambda^l) = \len(\lambda^r) = 1$ and $\len(\mu),\len(\nu) \leq 2$ we can use the Pieri rule to deduce $c^\mu_{\gamma \lambda^l}, c^\nu_{\gamma \lambda^r} \in \set{0,1}$ and calculate
\begin{align}
    c^\mu_{\gamma \lambda^l} \neq 0 &\text{ iff } \gamma \pt k \text{ and } \len(\gamma) \leq 2 \text{ and } \mu_2 \leq \gamma_1 \leq \mu_1 \text{ and } \gamma_2 \leq \mu_2, \nonumber  \\
    c^\nu_{\gamma \lambda^r} \neq 0 &\text{ iff } \gamma \pt k \text{ and } \len(\gamma) \leq 2 \text{ and } \nu_2 \leq \gamma_1 \leq \nu_1 \text{ and } \gamma_2 \leq \nu_2. \nonumber
\end{align}
Therefore the first sum in \cref{eq:Spq d=2 m intermediate} becomes
\begin{align}
    &\abs*{\set*{ (\gamma_1,\gamma_2) \pt k : \max(\mu_2,\nu_2) \leq \gamma_1 \leq \min(\mu_1,\nu_1), \gamma_2 \leq \min(\mu_2,\nu_2)}} \\
    &=\abs*{\set*{\gamma_2 \in \mathbb{Z} : \max \of*{0,k-\mu_1,k-\nu_1} \leq \gamma_2 \leq \min \of*{\floor*{\frac{k}{2}},k-\mu_2,k-\nu_2,\mu_2,\nu_2} }}. \nonumber
\end{align}
It can be rewritten as
\begin{equation}
\begin{aligned}[t]
    \label{eq:lem m d=2 sum1}
    \sum_{\gamma \pt k} c^\mu_{\gamma \lambda^l} c^\nu_{\gamma \lambda^r} =
    \max \Bigg( &\min \of*{\floor*{\frac{k}{2}},k-\mu_2,k-\nu_2,\mu_2,\nu_2} \\
    &
    - \max \of*{0,k-\mu_1,k-\nu_1} + 1,0 \Bigg).
\end{aligned}
\end{equation}
Similarly, the Littlewood--Richardson rule implies that $c^\mu_{\gamma (\lambda^l_1,1)}, c^\nu_{\gamma(\lambda^r_1,1)} \in \set{0,1}$. Furthermore,
\begin{align}
    c^\mu_{\gamma (\lambda^l_1,1)} \neq 0 &\text{ iff } \gamma \pt k-1 \text{ and } \len(\gamma) \leq 2 \text{ and } \mu_2 \leq \gamma_1 + 1 \leq \mu_1 \text{ and } \gamma_2 \leq \mu_2 - 1, \nonumber \\
    c^\nu_{\gamma (\lambda^r_1,1)} \neq 0 &\text{ iff } \gamma \pt k-1 \text{ and } \len(\gamma) \leq 2 \text{ and } \nu_2 \leq \gamma_1 + 1 \leq \nu_1 \text{ and } \gamma_2 \leq \nu_2 - 1. \nonumber
\end{align}
The second sum in \cref{eq:Spq d=2 m intermediate} now becomes
\begin{align}
    &\abs*{\set*{ (\gamma_1,\gamma_2) \pt k-1 : \max(\mu_2,\nu_2) \leq \gamma_1 + 1 \leq \min(\mu_1,\nu_1), \gamma_2 \leq \min(\mu_2,\nu_2) - 1 }} \nonumber \\
    &=\abs*{\set*{\gamma_2 \in \mathbb{Z} : \max \of*{0,k-\mu_1,k-\nu_1} \leq \gamma_2
    \leq \min \of*{\floor*{\frac{k - 1}{2}},k-\mu_2,k-\nu_2,\mu_2 - 1,\nu_2 - 1} }}.
\end{align}
It can be rewritten as
\begin{align}
    \label{eq:lem m d=2 sum2}
    \begin{aligned}[t]
    \sum_{\gamma \pt k-1} c^\mu_{\gamma (\lambda^l_1,1)} c^\nu_{\gamma (\lambda^r_1,1)} =
    \max \Bigg( &\min \of*{\floor*{\frac{k-1}{2}},k-\mu_2,k-\nu_2,\mu_2-1,\nu_2-1} \\
    &- \max \of*{0,k-\mu_1,k-\nu_1} + 1,0 \Bigg).
    \end{aligned}
\end{align}
From \cref{eq:Spq d=2 m intermediate,eq:lem m d=2 sum1,eq:lem m d=2 sum2} we clearly see that $m_{\mu,\nu}^\lambda(2) \in \set{0,1}$.
\end{proof}

\section{Proof of \texorpdfstring{\cref{lem:main_lem}}{the main lemma}}\label{apx:proof of main lemma}

\technical*

\begin{proof}
We will use the correspondence between a path $\T$ and a pair $(\tau,L)$ of a tableaux $\tau = (\tau^l, \tau^r)$ of shape $(\lambda^l, \lambda^r)$ and a tuple $L$ of pairs of numbers from the set $[p+q]$, see Theorem~1.11 of \cite{bchlls}. It follows from \cref{lem:additively central} that $J^\A_1 + \dotsb + J^\A_{p+q} \in \Z(\A_{p+q})$. Therefore it is enough to consider how this sum acts on any vector in the isotypic component $V^\lambda \x U^\lambda$ corresponding to the simple module labeled by $\lambda = (\lambda^l, \lambda^r) \in \Irr(\A^d_{p,q})$ in the mixed Schur--Weyl duality, see \cref{eq:schur_weyl_alt}. In particular, we can take a maximal vector
\begin{equation}
  \ket{t_{\tau, L}} \defeq \psi(y_\tau \bar{\sigma}_L) \ket{\beta_{\tau,L}} \in V^{p,q}
\end{equation}
from \cite[Definition~2.4]{bchlls}, where $y_\tau \in \C \of{ \S_p \times \S_q} \subseteq \B^d_{p,q}$ is the \emph{Young symmetrizer} for the tableaux $\tau$, $\bar{\sigma}_L \in \B^d_{p,q}$ is a diagram that contracts all pairs in $L$, and $\psi \defeq \psi^d_{p,q}$ is defined in \cref{eq:psi}. The Young symmetrizer $y_\tau$ is defined as $y_\tau \defeq y_{\tau^l} y_{\tau^r}$, the product of standard Young symmetrizers $y_{\tau^l}$ and $y_{\tau^r}$, see \cite[Eq.~2.2]{bchlls}. The Young symmetrizers $y_{\tau^l}$ and $y_{\tau^r}$ are products of \emph{column} and \emph{row symmetrizers} (the terms which correspond to \emph{column} and \emph{row groups} in \cite[Eq.~2.2]{bchlls}). The standard basis vector $\ket{\beta_{\tau,L}} \in V^{p,q}$ is defined as follows:
\begin{equation}
  \ket{\beta_{\tau,L}} \defeq \ket{u_{1}} \otimes \dotsb \otimes \ket{u_p} \otimes \ket{u_{p+1 }} \otimes \dotsb \otimes \ket{u_{p+q}}
\end{equation}
where each $u_i \in [d]$ and we distinguish two cases:
if $1 \leq i \leq p$ then
\begin{equation}
    u_{i} \defeq
        \begin{cases}
            j &\text{if $i$ belongs to the $j$-th row of $\tau^l$}, \\
            1 &\text{if $i \in L$},
        \end{cases}
\end{equation}
while if $p+1 \leq i \leq p+q$ then
\begin{equation}
    u_{i} \defeq
        \begin{cases}
            d-j+1 & \text{if $i$ belongs to the $j$-th row of $\tau^r$}, \\
            1 & \text{if $i \in L$}.
        \end{cases}
\end{equation}

Recall that $J^{\A}_{1}+\cdots+J^{\A}_{p+q} \in \Z(\A_{p+q})$, so
\begin{equation}
  \of*{J^{\A}_{1}+\cdots+J^{\A}_{p+q}} \ket{t_{\tau, L}}
  = \psi(y_\tau \bar{\sigma}_L)
    \of*{J^{\A}_{1}+\cdots+J^{\A}_{p+q}} \ket{\beta_{\tau,L}}.
\end{equation}
Since $J^{\A}_k \defeq \psi(J^{\B}_k)$ and $\psi$ is a homomorphism, we can use the definition of $J^{\B}_k$ from \cref{eq:JM for B} to write the right-hand side more explicitly:
\begin{align}
  \psi\of*{
    y_\tau \bar{\sigma}_L
    \of*{
      \sum_{\substack{ 1 \leq i < j \leq p \text{ or}\\ p+1 \leq i < j \leq p+q}} \sigma_{i,j} - \sum_{\mathclap{\substack{1 \leq i \leq p \\ p+1 \leq j \leq p+q}}}\bar{\sigma}_{i,j} + d \cdot q
    }
  }
  \ket{\beta_{\tau,L}}.
  \label{lem1:action_of_sum}
\end{align}
To simplify this, we need to evaluate all expressions of the form
$\psi(y_\tau \bar{\sigma}_L \sigma_{i,j}) \ket{\beta_{\tau,L}}$ and
$\psi(y_\tau \bar{\sigma}_L \bar{\sigma}_{i,j}) \ket{\beta_{\tau,L}}$,
where we distinguish between transpositions $\sigma_{i,j}$ and contractions $\bar{\sigma}_{i,j}$
which can be located in three possible positions relative to $L$:
\begin{enumerate}
  \item $i \notin L$, $j \notin L$,
  \item $i \in L$, $j \notin L$ or $i \notin L$, $j \in L$,
  \item $i \in L$, $j \in L$.
\end{enumerate}
We now proceed to consider each of these six cases separately (we will write~\emph{(1)} and~\emph{(\=1)} to distinguish between the cases with $\sigma_{i,j}$ and $\bar{\sigma}_{i,j}$).

\newcommand{\case}[2]{\vspace{5pt}\noindent\emph{Case (#1): #2.}}

\case{1}{$\sigma_{i,j}$ with $i \notin L$, $j \notin L$}
There are two sub-cases:
\begin{enumerate}[(a)]
    \item If $i$ and $j$ are in the same row of either $\tau^l$ or $\tau^r$, the row symmetrizer of $y_\tau$ does not change the resulting vector, i.e., $\psi(y_\tau \sigma_{i,j}) \ket{\beta_{\tau,L}} = \psi(y_\tau) \ket{\beta_{\tau,L}}$. Since $\bar{\sigma}_L$ and $y_\tau$ commute,
    \begin{equation}
        \psi(y_\tau \bar{\sigma}_L  \sigma_{i,j}) \ket{\beta_{\tau,L}} = \ket{t_{\tau, L}}.
    \end{equation}
    \item
    If $i$ and $j$ are in different rows of either $\tau^l$ or $\tau^r$, we denote the corresponding row numbers by $r_i$ and $r_j$. In this case the row symmetrizer of $y_\tau$ acting on $\psi(\sigma_{i,j}) \ket{\beta_{\tau,L}}$ produces a product of two ``$W$ states'' at positions defined by the rows $r_i$ and $r_j$ of $\tau^{l/r}$ with the numbers $i$ and $j$ swapped. The column antisymmetrizer of $y_\tau$ will then kill most terms, leaving only the terms with basis vectors $\ket{r_i}, \ket{r_j}$ in positions within the same column. There are $\lambda^{l/r}_{\max\{r_i,r_j\}}$ such terms, where $\lambda^{l/r}_{\max\{r_i,r_j\}}$ is the size of the smallest of the two rows $r_i,r_j$ within the corresponding left or right tableaux $\tau^{l/r}$. After this operation the resulting vector acquires a minus sign and a different factor compared to $\psi(y_\tau) \ket{\beta_{\tau,L}}$:
    \begin{equation}
        \psi(y_\tau \sigma_{i,j}) \ket{\beta_{\tau,L}}
        = - \frac{\psi(y_\tau) \ket{\beta_{\tau,L}}}{\lambda^{l/r}_{\max\{r_i,r_j\}}}.
    \end{equation}
    Again, since $\bar{\sigma}_L$ and $y_\tau$ commute,
    \begin{equation}
        \psi(y_\tau \bar{\sigma}_L \sigma_{i,j}) \ket{\beta_{\tau,L}}
        = - \frac{1}{\lambda^{l/r}_{\max\{r_i,r_j\}}}\ket{t_{\tau, L}}.
    \end{equation}
\end{enumerate}

\case{\=1}{$\bar{\sigma}_{i,j}$ with $i \notin L$, $j \notin L$}
In this case $\psi(\bar{\sigma}_{i,j}) \ket{\beta_{\tau,L}}$ can only be non-zero when $u_i = u_j$, which is equivalent to $r_i = d - r_j + 1$. Thus $r_i + r_j = d + 1$. But since $r_i + r_j \leq \len(\lambda^l) + \len(\lambda^r) \leq d$,
\begin{equation}
    \psi(y_\tau \bar{\sigma}_L \bar{\sigma}_{i,j}) \ket{\beta_{\tau,L}} = 0.
\end{equation}

\case{2}{$\sigma_{i,j}$ with $i \in L$, $j \notin L$ or $i \notin L$, $j \in L$}
Without loss of generality we can assume that $i \in L$ and $j \notin L$. There are two sub-cases:
\begin{enumerate}[(a)]
    \item If $u_j = 1$ then $\psi(\bar{\sigma}_L \sigma_{i,j}) \ket{\beta_{\tau,L}} = \psi(\bar{\sigma}_L) \ket{\beta_{\tau,L}}$. This happens when $j \in \tau^l$ and $r_j = 1$ or $j \in \tau^r$ and $r_j = d$. Therefore
    \begin{equation}
        \psi(y_\tau \bar{\sigma}_L \sigma_{i,j}) \ket{\beta_{\tau,L}} = \ket{t_{\tau, L}}.
    \end{equation}
    \item If $u_j \neq 1$ then $\psi(\bar{\sigma}_L \sigma_{i,j}) \ket{\beta_{\tau,L}} = 0$ because $\psi(\bar{\sigma}_L)$ would annihilate the vector $\psi(\sigma_{i,j}) \ket{\beta_{\tau,L}}$. Therefore
    \begin{equation}
        \psi(y_\tau \bar{\sigma}_L \sigma_{i,j}) \ket{\beta_{\tau,L}} = 0.
    \end{equation}
\end{enumerate}

\case{\=2}{$\bar{\sigma}_{i,j}$ with $i \in L$, $j \notin L$ or $i \notin L$, $j \in L$}
Assume again that $i \in L$ and $j \notin L$. There are two sub-cases:
\begin{enumerate}[(a)]
    \item If $u_j = 1$ then
    \begin{equation}
        \psi(\bar{\sigma}_L \bar{\sigma}_{i,j}) \ket{\beta_{\tau,L}}
        = \psi(\bar{\sigma}_L) \of*{\ket{\beta_{\tau,L}} + \sum_{k=2}^{d} \of[\big]{\dotsb \x \ket{k} \x \dotsb \x \ket{k} \x \dotsb}},
    \end{equation}
    where the basis vectors labeled by $k$ are in positions $i$ and $j$. But since $j \notin L$ all vectors $\dotsb \otimes \ket{k} \otimes \dotsb \otimes \ket{k} \otimes \dotsb$ for $k \geq 2$ are annihilated by $\bar{\sigma}_L$.
    Therefore in this case
    \begin{equation}
        \psi(y_\tau \bar{\sigma}_L \bar{\sigma}_{i,j}) \ket{\beta_{\tau,L}} = \ket{t_{\tau, L}}.
    \end{equation}
    \item If $u_j \neq 1$ then $\psi(\bar{\sigma}_{i,j}) \ket{\beta_{\tau,L}} = 0$ and therefore
    \begin{equation}
        \psi(y_\tau \bar{\sigma}_L \bar{\sigma}_{i,j}) \ket{\beta_{\tau,L}} = 0.
    \end{equation}
\end{enumerate}

\case{3}{$\sigma_{i,j}$ with $i \in L$, $j \in L$}
Since $i$ and $j$ must belong together either to $\tau^l$ or $\tau^r$, they cannot belong simultaneously to one pair of $L$. Since $u_i = u_j = 1$, it follows that $\psi(\sigma_{i,j}) \ket{\beta_{\tau,L}} = \ket{\beta_{\tau,L}}$ and therefore
\begin{equation}
    \psi(y_\tau \bar{\sigma}_L \sigma_{i,j}) \ket{\beta_{\tau,L}} = \ket{t_{\tau, L}}.
\end{equation}

\case{\=3}{$\bar{\sigma}_{i,j}$ with $i \in L$, $j \in L$}
In contrast to the previous case, two sub-cases can occur:
\begin{enumerate}[(a)]
    \item If $i$ and $j$ belong to the same pair in $L$ then
    \begin{equation}
        \psi(\bar{\sigma}_L \bar{\sigma}_{i,j}) \ket{\beta_{\tau,L}}
        = \psi(\bar{\sigma}_L) \of*{\sum_{k=1}^{d} \of[\big]{\dotsb \x \ket{k} \x \dotsb \x \ket{k} \x \dotsb}}
        = d \cdot \psi(\bar{\sigma}_L) \ket{\beta_{\tau,L}},
    \end{equation}
    where the basis vectors labeled by $k$ are in positions $i$ and $j$. Therefore
    \begin{equation}
        \psi(y_\tau \bar{\sigma}_L \bar{\sigma}_{i,j}) \ket{\beta_{\tau,L}}
        = d \cdot \ket{t_{\tau, L}}.
    \end{equation}
    \item If $i$ and $j$ belong to different pairs in $L$ then similarly to the previous case we can write
    \begin{equation}
        \psi(\bar{\sigma}_L \bar{\sigma}_{i,j}) \ket{\beta_{\tau,L}}
        = \psi(\bar{\sigma}_L) \of*{\sum_{k=1}^{d} \of[\big]{\dotsb \x \ket{k} \x \dotsb \x \ket{k} \x \dotsb}},
    \end{equation}
    where the basis vectors labeled by $k$ are in positions $i$ and $j$. But now since the positions $i$ and $j$ are not contracted with each other by $\psi(\bar{\sigma}_L)$, we will not acquire a factor of $d$:
    \begin{equation}
        \psi(y_\tau \bar{\sigma}_L \bar{\sigma}_{i,j}) \ket{\beta_{\tau,L}} = \ket{t_{\tau, L}}.
    \end{equation}
\end{enumerate}

Collecting everything together and separating the sums in \cref{lem1:action_of_sum} according to the cases above, we arrive at the following expression:
\begin{align}
    &\left( J^{\A}_{1}+\cdots+J^{\A}_{p+q} \right) \ket{t_{\tau, L}} = \psi  \bigg( y_\tau \bar{\sigma}_L \bigg( \sum_{\substack{ 1 \leq i < j \leq p \text{ or}\\ p+1 \leq i < j \leq p+q}} \sigma_{i,j} - \sum_{\mathclap{\substack{1 \leq i \leq p \\ p+1 \leq j \leq p+q}}}\bar{\sigma}_{i,j} + d \cdot q \bigg) \bigg) \ket{\beta_{\tau,L}} \nonumber \\
    &= \sum_{\substack{1 \leq i < j \leq p \text{ or}\\ p+1 \leq i < j \leq p+q \\ \text{case (1a)}}} \ket{t_{\tau,L}} +
    \sum_{\substack{ 1 \leq i < j \leq p \text{ or}\\ p+1 \leq i < j \leq p+q \\ \text{case (2a)}}} \ket{t_{\tau,L}} +
    \sum_{\substack{1 \leq i < j \leq p \text{ or}\\ p+1 \leq i < j \leq p+q \\ \text{case (3)}}} \ket{t_{\tau,L}} +
    d \cdot q \\
    &-\sum_{\substack{ 1 \leq i < j \leq p \text{ or}\\ p+1 \leq i < j \leq p+q\\ \text{case (1b)}}}  \frac{\ket{t_{\tau,L}}}{\lambda^{l/r}_{\max\{r_i,r_j\}}} -
    \sum_{\substack{1 \leq i \leq p \\ p+1 \leq j \leq p+q \\ \text{case (\=2a)}}} \ket{t_{\tau,L}} -
    \sum_{\substack{ 1 \leq i \leq p \\ p+1 \leq j \leq p+q \\ \text{case (\=3a)}}} d \cdot \ket{t_{\tau,L}} -
    \sum_{\substack{ 1 \leq i \leq p \\ p+1 \leq j \leq p+q \\ \text{case (\=3b)}}} \ket{t_{\tau,L}}. \nonumber
\end{align}
To simplify this we need to do some counting. Counting all possible pairs within each row of $\tau$ gives us
\begin{equation}
    \sum_{\substack{1 \leq i < j \leq p \text{ or}\\ p+1 \leq i < j \leq p+q \\ \text{case (1a)}}} \ket{t_{\tau,L}} = \left( \sum_{i=1}^{\len(\lambda^l)} \binom{\lambda^l_i}{2} + \sum_{i=1}^{\len(\lambda^r)} \binom{\lambda^r_i}{2} \right) \ket{t_{\tau,L}}.
\end{equation}
Pairing $i$ and $j$ across different rows gives us:
\begin{equation}
    \sum_{\substack{ 1 \leq i < j \leq p \text{ or}\\ p+1 \leq i < j \leq p+q\\ \text{case (1b)}}}  \frac{\ket{t_{\tau,L}}}{\lambda^{l/r}_{\max\{r_i,r_j\}}} = \left( \sum_{i=1}^{\len(\lambda^l)} \lambda^l_i (i-1) + \sum_{i=1}^{\len(\lambda^r)} \lambda^r_i (i-1) \right) \ket{t_{\tau,L}}.
\end{equation}
Note that for a single diagram $\lambda$ it is true that
\begin{equation}
     \sum_{i=1}^{\len(\lambda)} \binom{\lambda_i}{2} - \sum_{i=1}^{\len(\lambda)} \lambda_i (i-1) = \cont(\lambda),
\end{equation}
therefore
\begin{equation}
 \sum_{\substack{1 \leq i < j \leq p \text{ or}\\ p+1 \leq i < j \leq p+q \\ \text{case (1a)}}} \ket{t_{\tau,L}} - \sum_{\substack{ 1 \leq i < j \leq p \text{ or}\\ p+1 \leq i < j \leq p+q\\ \text{case (1b)}}}  \frac{\ket{t_{\tau,L}}}{\lambda^{l/r}_{\max\{r_i,r_j\}}} = \left( \cont(\lambda^l) + \cont(\lambda^r) \right) \ket{t_{\tau,L}}.
\end{equation}
Next, simple combinatorics gives us
\begin{align}
    &\sum_{\substack{ 1 \leq i < j \leq p \text{ or}\\ p+1 \leq i < j \leq p+q \\ \text{case (2a)}}} \ket{t_{\tau,L}} = \sum_{\substack{1 \leq i \leq p \\ p+1 \leq j \leq p+q \\ \text{case (\=2a)}}} \ket{t_{\tau,L}} = \left( \lambda_1^l \cdot \abs{L} +  \lambda_d^r \cdot \abs{L} \right) \ket{t_{\tau,L}}, \\
    &\sum_{\substack{1 \leq i < j \leq p \text{ or}\\ p+1 \leq i < j \leq p+q \\ \text{case (3)}}} \ket{t_{\tau,L}} = \sum_{\substack{ 1 \leq i \leq p \\ p+1 \leq j \leq p+q \\ \text{case (\=3b)}}} \ket{t_{\tau,L}} = 2 \cdot \binom{\abs{L}}{2} \ket{t_{\tau,L}},
\end{align}
so the corresponding sums cancel each other. Finally,
\begin{align}
    &\sum_{\substack{ 1 \leq i \leq p \\ p+1 \leq j \leq p+q \\ \text{case (\=3a)}}} d \cdot \ket{t_{\tau,L}} = d \cdot \abs{L} \cdot \ket{t_{\tau,L}}.
\end{align}

Using $q-\abs{L} = \size(\lambda^r)$ and combining everything together gives us the desired result:
\begin{equation}
    \left( J^{\A}_{1}+\cdots+J^{\A}_{p+q} \right) \ket{t_{\tau, L}} = \left( \cont(\lambda^l) + \cont(\lambda^r) + d \cdot \size(\lambda^r) \right) \ket{t_{\tau,L}}.
\end{equation}
Since $J^{\A}_{1}+\cdots+J^{\A}_{p+q} \in \Z(\A_{p+q})$ and $\varepsilon^\A_\T \ket{t_{\tau,L}} = \ket{t_{\tau,L}}$, we can draw the same conclusion for $\varepsilon^\A_\T$:
\begin{equation}
    \left( J^{\A}_{1}+\cdots+J^{\A}_{p+q} \right) \varepsilon^\A_\T = \left( \cont(\lambda^l) + \cont(\lambda^r) + d \cdot \size(\lambda^r) \right) \varepsilon^\A_\T
\end{equation}
which completes the proof.
\end{proof}

\section{Computing the blocks of \texorpdfstring{$\A^d_{p,q}$}{Apq} in Gelfand--Tsetlin basis}
\label{apx:blocks of Apq}

Here we propose an algorithm for computing the blocks of an arbitrary $\A^d_{p,q}$ algebra element in the Gelfand--Tsetlin basis.
In line with our philosophy, the algorithm is fully diagrammatic, namely, all computation takes place in $\B^d_{p,q}$ instead of $\A^d_{p,q}$.
Here we only sketch the reasoning behind the algorithm, and we leave it to future work to establish its correctness formally.
This paves one possible route for removing the additional symmetry assumption in \cref{thm:main} and thus extending our framework from linear to general semidefinite unitary-equivariant programs.

The natural $*$-algebra structure of $\B^d_{p,q}$ is the important ingredient in this section. Consider a random hermitian (with respect to the natural $*$-algebra structure) element of $\B^d_{p,q}$ given as a linear combination of diagrams with random real coefficients $b_{i}$:
\begin{equation}
  B \defeq \sum_{i=1}^{(p+q)!} b_{i} \sigma_i.
  \label{eq:symbolic B}
\end{equation}
Throughout this section we fix $\lambda \in \Irr(\A^d_{p,q})$ and choose an arbitrary ordering $i \in [d_\lambda]$ of all paths in $\Paths(\lambda)$.
Let $B_{ij}$ denote the $(i,j)$-th entry of block $\lambda$ when the matrix $\psi^d_{p,q}(B) \in \A^d_{p,q}$ is expressed in the Gelfand--Tsetlin basis.
That is, let $B_{ij} \defeq (A_\lambda)_{ij}$ for all $i,j \in [d_\lambda]$, where $A_\lambda$ is a matrix of size $d_\lambda \times d_\lambda$ that appears in the first register of the decomposition \eqref{rmk:eq_blocks}:
\begin{equation}
  U_{\textnormal{Sch}(p,q)} \, \psi^d_{p,q}(B) \, U_{\textnormal{Sch}(p,q)}\ct
  = \bigoplus_{\lambda \in \Irr(\A^d_{p,q})}
    \sof*{
      A_\lambda \x I_{m_\lambda}
    }.
\end{equation}
Note from \cref{eq:symbolic B} that $B_{ij}$ is a linear combination of the variables $b_k$.
Our goal is to determine $B_{ij}$ for all choices of $\lambda  \in \Irr(\A^d_{p,q})$ and $i,j \in [d_\lambda]$.

Let $\T = \lambda_0 \to \dots \to \lambda_{p+q}$ be the $i$-th path in the Bratteli diagram of $\B^d_{p,q}$ that goes from the root to the leaf $\lambda$.
Let us denote the preimage of $\varepsilon^\A_{\T}$ under $\psi^d_{p,q}$ by
\begin{equation}
  \varepsilon_i \defeq \prod_{k=1}^{p+q}
  \prod_{\mu : \lambda_{k-1} \to \mu \neq \lambda_{k}}
  \frac{J_k^\B - c_{\lambda_{k-1} \to \mu}}{c_{\lambda_{k-1} \to \lambda_{k}}-c_{\lambda_{k-1} \to \mu}}
  \in \B^d_{p,q},
\end{equation}
where, in contrast to \cref{eq:primitive_main_A_d}, the second product runs over edges in the Bratteli diagram of the family $\B$ instead of $\A$.
By construction, the block $\lambda$ of $\varepsilon_i$ is equal to $\proj{i}$ while all other blocks vanish:
\begin{equation}
  U_{\textnormal{Sch}(p,q)} \, \psi^d_{p,q}(\varepsilon_i) \, U_{\textnormal{Sch}(p,q)}\ct
  = \bigoplus_{\mu \in \Irr(\A^d_{p,q})}
    \sof*{
      \delta_{\lambda,\mu} \proj{i} \x I_{m_\mu}
    }.
  \label{eq:ii}
\end{equation}
Since $\psi(\varepsilon_i B \varepsilon_i) = B_{ii} \cdot \psi(\varepsilon_i)$,
knowing $\varepsilon_i$ allows us to diagrammatically extract $B_{ii}$ by computing
\begin{equation}
    B_{ii} = \frac{\Tr \of{B \varepsilon_i}}{\Tr \of{\varepsilon_i}},
\end{equation}
where $\Tr \of{\varepsilon_i} = m_\lambda$ for every $i \in [d_\lambda]$, see \cref{rmk:eq_blocks}.

To extract the off-diagonal entries $B_{ij}$, we would like to have operators $\varepsilon_{ij}$ that are analogous to $\varepsilon_i$ but instead of $\proj{i}$ have $\ketbra{i}{j}$, for any $i,j \in [d_\lambda]$, in block $\lambda$ of \cref{eq:ii}.
While we do not have an expression for $\varepsilon_{ij}$, knowing $\varepsilon_i$ and $\varepsilon_j$ is enough to diagrammatically extract $B_{ij}$.
This can be done via the following algorithm:
\begin{enumerate}
    \item For every $i \in [d_\lambda]$, diagrammatically compute $\varepsilon_1 B \varepsilon_i$ and $\varepsilon_i B \varepsilon_1$. Since $\psi(\varepsilon_1 B \varepsilon_i) \cdot \psi(\varepsilon_i B \varepsilon_1) = B_{1i} B_{i1} \cdot \psi(\varepsilon_1)$ we can diagrammatically compute
    \begin{equation}
        B_{1i}B_{i1} = \frac{\Tr\of[\big]{(\varepsilon_1 B \varepsilon_i) \cdot (\varepsilon_i B \varepsilon_1)}}{\Tr \of{\varepsilon_1}}.
        \label{eq:bilinear}
    \end{equation}
    \item Since the element $B$ is hermitian with real coefficients, $B_{1i} = B_{i1}$ as real numbers. From \cref{eq:bilinear} we can set 
    \begin{align}
        B_{1i} = B_{i1} \defeq \sqrt{\frac{\Tr\of[\big]{(\varepsilon_1 B \varepsilon_i) \cdot (\varepsilon_i B \varepsilon_1)}}{\Tr \of{\varepsilon_1}}}
    \end{align}
    \item For every $i \in [d_\lambda]$ we set
    \begin{align}
        \varepsilon_{1i} \defeq \frac{\varepsilon_1 B \varepsilon_i}{B_{1i}}, \qquad
        \varepsilon_{i1} \defeq \frac{\varepsilon_i B \varepsilon_1}{B_{i1}}.
    \end{align}
    \item Once we know all of the $\varepsilon_{1i}$ and $\varepsilon_{i1}$, we can diagrammatically compute
    \begin{equation}
        B_{ij} = \frac{\Tr \of[\big]{\varepsilon_{1i} B \varepsilon_{j1}}}{\Tr \of{\varepsilon_1}}
    \end{equation}
    for every $i,j \in [d_\lambda]$. We can also compute $\varepsilon_{ij} = \varepsilon_{i1}\varepsilon_{1j}$ for every $i,j \in [d_\lambda]$.
\end{enumerate}

Since the multiplication of two arbitrary linear combinations of diagrams has complexity $O\of[\big]{\dim(\B^d_{p,q})^2}$, the complexity $O\of[\big]{\dim(\B^d_{p,q})^2 \dim(\A^d_{p,q})}$ of the above algorithm does not depend on $d$ asymptotically since
\begin{equation}
    O\of[\big]{((p+q)!)^2 \dim(\A^d_{p,q})} \leq O\of[\big]{((p+q)!)^3},
\end{equation}
where we used
$\dim(\A^d_{p,q}) \leq \dim(\B^d_{p,q}) = (p+q)!$
(that is quite crude bound for small $d$).

\newpage
\section{Numerical values for the number of variables \texorpdfstring{$n$}{n}}
\label{apx:numeric tables}

\begin{table}[ht]
\hspace*{-2.75cm}
\begin{tabular}{|cc||c|c|c|c|c|c|c|c|c|c|c|c|c|c|c|c|c|c|c|c|c|c|c|c|c|}
\hline
\multicolumn{2}{|c||}{$\makecell{p \\ q}$} & $\makecell{1 \\ 1}$ & $\makecell{1 \\ 2}$ & $\makecell{1 \\ 3}$ & $\makecell{2 \\ 2}$ & $\makecell{1 \\ 4}$ & $\makecell{2 \\ 3}$ & $\makecell{1 \\ 5}$ & $\makecell{2 \\ 4}$ & $\makecell{3 \\ 3}$ & $\makecell{1 \\ 6}$ & $\makecell{2 \\ 5}$ & $\makecell{3 \\ 4}$ & $\makecell{1 \\ 7}$ & $\makecell{2 \\ 6}$ & $\makecell{3 \\ 5}$ & $\makecell{4 \\ 4}$ & $\makecell{1 \\ 8}$ & $\makecell{2 \\ 7}$ & $\makecell{3 \\ 6}$ & $\makecell{4 \\ 5}$ & $\makecell{1 \\ 9}$ & $\makecell{2 \\ 8}$ & $\makecell{3 \\ 7}$ & $\makecell{4 \\ 6}$ & $\makecell{5 \\ 5}$ \\ \hline \hline
\multicolumn{1}{|c|}{\multirow{9}{*}{$d$}} & $2$ & \multirow{9}{*}{$2$} & $2$ & $3$ & $3$ & $3$ & $3$ & $4$ & $4$ & $4$ & $4$ & $4$ & $4$ & $5$ & $5$ & $5$ & $5$ & $5$ & $5$ & $5$ & $5$ & $6$ & $6$ & $6$ & $6$ & $6$ \\ \cline{2-2} \cline{4-27}
\multicolumn{1}{|c|}{} & $3$ &  & \multirow{8}{*}{$3$} & $4$ & $5$ & $6$ & $6$ & $7$ & $8$ & $8$ & $9$ & $10$ & $10$ & $11$ & $12$ & $12$ & $13$ & $13$ & $14$ & $15$ & $15$ & $15$ & $17$ & $17$ & $18$ & $18$ \\ \cline{2-2} \cline{5-27}
\multicolumn{1}{|c|}{} & $4$ &  &  & \multirow{7}{*}{$5$} & \multirow{7}{*}{$6$} & $7$ & $8$ & $10$ & $12$ & $12$ & $13$ & $15$ & $16$ & $17$ & $21$ & $21$ & $23$ & $21$ & $25$ & $27$ & $28$ & $27$ & $32$ & $34$ & $37$ & $36$ \\ \cline{2-2} \cline{7-27}
\multicolumn{1}{|c|}{} & $5$ &  &  &  &  & \multirow{6}{*}{$8$} & \multirow{6}{*}{$9$} & $11$ & $14$ & $14$ & $16$ & $19$ & $21$ & $21$ & $27$ & $28$ & $31$ & $28$ & $35$ & $39$ & $41$ & $36$ & $46$ & $50$ & $56$ & $54$ \\ \cline{2-2} \cline{9-27}
\multicolumn{1}{|c|}{} & $6$ &  &  &  &  &  &  & \multirow{5}{*}{$12$} & \multirow{5}{*}{$15$} & \multirow{5}{*}{$15$} & $17$ & $21$ & $23$ & $24$ & $31$ & $33$ & $37$ & $32$ & $41$ & $47$ & $50$ & $43$ & $57$ & $63$ & $72$ & $70$ \\ \cline{2-2} \cline{12-27}
\multicolumn{1}{|c|}{} & $7$ &  &  &  &  &  &  &  &  &  & \multirow{4}{*}{$18$} & \multirow{4}{*}{$22$} & \multirow{4}{*}{$24$} & $25$ & $33$ & $35$ & $39$ & $35$ & $45$ & $52$ & $56$ & $47$ & $63$ & $71$ & $82$ & $80$ \\ \cline{2-2} \cline{15-27}
\multicolumn{1}{|c|}{} & $8$ &  &  &  &  &  &  &  &  &  &  &  &  & \multirow{3}{*}{$26$} & \multirow{3}{*}{$34$} & \multirow{3}{*}{$36$} & \multirow{3}{*}{$40$} & $36$ & $47$ & $54$ & $58$ & $50$ & $67$ & $76$ & $88$ & $86$ \\ \cline{2-2} \cline{19-27}
\multicolumn{1}{|c|}{} & $9$ &  &  &  &  &  &  &  &  &  &  &  &  &  &  &  &  & \multirow{2}{*}{$37$} & \multirow{2}{*}{$48$} & \multirow{2}{*}{$55$} & \multirow{2}{*}{$59$} & $51$ & $69$ & $78$ & $90$ & $88$ \\ \cline{2-2} \cline{23-27}
\multicolumn{1}{|c|}{} & $10$ &  &  &  &  &  &  &  &  &  &  &  &  &  &  &  &  &  &  &  &  & $52$ & $70$ & $79$ & $91$ & $89$ \\ \hline
\end{tabular}
\caption{The number of variables $n^d_{p,q}$ in a unitary-equivariant LP with full walled Brauer algebra symmetry, see \cref{eq:n full Brauer}.}
\label{table:FwBa}
\end{table}

\definecolor{mygray}{rgb}{0.87,0.87,0.87}
\newcommand{\cc}{\cellcolor{mygray}}
\newcommand{\rc}{\rowcolor{mygray}}

\begin{table}[ht]
\begin{tabular}{|c|c||ccccccccc|}
\hline
 &  & \multicolumn{9}{c|}{$d$} \\ \hhline{|~|~||-|-|-|-|-|-|-|-|-|}
\multirow{-2}{*}{$p$} & \multirow{-2}{*}{$q$} & \multicolumn{1}{c|}{\cc2} & \multicolumn{1}{c|}{3} & \multicolumn{1}{c|}{4} & \multicolumn{1}{c|}{5} & \multicolumn{1}{c|}{6} & \multicolumn{1}{c|}{7} & \multicolumn{1}{c|}{8} & \multicolumn{1}{c|}{9} & 10 \\ \hline \hline
\rc
1 & 1 & \multicolumn{9}{c|}{\cc2} \\ \hline
\rc
1 & 2 & \multicolumn{1}{c|}{\cc3} & \multicolumn{8}{c|}{\cc4} \\ \hline
\rc
1 & 3 & \multicolumn{1}{c|}{\cc4} & \multicolumn{1}{c|}{\cc6} & \multicolumn{7}{c|}{\cc7} \\ \hline
\rc
2 & 2 & \multicolumn{1}{c|}{\cc6} & \multicolumn{1}{c|}{\cc9} & \multicolumn{7}{c|}{\cc10} \\ \hline
\rc
1 & 4 & \multicolumn{1}{c|}{\cc5} & \multicolumn{1}{c|}{\cc9} & \multicolumn{1}{c|}{\cc11} & \multicolumn{6}{c|}{\cc12} \\ \hline
\rc
2 & 3 & \multicolumn{1}{c|}{\cc7} & \multicolumn{1}{c|}{\cc14} & \multicolumn{1}{c|}{\cc17} & \multicolumn{6}{c|}{\cc18} \\ \hline
\rc
1 & 5 & \multicolumn{1}{c|}{\cc6} & \multicolumn{1}{c|}{\cc12} & \multicolumn{1}{c|}{\cc16} & \multicolumn{1}{c|}{\cc18} & \multicolumn{5}{c|}{\cc19} \\ \hline
\rc
2 & 4 & \multicolumn{1}{c|}{\cc10} & \multicolumn{1}{c|}{\cc22} & \multicolumn{1}{c|}{\cc30} & \multicolumn{1}{c|}{\cc33} & \multicolumn{5}{c|}{\cc34} \\ \hline
3 & 3 & \multicolumn{1}{c|}{\cc10} & \multicolumn{1}{c|}{25} & \multicolumn{1}{c|}{34} & \multicolumn{1}{c|}{37} & \multicolumn{5}{c|}{38} \\ \hline
\rc
1 & 6 & \multicolumn{1}{c|}{\cc7} & \multicolumn{1}{c|}{\cc16} & \multicolumn{1}{c|}{\cc23} & \multicolumn{1}{c|}{\cc27} & \multicolumn{1}{c|}{\cc29} & \multicolumn{4}{c|}{\cc30} \\ \hline
\rc
2 & 5 & \multicolumn{1}{c|}{\cc11} & \multicolumn{1}{c|}{\cc30} & \multicolumn{1}{c|}{\cc44} & \multicolumn{1}{c|}{\cc52} & \multicolumn{1}{c|}{\cc55} & \multicolumn{4}{c|}{\cc56} \\ \hline
3 & 4 & \multicolumn{1}{c|}{\cc13} & \multicolumn{1}{c|}{39} & \multicolumn{1}{c|}{60} & \multicolumn{1}{c|}{70} & \multicolumn{1}{c|}{73} & \multicolumn{4}{c|}{74} \\ \hline
\rc
1 & 7 & \multicolumn{1}{c|}{\cc8} & \multicolumn{1}{c|}{\cc20} & \multicolumn{1}{c|}{\cc31} & \multicolumn{1}{c|}{\cc38} & \multicolumn{1}{c|}{\cc42} & \multicolumn{1}{c|}{\cc44} & \multicolumn{3}{c|}{\cc45} \\ \hline
\rc
2 & 6 & \multicolumn{1}{c|}{\cc14} & \multicolumn{1}{c|}{\cc41} & \multicolumn{1}{c|}{\cc67} & \multicolumn{1}{c|}{\cc82} & \multicolumn{1}{c|}{\cc90} & \multicolumn{1}{c|}{\cc93} & \multicolumn{3}{c|}{\cc94} \\ \hline
3 & 5 & \multicolumn{1}{c|}{\cc16} & \multicolumn{1}{c|}{56} & \multicolumn{1}{c|}{96} & \multicolumn{1}{c|}{119} & \multicolumn{1}{c|}{129} & \multicolumn{1}{c|}{132} & \multicolumn{3}{c|}{133} \\ \hline
4 & 4 & \multicolumn{1}{c|}{\cc19} & \multicolumn{1}{c|}{66} & \multicolumn{1}{c|}{116} & \multicolumn{1}{c|}{143} & \multicolumn{1}{c|}{154} & \multicolumn{1}{c|}{157} & \multicolumn{3}{c|}{158} \\ \hline
\rc
1 & 8 & \multicolumn{1}{c|}{\cc9} & \multicolumn{1}{c|}{\cc25} & \multicolumn{1}{c|}{\cc41} & \multicolumn{1}{c|}{\cc53} & \multicolumn{1}{c|}{\cc60} & \multicolumn{1}{c|}{\cc64} & \multicolumn{1}{c|}{\cc66} & \multicolumn{2}{c|}{\cc67} \\ \hline
\rc
2 & 7 & \multicolumn{1}{c|}{\cc15} & \multicolumn{1}{c|}{\cc52} & \multicolumn{1}{c|}{\cc91} & \multicolumn{1}{c|}{\cc119} & \multicolumn{1}{c|}{\cc134} & \multicolumn{1}{c|}{\cc142} & \multicolumn{1}{c|}{\cc145} & \multicolumn{2}{c|}{\cc146} \\ \hline
3 & 6 & \multicolumn{1}{c|}{\cc19} & \multicolumn{1}{c|}{79} & \multicolumn{1}{c|}{148} & \multicolumn{1}{c|}{195} & \multicolumn{1}{c|}{219} & \multicolumn{1}{c|}{229} & \multicolumn{1}{c|}{232} & \multicolumn{2}{c|}{233} \\ \hline
4 & 5 & \multicolumn{1}{c|}{\cc22} & \multicolumn{1}{c|}{97} & \multicolumn{1}{c|}{189} & \multicolumn{1}{c|}{253} & \multicolumn{1}{c|}{282} & \multicolumn{1}{c|}{293} & \multicolumn{1}{c|}{296} & \multicolumn{2}{c|}{297} \\ \hline
\rc
1 & 9 & \multicolumn{1}{c|}{\cc10} & \multicolumn{1}{c|}{\cc30} & \multicolumn{1}{c|}{\cc53} & \multicolumn{1}{c|}{\cc71} & \multicolumn{1}{c|}{\cc83} & \multicolumn{1}{c|}{\cc90} & \multicolumn{1}{c|}{\cc94} & \multicolumn{1}{c|}{\cc96} & 97 \\ \hline
\rc
2 & 8 & \multicolumn{1}{c|}{\cc18} & \multicolumn{1}{c|}{\cc66} & \multicolumn{1}{c|}{\cc126} & \multicolumn{1}{c|}{\cc172} & \multicolumn{1}{c|}{\cc201} & \multicolumn{1}{c|}{\cc216} & \multicolumn{1}{c|}{\cc224} & \multicolumn{1}{c|}{\cc227} & 228 \\ \hline
3 & 7 & \multicolumn{1}{c|}{\cc22} & \multicolumn{1}{c|}{102} & \multicolumn{1}{c|}{213} & \multicolumn{1}{c|}{298} & \multicolumn{1}{c|}{347} & \multicolumn{1}{c|}{371} & \multicolumn{1}{c|}{381} & \multicolumn{1}{c|}{384} & 385 \\ \hline
4 & 6 & \multicolumn{1}{c|}{\cc28} & \multicolumn{1}{c|}{139} & \multicolumn{1}{c|}{306} & \multicolumn{1}{c|}{434} & \multicolumn{1}{c|}{505} & \multicolumn{1}{c|}{535} & \multicolumn{1}{c|}{546} & \multicolumn{1}{c|}{549} & 550 \\ \hline
5 & 5 & \multicolumn{1}{c|}{\cc28} & \multicolumn{1}{c|}{149} & \multicolumn{1}{c|}{332} & \multicolumn{1}{c|}{478} & \multicolumn{1}{c|}{556} & \multicolumn{1}{c|}{587} & \multicolumn{1}{c|}{598} & \multicolumn{1}{c|}{601} & 602 \\ \hline
\end{tabular}
\caption{The number of variables $n^d_{p,q}$ in a unitary-equivariant SDP with an $\S_p \times \S_q$ symmetry, see \cref{eq:n Spq}. Such SDP reduces to an LP when $\min(p,q) \leq 2$ or $d=2$, see \cref{apx:restriction to Spq}. These cases are highlighted in grey.}
\label{table:SpSq}
\end{table}

\begin{table}[ht]
\begin{tabular}{|c|c||c|c|c|c|c|c|c|c|c|}
\hline
\multicolumn{2}{|c||}{$p+q$} & $2$ & $3$ & $4$ & $5$ & $6$ & $7$ & $8$ & $9$ & $10$ \\ \hline \hline
\multirow{9}{*}{$d$} & 2 & \multirow{9}{*}{$2$} & $3$ & $6$ & $10$ & $20$ & $35$ & $70$ & $126$ & $252$ \\ \cline{2-2} \cline{4-11}
 & 3 &  & \multirow{8}{*}{$4$} & $9$ & $21$ & $51$ & $127$ & $323$ & $835$ & $\num{2188}$ \\ \cline{2-2} \cline{5-11}
 & 4 &  &  & \multirow{7}{*}{$10$} & $25$ & $70$ & $196$ & $588$ & $\num{1764}$ & $\num{5544}$ \\ \cline{2-2} \cline{6-11}
 & 5 &  &  &  & \multirow{6}{*}{$26$} & $75$ & $225$ & $715$ & $\num{2347}$ & $\num{7990}$ \\ \cline{2-2} \cline{7-11}
 & 6 &  &  &  &  & \multirow{5}{*}{$76$} & $231$ & $756$ & $\num{2556}$ & $\num{9096}$ \\ \cline{2-2} \cline{8-11}
 & 7 &  &  &  &  &  & \multirow{4}{*}{$232$} & $763$ & $\num{2611}$ & $\num{9415}$ \\ \cline{2-2} \cline{9-11}
 & 8 &  &  &  &  &  &  & \multirow{3}{*}{$764$} & $\num{2619}$ & $\num{9486}$ \\ \cline{2-2} \cline{10-11}
 & 9 &  &  &  &  &  &  &  & \multirow{2}{*}{$\num{2620}$} & $\num{9495}$ \\ \cline{2-2} \cline{11-11}
 & 10 &  &  &  &  &  &  &  &  & $\num{9496}$ \\ \hline
\end{tabular}
\caption{The number of variables $n^d_{p,q}$ in a unitary-equivariant LP with Gelfand--Tsetlin symmetry, see \cref{eq:n GT}.}
\label{table:GT}
\end{table}

\begin{table}[ht]
\begin{tabular}{|c|c||c|c|c|c|c|c|c|c|c|} \hline
\multicolumn{2}{|c||}{$p+q$} & $2$ & $3$ & $4$ & $5$ & $6$ & $7$ & $8$ & $9$ & $10$ \\ \hline \hline
\multirow{9}{*}{$d$} & 2 & \multirow{9}{*}{$\num{2}$} & $\num{5}$ & $\num{14}$ & $\num{42}$ & $\num{132}$ & $\num{429}$ & $\num{1430}$ & $\num{4862}$ & $\num{16796}$ \\ \cline{2-2} \cline{4-11}
 & 3 &  & \multirow{8}{*}{$\num{6}$} & $\num{23}$ & $\num{103}$ & $\num{513}$ & $\num{2761}$ & $\num{15767}$ & $\num{94359}$ & $\num{586590}$ \\ \cline{2-2} \cline{5-11}
 & 4 &  &  & \multirow{7}{*}{$\num{24}$} & $\num{119}$ & $\num{694}$ & $\num{4582}$ & $\num{33324}$ & $\num{261808}$ & $\num{2190688}$ \\ \cline{2-2} \cline{6-11}
 & 5 & & &  & \multirow{6}{*}{$\num{120}$} & $\num{719}$ & $\num{5003}$ & $\num{39429}$ & $\num{344837}$ & $\num{3291590}$ \\ \cline{2-2} \cline{7-11}
 & 6 &  &  &  &  & \multirow{5}{*}{$\num{720}$} & $\num{5039}$ & $\num{40270}$ & $\num{361302}$ & $\num{3587916}$ \\ \cline{2-2} \cline{8-11}
 & 7 &  &  &  &  &  & \multirow{4}{*}{$\num{5040}$} & $\num{40319}$ & $\num{362815}$ & $\num{3626197}$ \\ \cline{2-2} \cline{9-11}
 & 8 &  &  &  &  &  &  & \multirow{3}{*}{$\num{40320}$} & $\num{362879}$ & $\num{3628718}$ \\ \cline{2-2} \cline{10-11}
 & 9 &  &  &  &  &  &  &  & \multirow{2}{*}{$\num{362880}$} & $\num{3628799}$ \\ \cline{2-2} \cline{11-11}
 & 10 &  &  &  &  &  &  &  &  & $\num{3628800}$ \\ \hline
\end{tabular}
\caption{The number of variables $n^d_{p,q}  = \dim(\A^d_{p,q})$ in a unitary-equivariant SDP with no additional symmetry.}
\label{table:SDP}
\end{table}

\begin{table}[ht]
\begin{tabular}{|c|c||c|c|c|c|c|c|c|c|c|} \hline
\multicolumn{2}{|c||}{$p+q$} & $2$ & $3$ & $4$ & $5$ & $6$ & $7$ & $8$ & $9$ & $10$ \\ \hline \hline
\multirow{9}{*}{$d$} & 2 & $\num{1.20}$ & $\num{1.81}$ & $\num{2.41}$ & $\num{3.01}$ & $\num{3.61}$ & $\num{4.21}$ & $\num{4.82}$ & $\num{5.42}$ & $\num{6.02}$ \\ \cline{2-11}
 & 3 & $\num{1.91}$ & $\num{2.86}$ & $\num{3.82}$ & $\num{4.77}$ & $\num{5.73}$ & $\num{6.68}$ & $\num{7.63}$ & $\num{8.59}$ & $\num{9.54}$ \\ \cline{2-11}
 & 4 & $\num{2.41}$ & $\num{3.61}$ & $\num{4.82}$ & $\num{6.02}$ & $\num{7.22}$ & $\num{8.43}$ & $\num{9.63}$ & $\num{10.84}$ & $\num{12.04}$ \\ \cline{2-11}
 & 5 & $\num{2.80}$ & $\num{4.19}$ & $\num{5.59}$ & $\num{6.99}$ & $\num{8.39}$ & $\num{9.79}$ & $\num{11.18}$ & $\num{12.58}$ & $\num{13.98}$ \\ \cline{2-11}
 & 6 & $\num{3.11}$ & $\num{4.67}$ & $\num{6.23}$ & $\num{7.78}$ & $\num{9.34}$ & $\num{10.89}$ & $\num{12.45}$ & $\num{14.01}$ & $\num{15.56}$ \\ \cline{2-11}
 & 7 & $\num{3.38}$ & $\num{5.07}$ & $\num{6.76}$ & $\num{8.45}$ & $\num{10.14}$ & $\num{11.83}$ & $\num{13.52}$ & $\num{15.21}$ & $\num{16.90}$ \\ \cline{2-11}
 & 8 & $\num{3.61}$ & $\num{5.42}$ & $\num{7.22}$ & $\num{9.03}$ & $\num{10.84}$ & $\num{12.64}$ & $\num{14.45}$ & $\num{16.26}$ & $\num{18.06}$ \\ \cline{2-11}
 & 9 & $\num{3.82}$ & $\num{5.73}$ & $\num{7.63}$ & $\num{9.54}$ & $\num{11.45}$ & $\num{13.36}$ & $\num{15.27}$ & $\num{17.18}$ & $\num{19.08}$ \\ \cline{2-11}
 & 10 & $\num{4.00}$ & $\num{6.00}$ & $\num{8.00}$ & $\num{10.00}$ & $\num{12.00}$ & $\num{14.00}$ & $\num{16.00}$ & $\num{18.00}$ & $\num{20.00}$ \\ \hline
\end{tabular}
\caption{The logarithm of the number of variables $\log_{10}(d^{2(p+q)})$ in a naive implementation of the SDP~\eqref{eq:input SDP}.}
\label{table:SDP_naive}
\end{table}

\end{document}